\pdfoutput=1
\documentclass[a4paper,american,floatfix,pdftex,superscriptaddress,twoside,%
aps,pra,%
reprint,twocolumn,final
]{revtex4-2}%
\usepackage{algorithm}
\usepackage{amsfonts,amsmath,amssymb}
\usepackage{booktabs}
\usepackage{enumitem}
\usepackage[T1]{fontenc}
\usepackage{graphicx}%
\usepackage[utf8]{inputenc}
\usepackage{mathtools}%
\usepackage{tabularx}
\usepackage{hyperref, hypernat}
\usepackage[todos]{CMImacros}

\newcolumntype{R}{>{\raggedleft\arraybackslash}X}
\graphicspath{{./}{./fig/}}

\makeatletter
\renewcommand{\p@subsection}{}
\renewcommand{\p@subsubsection}{}
\makeatother

\newcommand*{\cminject}{\textit{CMInject}\xspace}
\newcommand*{\cminjectvis}{\texttt{cminject\_visualize}\xspace}
\newcommand*{\numpy}{\textit{NumPy}\xspace}

\newcommand*{\numba}{\textit{Numba}\xspace}
\newcommand*{\cython}{\textit{Cython}\xspace}

\newcommand{\cfeldesy}{\affiliation{Center for Free-Electron Laser Science, Deutsches
      Elektronen-Synchrotron DESY, Notkestraße 85, 22607 Hamburg, Germany}}%

\newcommand{\uhhcui}{\affiliation{Center for Ultrafast Imaging, Universität Hamburg, Luruper
      Chaussee 149, 22761 Hamburg, Germany}}%
\newcommand{\uhhphys}{\affiliation{Department of Physics, Universität Hamburg, Luruper Chaussee 149,
      22761 Hamburg, Germany}}%
\newcommand{\groco}{\affiliation{Department of Sciences, University College Groningen,
      University~of~Groningen, Hoendiepskade 23/24, 9718 BG Groningen, Netherlands}}%
\newcommand{\groinst}{\affiliation{Groningen Biomolecular Sciences and Biotechnology Institute,
      University~of~Groningen, Nijenborgh 4, 9747 AG Groningen, Netherlands}}%

\newcommand{\jkemail}{\email[Email:~]{jochen.kuepper@cfel.de}}%
\newcommand{\cmiweb}{\homepage[website:~]{https://www.controlled-molecule-imaging.org}}%

\begin{document}
\hyphenation{Brownian-Motion-Property-Updater}
\hyphenation{Simple-Z-Detector}\
\hyphenation{Stokes-Drag-Force-Field}
\hyphenation{Property-Updater}
\hyphenation{Result-Storage}
\hyphenation{CMInject}
\hyphenation{Ex-pe-ri-men-ta-lists}
\hyphenation{acc-cel-er-a-tions}

\title{\cminject: Python framework for the numerical simulation \\ of nanoparticle injection pipelines}%
\author{Simon~Welker}\cfeldesy%
\author{Muhamed~Amin}\cfeldesy\groco\groinst%
\author{Jochen~Küpper}\jkemail\cmiweb\cfeldesy\uhhcui\uhhphys%
\date{\today}
\begin{abstract}\noindent%
   \cminject simulates nanoparticle injection experiments of particles with diameters in the
   micrometer to nanometer-regime, \eg, for single-particle-imaging experiments. Particle-particle
   interactions and particle-induced changes in the surrounding fields are disregarded, due to low
   nanoparticle concentration in these experiments. \cminject's focus lies on the correct modeling
   of different forces on such particles, such as fluid-dynamics or light-induced interactions, to
   allow for simulations that further the scientific development of nanoparticle injection
   pipelines. To provide a usable basis for this framework and allow for a variety of experiments to
   be simulated, we implemented first specific force models: fluid drag forces, Brownian motion, and
   photophoretic forces. For verification, we benchmarked a drag-force-based simulation against a
   nanoparticle focusing experiment. We envision its use and further development by
   experimentalists, theorists, and software developers.
   \\[1ex]
   \\[1ex]
   \noindent
   \textbf{PROGRAM SUMMARY}\\
   \begin{footnotesize}\noindent%
      \emph{Program Title:} CMInject\\
      \emph{CPC Library link to program files:} (to be added by Technical Editor)\\
      \emph{Developer's repository link:} \url{https://github.com/cfel-cmi/cminject}\\
      \emph{Code Ocean capsule:} (to be added by Technical Editor)\\
      \emph{Licensing provisions:} GPLv3\\
      \emph{Programming language:} Python 3\\
      \emph{Supplementary material:} Code to reproduce and analyze simulation results, example input
      and output data, video files of trajectory movies
      \\
      \emph{Nature of problem:} Well-defined, reproducible, and interchangeable simulation setups of
      experimental injection pipelines for biological and artificial nanoparticles, in particular
      such pipelines that aim to advance the field of single-particle imaging.
      \\
      \emph{Solution method:} The definition and implementation of an extensible \emph{Python 3}
      framework to model and execute such simulation setups based on object-oriented software
      design, making use of parallelization facilities and modern numerical integration routines.
      \\
      \emph{Additional comments including restrictions and unusual features:} Supplementary
      executable scripts for quantitative and visual analyses of result data are also part of the
      framework.
   \end{footnotesize}
\end{abstract}
\maketitle

\section{Introduction}
Single-particle imaging (SPI) with x-ray beams is a relatively new
technique~\cite{Neutze:Nature406:752, Spence:PTRSB369:20130309} for the imaging of small particles
down to the size of single macromolecules, promising to image nanometer-sized particles without the
need for crystallization. In this context a ``particle'' can be anything from a small molecule to an
entire protein or an artificial nanoparticle. In SPI, a beam of x-rays illuminates single particles
in flight, with each particle hit by the x-ray pulse producing a diffraction pattern. From a
collection of such patterns gathered from many identical particles, the particle 3D structure can be
approximated. Substantial advances were made on the capabilities of x-ray free-electron lasers
(XFELs) in recent years~\cite{Emma:NatPhoton4:641, Decking:NatPhot14:391}, offering brilliant and
collimated ultra-short pulsed x-ray beams that can outrun radiation damage to the
sample~\cite{Neutze:Nature406:752, Lorenz:PRE86:051911} and allow for time-resolved imaging on
femtosecond timescales~\cite{Barty:ARPC64:415, Pande:Science352:725}.

There are multiple factors to consider for collecting and reconstructing electron densities and
molecular structures with high resolution: Incident x-ray intensity, experimental repetition rate,
and particle density in the interaction region. They all affect the quality of the reconstructed
structure: increasing the incident intensity results in more signal in each diffraction pattern, and
increasing the repetition rate or particle density results in more diffraction patterns being
collected in the same timespan. It was suggested that incident laser intensity is not the limiting
factor~\cite{Ourmazd:NatMeth16:941}, which was corroborated by showing that the level of signal
contained in collected patterns can be reduced drastically while maintaining good reconstruction
quality~\cite{Ayyer:OptExp27:37816}. However, a large number of good hits, \ie, diffraction patterns
of single particles inside the focus of the x-ray pulse, need to be collected in any case. It was
previously noted in the literature that ``different injection strategies to extend XFEL imaging to
smaller targets, such as single proteins'' are needed~\cite{Hantke:IUCr5:673}, and that
``improvements could be made through optimized focusing for the targeted size distribution or
cryogenic injection systems that additionally allow conformational
selection''~\cite{Ayyer:Optica8:15}. Therefore, there is an urgent need for novel optimized particle
injection systems.

To recover the 3D structure of the imaged particles from their 2D diffraction patterns,
sophisticated computer algorithms are used~\cite{Fung:NatPhys5:64, Bortel:JStructBiol:158,
   Ayyer:OptExp27:37816, Ayyer:Optica8:15}. These algorithms use diffraction patterns from
structurally identical particles. Thus, it is important to understand how the variation in
particles' sizes/shapes and structural conformations will affect their trajectories in the injection
system. These trajectories are also dependent on several experimental parameters, \eg, the geometry
of the injection system, the temperature and pressure of the guiding aerosols, and the initial phase
space distribution of the injected particles~\cite{Samanta:StructDyn7:024304, Roth:JAS124:17}.
Accordingly, selecting specific particle species, \eg, through the use of inhomogeneous electric
fields~\cite{Chang:IRPC34:557}, are an advanced topic for creating a high-quality particle beam.

A simulation framework provides a quick and efficient tool for searching the experimental
parameters' space and to produce optimized molecular/nanoparticle beams. Furthermore, the feedback
loop between simulation and experiment offers a road to progress in both theoretical and
experimental physics. Simulations are repeatedly used as a basis, supplementary, guiding, or
verification method in SPI research. Examples for this are (1) optimization of experimentally
verified aerodynamical injector designs for a variety of specific particle sizes and
materials~\cite{Liu:AST22:293, Wang:IJMS258:30, Wang:AST39:624, Roth:JAS124:17, Wang:AST40:320,
   Worbs:geomopt:inprep}, (2) exploration of the effects of experimental injection
parameters~\cite{Sobolev:CommPhys3:97} and types of injectors~\cite{Bielecki:SciAdv5:eaav8801} on
diffraction patterns, and (3) control of shock frozen isolated particles of both biological and
artificial origin~\cite{Samanta:StructDyn7:024304}. Progress in all of these areas was the
foundation of recent significant improvements of the amount of data that can be collected in a given
timeframe in SPI experiments~\cite{Ayyer:Optica8:15}.

\begin{figure}
   \includegraphics[width=\linewidth]{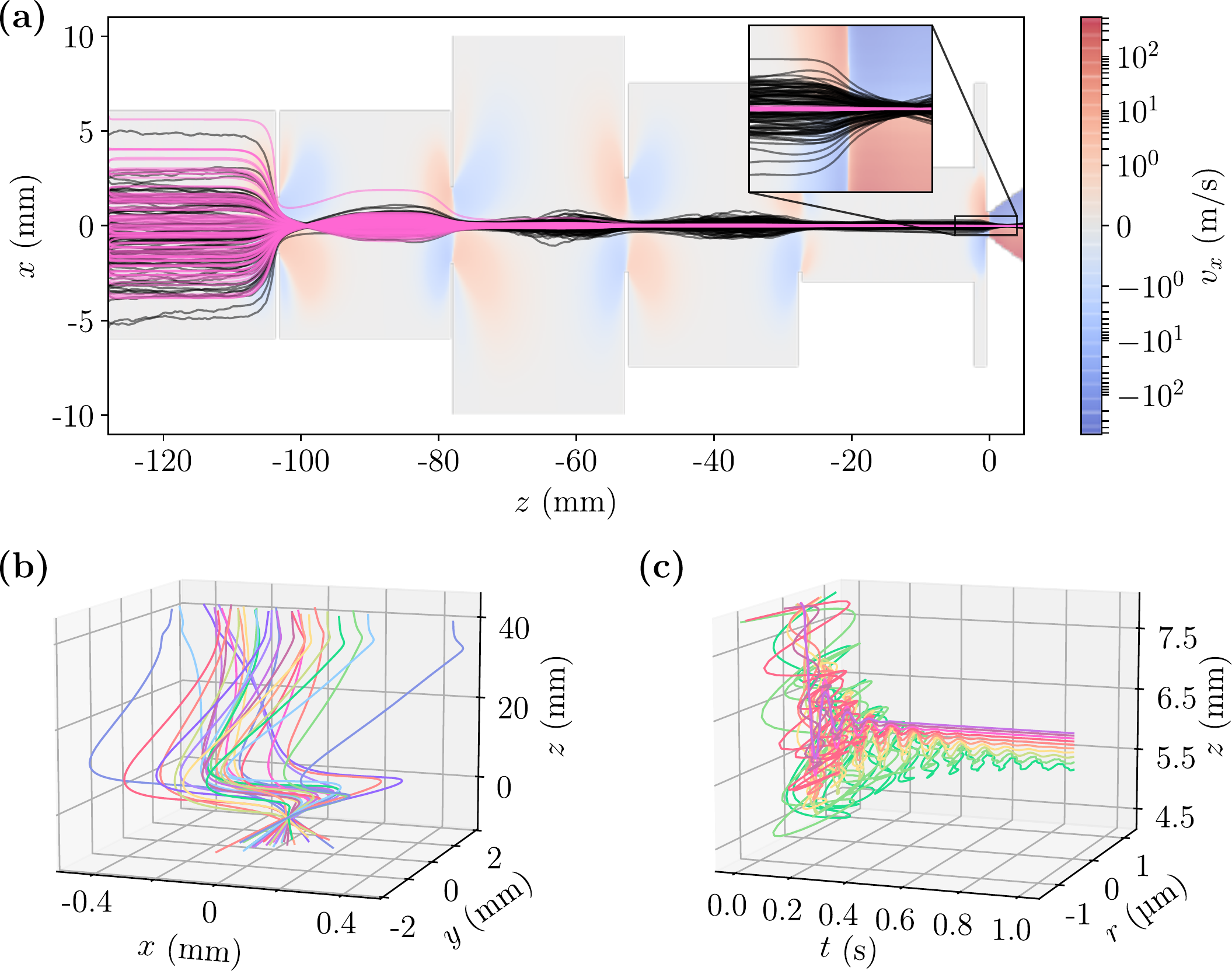}
   \caption{Example trajectory plots of experiments simulated with \cminject: \textbf{(a)} 2D
      trajectories from a simulated focusing experiment using an axisymmetric aerodynamic lens stack
      (ALS) to focus $d=27\,\text{nm}$ gold nanoparticles~\cite{Worbs:geomopt:inprep}. The
      simulations include (black) or disregard (pink) Brownian motion. The background shows the
      fluid's velocity in $x$ direction. \textbf{(b)} 3D trajectories of a focusing
      experiment~\cite{Samanta:StructDyn7:024304}, where a cooled buffer gas cell (BGC) focuses
      $d=490\,\text{nm}$ polystyrene nanoparticles by a flowing carrier gas at a cryogenic
      temperature (4$\,$K). The used force model is a new theoretical development for particles
      moving in the molecular flow regime and at low temperatures, see
      \autoref{sec:physmodels:microscopic-drag-force}~\cite{Roth:microscopic-drag-force:inprep}.
      \textbf{(c)} A qualitative reproduction of an optical trapping and levitation
      experiment~\cite{Eckerskorn:PRAppl4:064001}, showing the interplay of photophoretic
      forces~\cite{Desyatnikov:OptExp17:8201}, gravity, and air resistance. Particles with different
      masses settle into an equilibrium position over time.}
   \label{fig:overview-plots}
\end{figure}

Here, we introduce and describe \cminject, a computational framework that aims to be an extensible
basis for such simulations. \autoref{fig:overview-plots} depicts examples of simulation results,
indicating that recent developments, as well as future ideas, are supported by our framework.

\section{Problem description}
Creating high-quality nanoparticle beams poses diverse technical and scientific
challenges~\cite{Roth:JAS124:17, Bielecki:SciAdv5:eaav8801, Daurer:IUCrJ4:3,
  Samanta:StructDyn7:024304}. The development of improved or sample-adjusted injection
pipelines~\cite{Roth:JAS124:17} needs to be supported by a flexible and extensible simulation
package, which enables quantitative predictions of arbitrary nanoparticle injection pipelines.
Possibilities to easily implement additional virtual detectors, particle types, and force fields are
crucial for the usability in a wider scientific context. Capabilities for the subsequent
visualization and analysis of simulation results, on their own or in comparison with experimental
data, are also important.

Reasonable assumptions within the set of possible simulated experiments were made when designing the
initial computational framework presented here: (1) particles do not interact with each other, nor do
they affect the surrounding environment. (2) experiments have a designated, finite, spatial dimension
that the particle beam travels along.

Further points of interest for nanoparticle injection are the separation of species, \eg, by quantum
state, conformation, or enantiomer~\cite{Chang:IRPC34:557, Yachmenev:PRL123:243202}, the alignment
or orientation in space~\cite{Stapelfeldt:RMP75:543, Spence:PRL92:198102, Holmegaard:PRL102:023001,
   Kuepper:PRL112:083002, Karamatskos:NatComm10:3364}, or the preservation of native biological
structures~\cite{Miao:ARPC59:387, Miao:Nature400:342, Samanta:StructDyn7:024304}. These are not yet
implemented in \cminject and will not be discussed further in this paper, but we designed our
framework foreseeing corresponding as well as unforeseen extensions.

Furthermore, the framework must be usable by theorists and experimentalists alike in order to
evaluate and exchange ideas and experiments for nanoparticle injection. The framework should strike
a balance between expressiveness and processing requirements: a long procedural script, written with
optimized functions, might run simulations very quickly, but is likely incomprehensible to most
potential developers and users. A very general framework, while intuitively usable by users and
developers, might in turn require so much dynamicism in its implementation that simulations become
unsuitably slow.

\section{Framework description}
\cminject enables theorists and experimentalists to work together toward inventing or optimizing
nanoparticle injection pipelines~\cite{Welker:thesis:2019}. \cminject is written in Python~3 and its
design follows an object-oriented paradigm. Most objects in this framework represent real-world
counterparts that are present in actual experiments. For example, a user might create a
\texttt{Setup} instance, passing along one or more \texttt{Source} instances that generate
particles, and one or more \texttt{Device} instances that affect particles throughout their
simulated trajectories by simulating physical forces. The user can \texttt{run()} a concrete
\texttt{Experiment} constructed by the \texttt{Setup} and observe the returned results: a list of
\texttt{Particle} instances that have been updated and, if desired, tracked along each particle's
trajectory.

\cminject does not impose many explicit constraints on how specific objects need to behave, it only
requires that all parts of an \texttt{Experiment} work with each other in a well-defined way. For
example, while all currently implemented sub-types of \texttt{Particle} are spherical objects,
\cminject is in principle agnostic to the particle shape. If someone wished to, for example,
simulate elliptical articles in a fluid flow, they could do so by (i) defining a \texttt{Particle}
subclass \texttt{EllipticalParticle} with additional shape parameters, \eg, $r_x$, $r_y$, $r_z$ for
an ellipsoid and (ii) deriving an implementation of the \texttt{FluidFlowField} class to be able to
handle these new particles by an appropriate force model.

Going further, one could even implement the manipulation of molecules by electric fields using the
quantum-mechanical Stark effect~\cite{Filsinger:JCP131:064309, Chang:IRPC34:557}, something we are
foreseeing for the near future.

\subsection{Framework structure}
\cminject's framework structure consists of:
\begin{enumerate}
\item a set of abstract definitions corresponding to real-world experimental objects, with a
   prescribed way of constructing a \emph{virtual experiment} out of these objects.
\item a parallelized routine that uses numerical integration to generate particle trajectories
   through a virtual experiment.
\item supplementary executable scripts, mostly for the analysis of result data.
\item implementations of the abstract definitions for the concrete physical models listed in
   \autoref{sec:physmodels}.
\end{enumerate}

\subsubsection{Base class definitions}
\label{sec:base-class-definitions}
The following list provides the base classes~\cite{python:website:PEP3119} of \cminject implemented
in the \texttt{cminject.base} and \texttt{cminject.experiment} modules, including brief versions of
their documentation. The full documentation is attached in the supplementary materials and updated
versions are available at \url{https://cminject.readthedocs.org}.

\textit{\texttt{cminject.base}}:
\begin{itemize}
\item \texttt{Particle}: A particle whose trajectory we want to simulate. First and foremost a
   simple data container.
\item \texttt{Field}: An acceleration field acting on \texttt{Particle}s.
\item \texttt{Action}: Updates the properties of a \texttt{Particle} after each integration step.
   Useful for changes over time that are not described by the ordinary differential equations in
   \autoref{sec:newton-motion}.
\item \texttt{Boundary}: A spatial boundary, evaluates if a \texttt{Particle} is inside its spatial
   extend.
\item \texttt{Device}: A combination of \texttt{Field}s, \texttt{Action}s, and a \texttt{Boundary},
   modeling real-world experimental devices. Applies the effects of its \texttt{Field}s and
   \texttt{Action}s only if a particle is inside of its \texttt{Boundary}; otherwise does not affect
   the particle in any way.
\item \texttt{Detector}: Evaluates if and where a \texttt{Particle} interacted with it.
\item \texttt{ResultStorage}: Stores experiment results to disk, and offers convenience methods to
   read them back into memory later.
\item \texttt{Setup}: Akin to a laboratory experimental setup with changeable pieces and parameters.
   Exposes a set of parameters that can be changed by the user, and constructs an
   \texttt{Experiment} instance from them that can then be simulated.
\end{itemize}

\textit{\texttt{cminject.experiment}}:
\begin{itemize}
\item \texttt{Experiment}: Akin to a real-world experiment which has a fixed set of sources,
   devices, detectors, and experimental parameters. Contains one or more instances of all of the
   classes from \texttt{cminject.base} listed above (except for \texttt{Setup}, which constructs
   \texttt{Experiment} instances). Constitutes the entry point for simulation, and returns the
   results.
\end{itemize}

\subsubsection{Numerical and technical implementation}
\label{sec:numerical-technical-impl}
To numerically solve the particle trajectories for any virtual experiment, we used the numerical
integration routine LSODA~\cite{Petzold:SIAM:4:136} as offered by the \texttt{scipy.ode}
module~\cite{Virtanen:NatMeth17:261}. This routine was chosen for its automatic method switching for
stiff and non-stiff problems~\cite{Petzold:SIAM:4:136}, which is very useful in our generic
multiphysics framework where various forces make up an ODE system that can exhibit different degrees
of stiffness at different positions in space of the same experiment.

These integration calculations are, as well as most other calculations in \cminject, heavily based
on \numpy arrays~\cite{vanderWalt:CSE13:22}. We wrote a parallel implementation based on the
\texttt{multiprocessing} module offered by the Python~3 core library, letting simulated particles be
processed in parallel by a pool of $\omega\in\mathbb{N}$ worker processes, where by default $\omega$
is the number of available CPU cores. We use the automatic optimization library
\numba~\cite{Lam:LLVM15:7} as well as the compiled language \cython~\cite{Behnel:CompSciEng13:31},
for automatic and manual optimization of the calculation functions, respectively.

\subsubsection{Executable scripts}
\label{sec:executables}
\cminject is supported by a collection of executable scripts. The main program, \texttt{cminject},
simulates a specified setup, passing along mandatory and optional parameters and providing
documentation for them if needed, and writes the results to a specified output (HDF5) file.
\cminjectvis and \texttt{cminject\_analyze-asymmetry} support the user's analysis of the result
files: They provide visualization and metrics of beam profile asymmetry, respectively. Documentation
for all utility programs is provided with the software.

\subsection{Program flow of simulation runs}
\label{sec:program-flow}
To provide a foundation for further discussion of the generality and possible improvements, we
provide a description of the general program flow of a \cminject simulation. A listing of the steps
involved in the current implementation is given in \autoref{algo:program-flow}. To clarify the short
descriptions given there, we note the following: A particle is considered ``done'' if it is outside
of all \texttt{Boundary} objects, or if its current time is outside of the simulation
time-window. Whether integration is successful is determined by the integrator. When a
\texttt{Detector} detects a given particle, it stores an detection event on the \texttt{Particle},
so this event is stored in the result list. \texttt{Action}s can change a particle's phase space
position, and if this happens, it is taken into account for the integration routine by resetting the
integrator accordingly.
\begin{algorithm}[H]
\caption{Program flow of a \cminject simulation.}\label{algo:program-flow}
\begin{enumerate}[leftmargin=1.5em]
  \item Get particles from all \texttt{Source}s, merge into one list
  \item Initialize an empty result list
  \item For each particle, parallelized via \texttt{multiprocessing.Pool}:
    \begin{enumerate}[leftmargin=1.5em]
      \item Initialize integrator: $t=t_0, x=x_0$
      \item If particle ``not done'' and integration successful:
        \begin{itemize}[leftmargin=1.5em]
          \item Update particle phase space position from integrator
          \item Update done-ness of particle using every \texttt{Boundary}
          \item Let each \texttt{Action} update the particle
          \item If particle position changed, reset the integrator
          \item Let each \texttt{Detector} try to detect the particle
          \item Update $t$, by incrementing it by the time step $dt$
          \item Run the integrator until $t$. At each evaluated point:
            \begin{itemize}[leftmargin=1.5em]
              \item Consult each \texttt{Device}'s applicable \texttt{Field}s
              \item Sum all accelerations
            \end{itemize}
          \item Go to (b)
        \end{itemize}
      \item Store fully simulated particle in the result list
    \end{enumerate}
  \item Store the result list as an HDF5 file
\end{enumerate}
\end{algorithm}

\begin{figure}[b]
   \includegraphics[width=\linewidth]{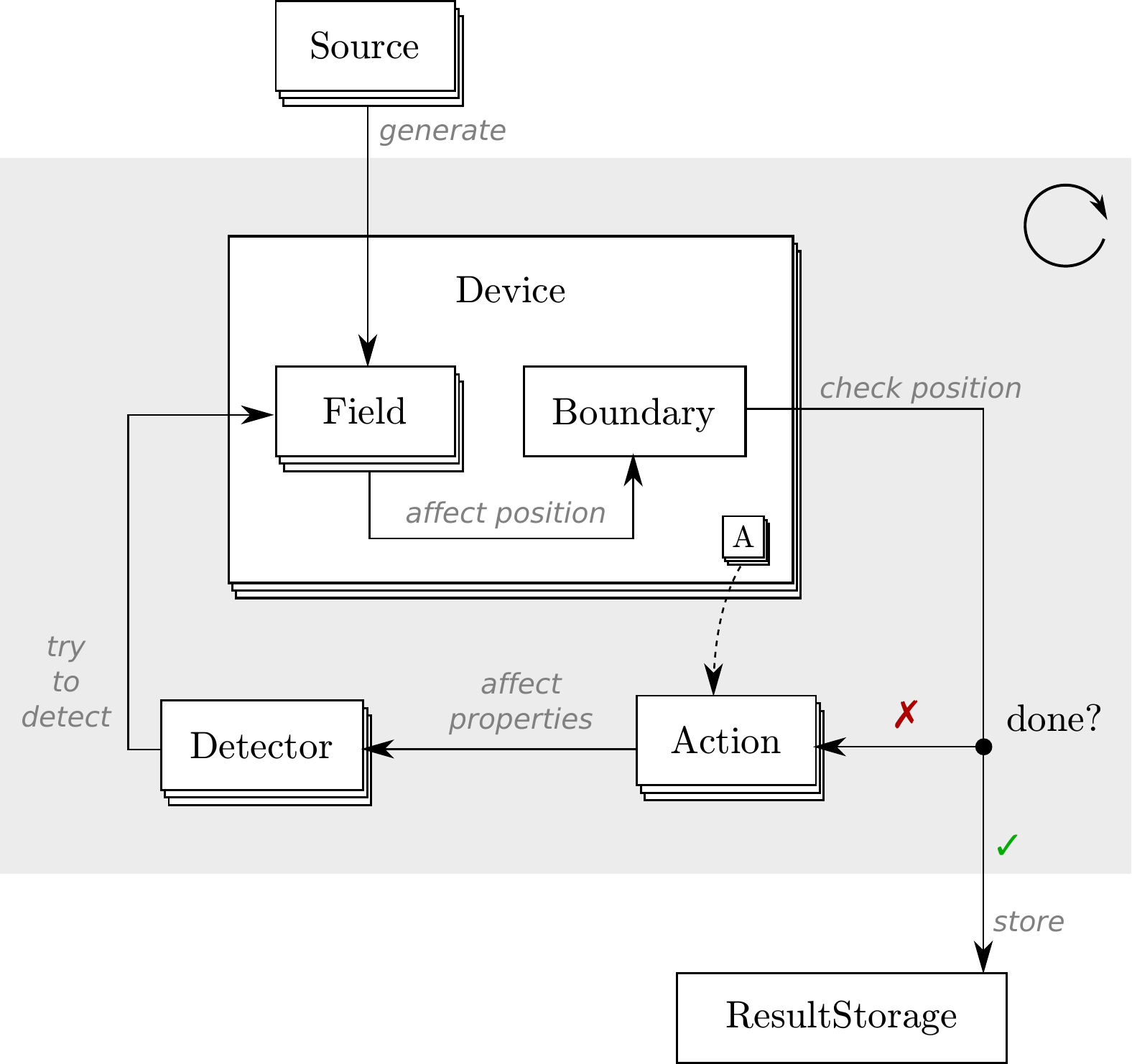}
   \caption{Conceptual program flow of a particle simulation with \cminject, following a single
     particle through a collection of objects instantiated from the classes provided by the
     framework. Solid arrows follow the particle's path; their grey annotations show the effect each
     object can have on the particle. The shaded background indicates the integration loop, which is
     repeated until the particle's simulation is considered ``done''. Classes displayed as a stack
     of layered rectangles, like \texttt{Source}, imply that a simulation can contain more than one
     object of such a class. The stack simply labeled ``A'' and the dashed arrow to the
     \texttt{Action} stack indicate that each \texttt{Device} can contain \texttt{Action}s, which
     only affect particles inside that \texttt{Device}.}
   \label{fig:objflow}
\end{figure}
\autoref{fig:objflow} simplifies the description given in the step-by-step listing,
\autoref{algo:program-flow} to a higher-level form and omits implementation details in favor of
general concept. We anticipate that the community will discuss and optimize, or even replace, the
concrete implementation further, while keeping the conceptual program flow as illustrated in
\autoref{fig:objflow} fairly consistent across future versions of \cminject.

\subsection{Physics models}
\label{sec:physmodels}
This first release of \cminject provides several physical models that are briefly described in the
following.

\subsubsection{Newton's equations of motion}
\label{sec:newton-motion}
We treat particle trajectories as a collection of incremental numerical solutions to the initial
value problem:
\begin{equation}
   \begin{aligned}
      \phi'(t) &= f(t, \phi(t))\\
      \phi(t) &\coloneqq (x, y, z, v_x, v_y, v_z)^T(t)\\
      \phi(0) &\coloneqq \phi_0 = (x_0, y_0, z_0, v_{x,0}, v_{y,0}, v_{z,0})^T\\
      f(t, \phi(t)) &\coloneqq (v_x, v_y, v_z, a_x, a_y, a_z)^T(t)
   \end{aligned}
\end{equation}
$\phi$ is a time-dependent vector in $(2n)$-dimensional phase-space, with $n=3$ in the general case
or $n=2$ for axially symmetrical simulations. $v_i$ are the velocities and $a_i$ the accelerations
corresponding to spatial dimension $i$. $\vec{a}=\vec{F}/m_p$, with the total force $\vec{F}$
exerted on a particle having mass $m_p$.

\subsubsection{Stokes' drag force}
\label{sec:physmodels:stokes-drag-force}
We use Stokes' model for the drag force of an isolated spherical particle embedded in a flowing
medium~\cite{Stokes:TCaPS9:8} for very small Reynolds numbers $\mathrm{Re}\ll1$, which is essential
to our simulations of aerodynamical focusing. It is formulated in terms of the fluid dynamic
viscosity $\mu$, particle radius $r$, particle mass $m$, difference in velocity between fluid and
particle $\Delta{v}$, and a Cunningham slip-correction factor $C_c$~\cite{Cunningham:PRSA83:357}.
\begin{equation}
   \vec{F}_{\text{Stokes}} = \frac{6\pi\mu r \vec{\Delta{v}}}{C_c}
   \label{eq:stokes-force}
\end{equation}

\subsubsection{Brownian motion}
\label{sec:physmodels:brownian-motion}
Since we model nanoparticle injection, Brownian motion becomes non-negligible, especially for
smaller nanoparticles. The model for Brownian motion used is that of a Gaussian white-noise random
process with a spectral intensity $S_0$ taken from Li and Ahmadi~\cite{Li:AST16:209}.
\begin{align}
  \vec{a}_{\text{Brown}} &= \vec{\mathcal{N}}(0, 1, k)\sqrt{\frac{\pi S_0}{\Delta{t}}} \\
  S_0 &= \frac{216\mu{}k_BT}{\pi^2(2r)^5\rho^2C_c} \notag
\end{align}
$\vec{\mathcal{N}}(0,1,k)$ denotes a vector of $k$ entries, each being independently and randomly
drawn from a zero mean unit variance normal distribution. $\Delta{t}$ is the duration of the
time-interval over the force should be calculated, which is the time increment of each integration
step. $r, m, \mu$ and $C_c$ are the same quantities as defined in
\autoref{sec:physmodels:stokes-drag-force}. $k_B$ is the Boltzmann constant, $T$ is the temperature
of the fluid, and $\rho$ is the density of the particle material.

\subsubsection{Microscopic drag force}
\label{sec:physmodels:microscopic-drag-force}
For the simulation of nanoparticles moving through fluids with a wide range of pressures,
velocities, and temperatures, Stokes' drag force is often not well applicable. Thus, a new drag
force model based on the kinetic theory of gases was
developed~\cite{Roth:microscopic-drag-force:inprep}. The original
formulation~\cite{Epstein:PR23:710} of this model was extended to broad sets of conditions
encountered in novel nanoparticle injection experiments, for instance, temperatures as low as
4~K~\cite{Samanta:StructDyn7:024304}. This force is defined as a combination of $10~\%$ specular
reflections and $90~\%$ diffuse reflections and takes into account the time-dependent temperature
difference between the injected particles and the fluid. An accompanying model for Brownian motion
was also provided~\cite{Roth:microscopic-drag-force:inprep}.

\subsubsection{Photophoretic force}
\label{sec:physmodels:photophoretic-force}
Furthermore, a model of the photophoretic force, \ie, the force of the surrounding gas exerted on an
anisotropically radiatively-heated particle. This has found various applications in the physical and
biological sciences~\cite{Bowman:RPP76:026401} and has also been exploited for controlling and
focusing particle beams~\cite{Eckerskorn:thesis:2016, Shvedov:OptExp17:5743, Shvedov:OptExp19:17350,
   Eckerskorn:PRAppl4:064001, Awel:optical-funnel:inprep}. A full theoretical description is not
available~\cite{Desyatnikov:OptExp17:8201} and we have implemented an approximate force model
described and benchmarked before~\cite{Desyatnikov:OptExp17:8201}. It assumes a Laguerre-Gaussian
laser beam of order 1, and uses a phenomenological constant $\kappa$ to model the axial and
transverse components separately. A description of how we implemented this model, which closely
follows the publication by Desyatnikov, is given in the supplementary information.

\section{Simulation results and comparison with experiment}
\label{sec:results}
To verify baseline correctness of our framework, we benchmarked it against particle distributions
from experiment~\cite{Worbs:geomopt:inprep}. There, $d=27~\text{nm}$ gold spheres were injected into
vacuum in an electrospray-ionization setup, passed through a differential pumping stage to remove
background gas, and then guided into an optimized aerodynamic lens stack (ALS)~\cite{Roth:JAS124:17,
   Worbs:geomopt:inprep}. The 1D position distributions, an arbitrary cross-section of the true 2D
distribution assuming axisymmetry around the $z$ axis, was measured at various distances from the
exit of the ALS along the propagation axis $z$.

To simulate this experiment we used the models for the drag force and Brownian motion described in
\autoref{sec:physmodels}. We modeled the ALS using its known geometry and experimentally recorded
pressures at fixed points in the system. We then solved for a laminar flow through this geometry
using a finite-element solver~\cite{Comsol:Multiphysics:5.5} and exported a regular grid of the
quantities flow velocity $\vec{v}$ and gas pressure $p$ throughout the ALS. We defined one
\texttt{FluidFlowDevice} and nine \texttt{SimpleZDetector}s at the distances from the ALS exit where
the experimental measurements were made. Then we let \cminject read in the flow field and run a
simulation for $10^5$ gold spheres with $d=27$~nm. The code to reproduce these results is provided
in the supplementary materials.

\begin{figure}
   \includegraphics[width=0.8\linewidth]{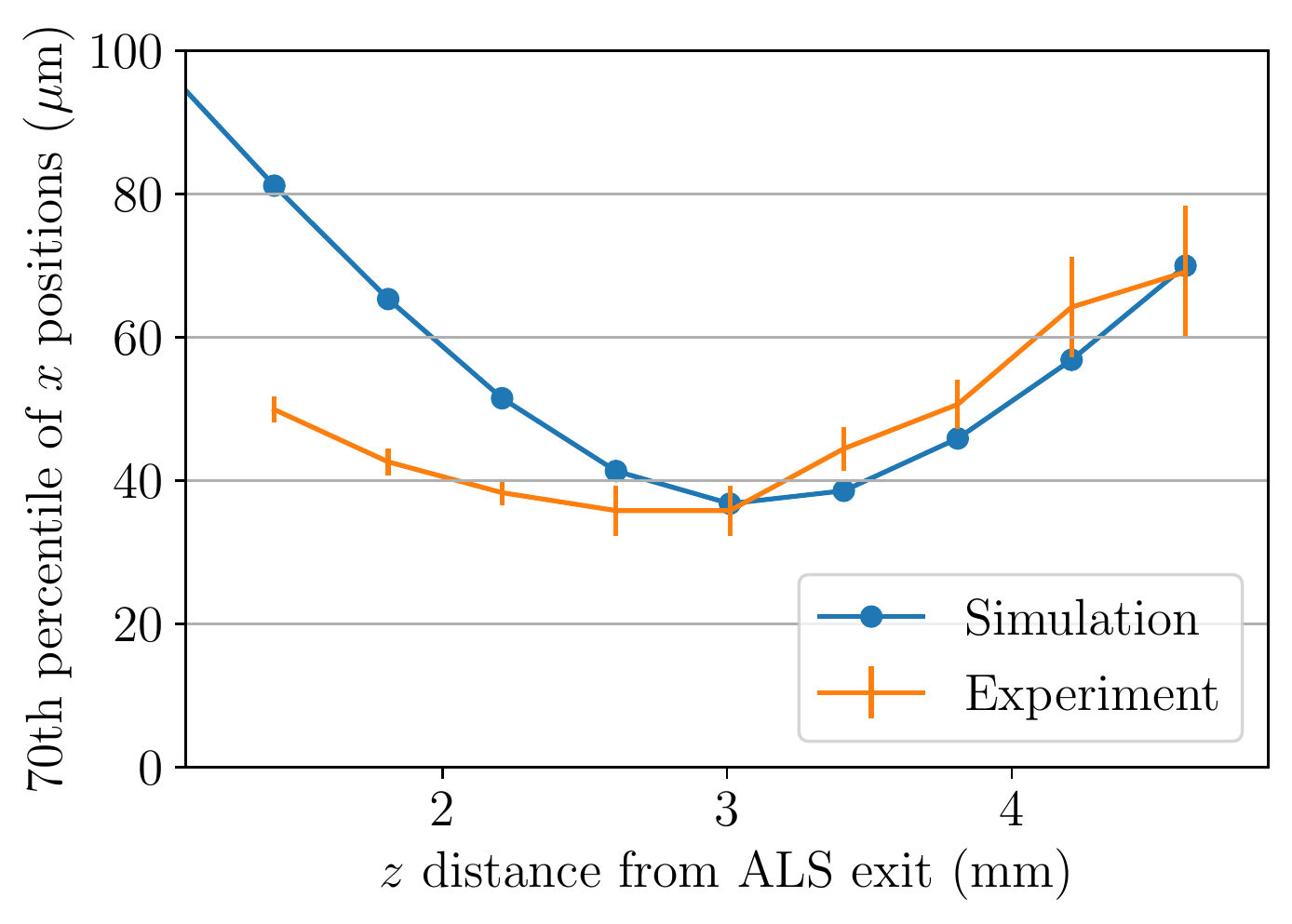}
   \caption{Focus curves determined by experiment (orange) and simulation (blue) for 27nm gold
     particles, moving through an ALS~\cite{Worbs:geomopt:inprep}. We measure the $x$ positions of
     all particles relative to the origin, and take the 70\% quantile of these positions as our
     measure of focus size. The results agree well on the minimum focus size and position, \ie, a
     $38\,$µm focus at $z=3\,$mm after the ALS exit, and also agree on the defocusing behavior after
     this minimum.}
   \label{fig:focus-curve-gold}
\end{figure}
To get a comparable measure for the quality of the particle beam's focus that does not depend on
fitting any particular beam shape, we calculated the distance from the origin in $X$ to which 70~\%
of the particles were detected, both for the simulated and the measured data. The results are shown
in \autoref{fig:focus-curve-gold}. One can see that there is good agreement regarding the minimum
focus size, $\ordsim35~\um$ at $z=3$~mm and the defocusing behavior after $z=3$~mm. The focusing at
$z<35$~mm is in fair agreement, but deviations are clearly visible and tentatively ascribed to the
neglect of gas-particle interactions in the initial space outside the ALS. Nevertheless, the
position and size of the minimum focus are the most important results for an injection pipeline used
for single-particle X-ray imaging, which our simulations model very well.

\begin{figure}
   \includegraphics[width=\linewidth]{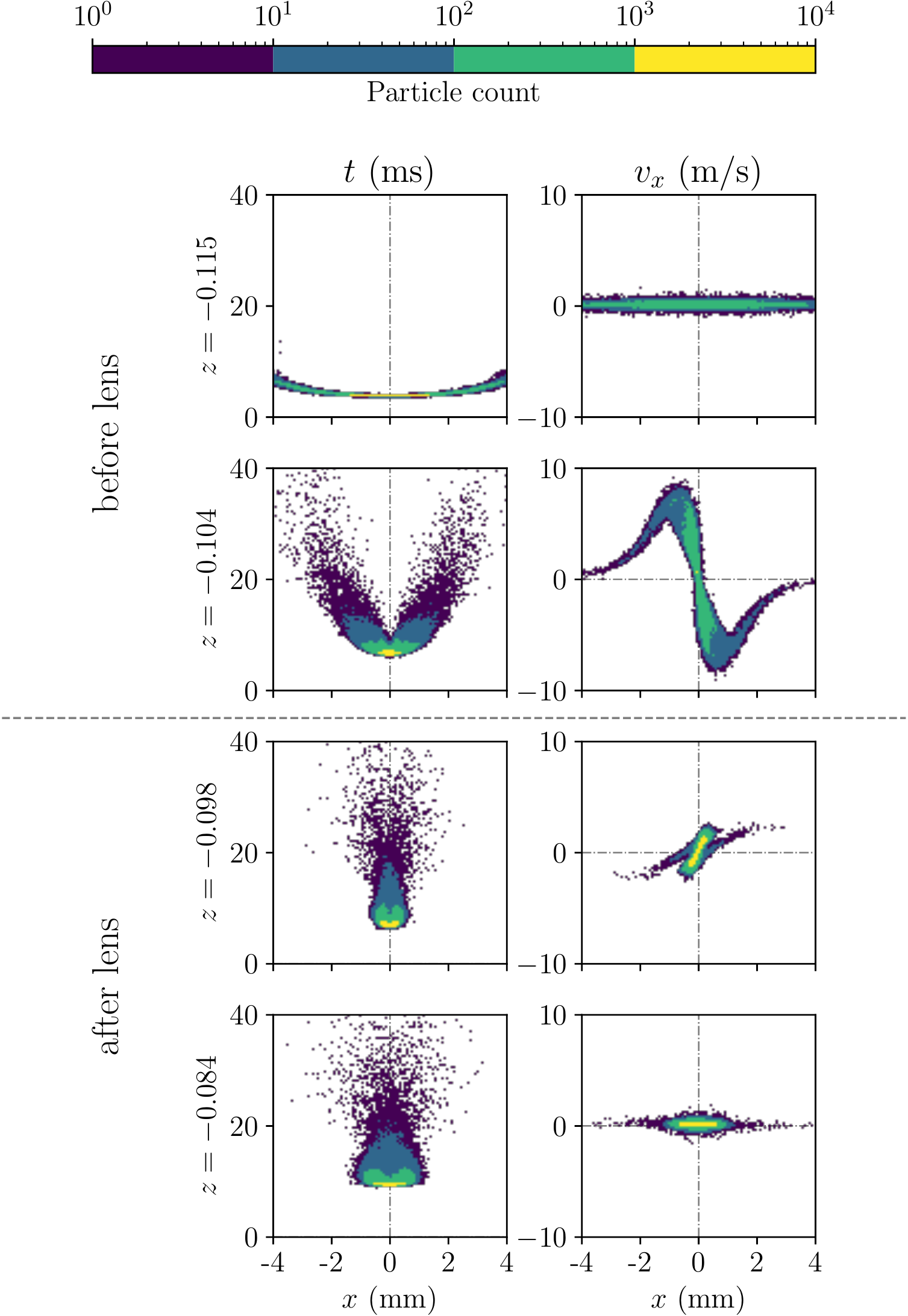}
   \caption{Phase-space histograms of $10^5$ simulated particles, at four detectors in an ALS. Two
      detectors are positioned before the aerodynamic lens (top), and two after (bottom). The left
      column shows the $t / x$ distribution evolving, the right one the $v_x / x$ distribution. In
      the left column, one can see that particles with a larger initial $x$ deviation take longer to
      arrive at the lens: the slowest particles arrive more than $\mathtt{\sim}$30\,ms later at the
      lens than the fastest ones (see 2nd row). The right column shows the evolving, initially
      Gaussian, $v_x / x$ distribution: it is strongly focused just before and slightly defocused
      just after the lens (2nd and 3rd row), and finally turns into a more focused, collimated
      particle beam (4th row).}
   \label{fig:phase-space-hist2ds}
\end{figure}
To better understand the focusing and defocusing behavior, we visually examined the results. For
instance, \autoref{fig:phase-space-hist2ds} shows 2D histograms of useful quantity pairs at
different $z$ positions in the experiment described above. This allows for a visual, somewhat
intuitive, disentangling of the evolving ensemble of particles. Such plots can be generated with the
provided \cminjectvis tool using the \texttt{-H} option. Alternatively, a qualitative visual
analysis can be obtained by plotting and inspecting particle trajectories as lines, using the
\texttt{-T} option, as shown for this and other experiments in \autoref{fig:overview-plots}. Less
congested visualizations are obtained by animated trajectory evolutions, using the \texttt{-M}
option, where time-dependent snapshots of the trajectories provide a visualization of the particles
positions and velocities. Examples are provided as video files in the supplementary materials.

\section{Program performance}
\label{sec:program-performance}
The achievable simulation performance was benchmarked on modern multicore computers, specifically
nodes of the Maxwell compute cluster at DESY. The nodes we used are equipped with ``Intel(R) Xeon(R)
CPU E5-2698 v3'' or ``Intel(R) Xeon(R) E5-2640'' CPUs, offering 32/64 and 16/32 cores/hyperthreads,
respectively.

We note that performance may improve or degrade substantially compared to what is shown here when
different force models, experiment sizes, or time steps are used. Here, we benchmarked the fluid
dynamics simulation described in \autoref{sec:results}, involving only Stokes' drag force and
Brownian motion at a time-step of $10$~\us and an experiment length of $\ordsim13$~cm.
\begin{figure}
  \includegraphics[width=\linewidth]{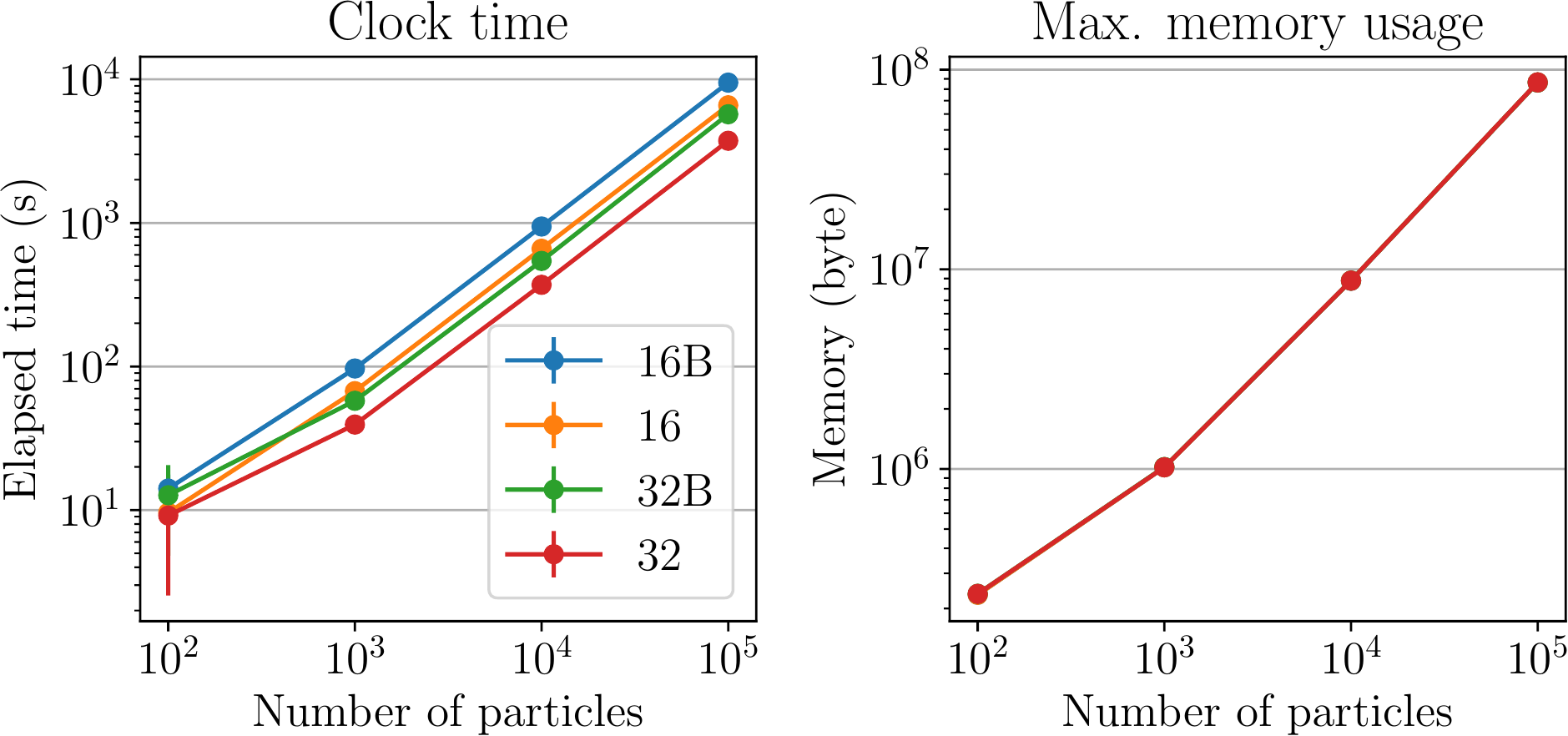}
  \caption{Scaling behavior of clock time and memory usage for the simulation described in
     \autoref{sec:program-performance}. ``32'' and ``16'' refer to a Intel Xeon E5-2698 (32 cores)
     and a Intel Xeon E5-2640 CPU (16 cores), respectively. ``B'' indicates that Brownian motion was
     enabled, whereas it wasn't otherwise. Note that the variance in memory usage is very low for a
     fixed number of particles and all curves look like one. Besides initial setup overhead, linear
     scaling of both clock time and memory is clearly visible.}
   \label{fig:performance-scaling}
\end{figure}

\begin{figure}
  \includegraphics[width=0.85\linewidth]{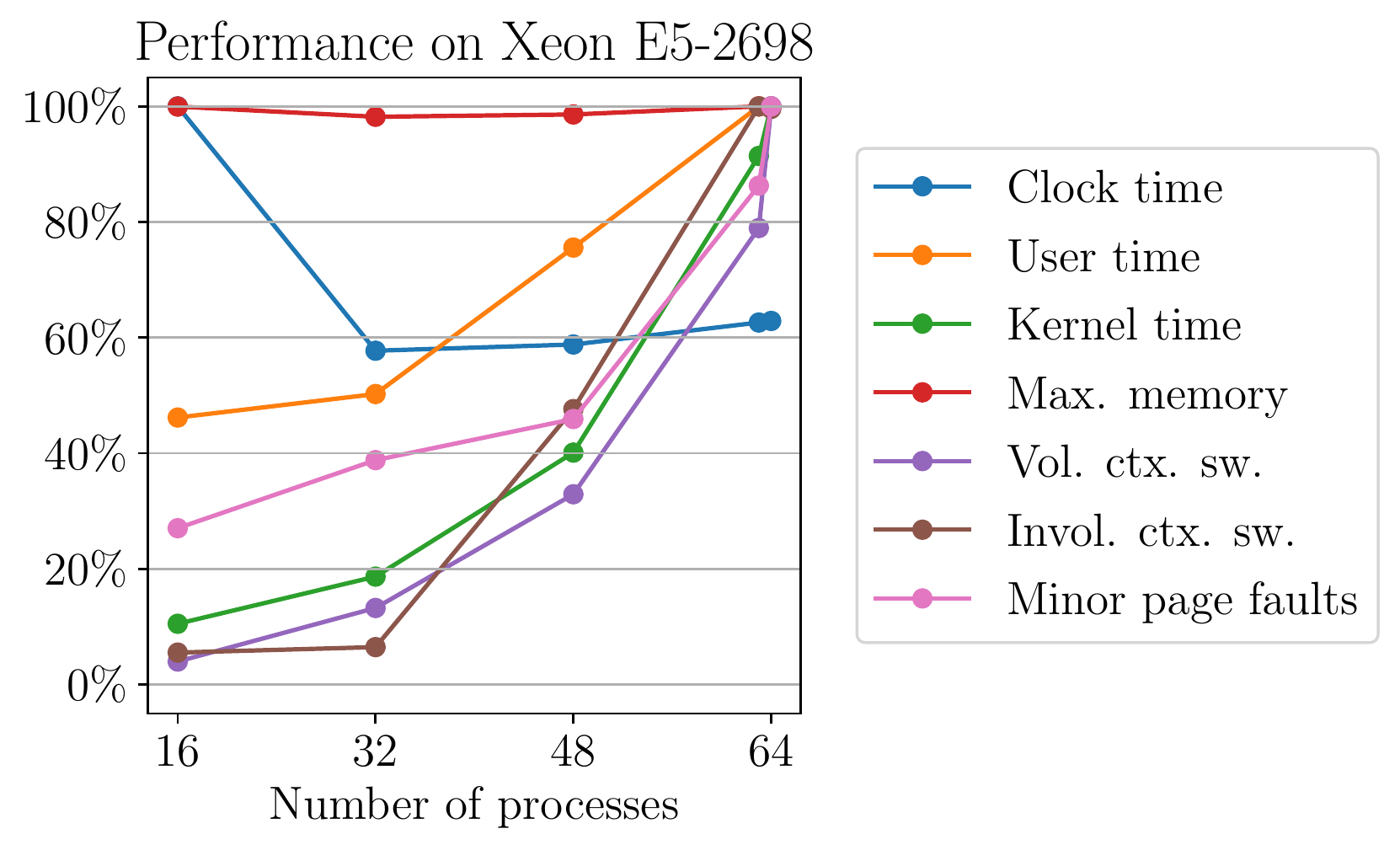}%
   \caption{Performance measurements made on an Intel Xeon E5-2698 CPU with 32 physical cores for
      the same simulation with different numbers of threads. The maximum value for each measurement
      is set to be 100~\%, and the other values displayed in relation to it. ``Vol.''/``Invol.'' are
      shorthand for ``voluntary''/``involuntary'', and ``ctx. sw.'' is shorthand for ``context
      switches''.}
   \label{fig:cores-vs-threads}
\end{figure}
%
\autoref{fig:performance-scaling} shows runtime and memory requirements for this simulation when
varying the number of particles, and demonstrates that both scale linearly with the number of
particles---as expected from a Monte Carlo simulation with no particle-particle interactions, which
is trivially parallel.

In \autoref{fig:cores-vs-threads}, we analyze multiple performance metrics as functions of the
number of parallel computation processes. The optimal runtime is reached when we use exactly as many
processes as there are physical CPU cores. When we use more processes, runtime performance degrades
significantly, together with several other performance metrics. This is observed even though the CPU
offers up to 64 available hyperthreads, which points to our current implementation not being
well-suited to gain performance from hyperthreading. In line with previous literature on this topic,
we assume the reason to be that hyperthreading increases competition for resources in the memory
hierarchy~\cite{Saini:ICHPC2011:1}. If this is indeed the reason, it could mean that our current
implementation performs a significant number of memory accesses that under-utilize caches.

Graphics-processing units (GPUs) are particularly well-suited to trivially parallelizable
calculations that largely consist of repeated, similar floating-point operations. They offer much
higher internal bandwidths in their memory hierarchy than the bandwidths between CPU and main
memory~\cite{NVIDIA:website:CUDAcppbestpractices}, and as such should have less trouble maintaining
reasonable performance even when faced with many cache misses. Therefore, they should exhibit
significantly better runtime-performance scaling at a much larger number of parallel threads. We had
also discussed other reasons why GPUs could offer significant speedups for our
calculations~\cite[ch.~7]{Welker:thesis:2019}. With recent developments in the automatic
optimization library \numba~\cite{Lam:LLVM15:7} making GPU calculations in Python more accessible,
GPUs could be effectively utilized in future versions of our object-oriented framework.

\section{Summary and outlook}
We introduced, described, and benchmarked a new Python framework for the simulation of
nanoparticle-injection pipelines. We hope that it will not only improve the sample delivery in
single-particle x-ray imaging~\cite{Ayyer:Optica8:15}, but also other
isolated-nanoparticle experiments~\cite{Seiffert:NatPhys13:766, Aquila:StructDyn2:041701}.

The force models already implemented in \cminject enable simulation-based development and exchange
of improved and novel injector designs, help to understand the effects of Brownian motion and how to
control it better, and facilitate scientific development for injecting single, noncrystalline
proteins, \eg, for single particle/molecule imaging experiments. Improvements directly relevant to
scientific applications could be made through the systematic derivation and implementation as well
as experimental comparison of models for novel techniques, \eg, acoustic~\cite{Li:PRAppl11:064036}
or photophoretic \cite{Eckerskorn:PRAppl4:064001, Awel:optical-funnel:inprep} focusing. This would
open up possibilities to explore these new and exciting pathways toward higher-quality particle
beams with \cminject, pushing the limits of the imaging of chemical and biochemical processes with
atomic resolution.

From a software perspective, development effort should be well-invested to make MPI bindings and
GPUs available for users of \cminject, \eg, by use of the mpi4py
library~\cite{Dalcin:JParDistComp:2005} or the CUDA bindings in the automatic optimization library
\numba~\cite{Lam:LLVM15:7}, which should significantly improve simulation
runtimes~\cite{Welker:thesis:2019} and is foreseen for future versions of \cminject.

To facilitate fast availability of improvements as well as community contributions to the
development, the framework has been published at \url{https://github.com/cfel-cmi/cminject} under a
modified GPLv3 license requiring attribution, \eg, through referencing of this publication. Up to
date documentation is available at \url{https://cminject.readthedocs.org}. Additional forces and
experiments will be modeled and open problems that were discussed here and
elsewhere~\cite{Welker:thesis:2019} will be resolved, in close exchange with the user community.

\section*{Declaration of interests}
The authors declare that they have no known competing financial interests or personal relationships
that could have appeared to influence the work reported in this paper.

\section*{Credit authorship contribution statement}
\textbf{Simon~Welker:} Methodology, Software, Validation, Formal analysis, Investigation,
Visualization, Writing - Original draft, Writing - Review \& Editing; \textbf{Muhamed~Amin:}
Conceptualization, Writing - Review \& Editing, Supervision; \textbf{Jochen~Küpper:}
Conceptualization, Resources, Writing - Review \& Editing, Supervision, Project administration

\begin{acknowledgments}
   We gratefully acknowledge useful discussions with Andrei Rode on photophoretic forces. We thank
   Lena Worbs for many useful comments and the CMI COMOTION team for tests and feedback of the
   software package.

   This work has been supported by the European Research Council under the European Union's Seventh
   Framework Programme (FP7/2007-2013) through the Consolidator Grant COMOTION (ERC-CoG 614507) and
   by the Deutsche Forschungsgemeinschaft through the Cluster of Excellence ``Advanced Imaging of
   Matter'' (AIM, EXC~2056, ID~390715994).

   This research was supported in part through the Maxwell computational resources operated at
   Deutsches Elektronen-Synchrotron DESY, Hamburg, Germany.
\end{acknowledgments}

\bibliography{string,cmi}

\begin{thebibliography}{59}%
\makeatletter
\providecommand \@ifxundefined [1]{%
 \@ifx{#1\undefined}
}%
\providecommand \@ifnum [1]{%
 \ifnum #1\expandafter \@firstoftwo
 \else \expandafter \@secondoftwo
 \fi
}%
\providecommand \@ifx [1]{%
 \ifx #1\expandafter \@firstoftwo
 \else \expandafter \@secondoftwo
 \fi
}%
\providecommand \natexlab [1]{#1}%
\providecommand \enquote  [1]{``#1''}%
\providecommand \bibnamefont  [1]{#1}%
\providecommand \bibfnamefont [1]{#1}%
\providecommand \citenamefont [1]{#1}%
\providecommand \href@noop [0]{\@secondoftwo}%
\providecommand \href [0]{\begingroup \@sanitize@url \@href}%
\providecommand \@href[1]{\@@startlink{#1}\@@href}%
\providecommand \@@href[1]{\endgroup#1\@@endlink}%
\providecommand \@sanitize@url [0]{\catcode `\\12\catcode `\$12\catcode
  `\&12\catcode `\#12\catcode `\^12\catcode `\_12\catcode `\%12\relax}%
\providecommand \@@startlink[1]{}%
\providecommand \@@endlink[0]{}%
\providecommand \url  [0]{\begingroup\@sanitize@url \@url }%
\providecommand \@url [1]{\endgroup\@href {#1}{\urlprefix }}%
\providecommand \urlprefix  [0]{URL }%
\providecommand \Eprint [0]{\href }%
\providecommand \doibase [0]{https://doi.org/}%
\providecommand \selectlanguage [0]{\@gobble}%
\providecommand \bibinfo  [0]{\@secondoftwo}%
\providecommand \bibfield  [0]{\@secondoftwo}%
\providecommand \translation [1]{[#1]}%
\providecommand \BibitemOpen [0]{}%
\providecommand \bibitemStop [0]{}%
\providecommand \bibitemNoStop [0]{.\EOS\space}%
\providecommand \EOS [0]{\spacefactor3000\relax}%
\providecommand \BibitemShut  [1]{\csname bibitem#1\endcsname}%
\let\auto@bib@innerbib\@empty
\bibitem [{\citenamefont {Neutze}\ \emph {et~al.}(2000)\citenamefont {Neutze},
  \citenamefont {Wouts}, \citenamefont {van~der Spoel}, \citenamefont
  {Weckert}, and\ \citenamefont {Hajdu}}]{Neutze:Nature406:752}%
  \BibitemOpen
  \bibfield  {author} {\bibinfo {author} {\bibfnamefont {R.}~\bibnamefont
  {Neutze}}, \bibinfo {author} {\bibfnamefont {R.}~\bibnamefont {Wouts}},
  \bibinfo {author} {\bibfnamefont {D.}~\bibnamefont {van~der Spoel}}, \bibinfo
  {author} {\bibfnamefont {E.}~\bibnamefont {Weckert}}, and\ \bibinfo {author}
  {\bibfnamefont {J.}~\bibnamefont {Hajdu}},\ }\bibfield  {title} {\bibinfo
  {title} {Potential for biomolecular imaging with femtosecond x-ray pulses},\
  }\href {https://doi.org/10.1038/35021099} {\bibfield  {journal} {\bibinfo
  {journal} {Nature}\ }\textbf {\bibinfo {volume} {406}},\ \bibinfo {pages}
  {752} (\bibinfo {year} {2000})}\BibitemShut {NoStop}%
\bibitem [{\citenamefont {Spence} and\ \citenamefont
  {Chapman}(2014)}]{Spence:PTRSB369:20130309}%
  \BibitemOpen
  \bibfield  {author} {\bibinfo {author} {\bibfnamefont {J.~C.~H.}\
  \bibnamefont {Spence}} and\ \bibinfo {author} {\bibfnamefont {H.~N.}\
  \bibnamefont {Chapman}},\ }\bibfield  {title} {\bibinfo {title} {The birth of
  a new field},\ }\href {https://doi.org/10.1098/rstb.2013.0309} {\bibfield
  {journal} {\bibinfo  {journal} {Phil. Trans. R. Soc. B}\ }\textbf {\bibinfo
  {volume} {369}},\ \bibinfo {pages} {20130309} (\bibinfo {year}
  {2014})}\BibitemShut {NoStop}%
\bibitem [{\citenamefont {Emma}\ \emph {et~al.}(2010)\citenamefont {Emma},
  \citenamefont {Akre}, \citenamefont {Arthur}, \citenamefont {Bionta},
  \citenamefont {Bostedt}, \citenamefont {Bozek}, \citenamefont {Brachmann},
  \citenamefont {Bucksbaum}, \citenamefont {Coffee}, \citenamefont {Decker},
  \citenamefont {Ding}, \citenamefont {Dowell}, \citenamefont {Edstrom},
  \citenamefont {Fisher}, \citenamefont {Frisch}, \citenamefont {Gilevich},
  \citenamefont {Hastings}, \citenamefont {Hays}, \citenamefont {Hering},
  \citenamefont {Huang}, \citenamefont {Iverson}, \citenamefont {Loos},
  \citenamefont {Messerschmidt}, \citenamefont {Miahnahri}, \citenamefont
  {Moeller}, \citenamefont {Nuhn}, \citenamefont {Pile}, \citenamefont
  {Ratner}, \citenamefont {Rzepiela}, \citenamefont {Schultz}, \citenamefont
  {Smith}, \citenamefont {Stefan}, \citenamefont {Tompkins}, \citenamefont
  {Turner}, \citenamefont {Welch}, \citenamefont {White}, \citenamefont {Wu},
  \citenamefont {Yocky}, and\ \citenamefont {Galayda}}]{Emma:NatPhoton4:641}%
  \BibitemOpen
  \bibfield  {author} {\bibinfo {author} {\bibfnamefont {P.}~\bibnamefont
  {Emma}}, \bibinfo {author} {\bibfnamefont {R.}~\bibnamefont {Akre}}, \bibinfo
  {author} {\bibfnamefont {J.}~\bibnamefont {Arthur}}, \bibinfo {author}
  {\bibfnamefont {R.}~\bibnamefont {Bionta}}, \bibinfo {author} {\bibfnamefont
  {C.}~\bibnamefont {Bostedt}}, \bibinfo {author} {\bibfnamefont
  {J.}~\bibnamefont {Bozek}}, \bibinfo {author} {\bibfnamefont
  {A.}~\bibnamefont {Brachmann}}, \bibinfo {author} {\bibfnamefont
  {P.}~\bibnamefont {Bucksbaum}}, \bibinfo {author} {\bibfnamefont
  {R.}~\bibnamefont {Coffee}}, \bibinfo {author} {\bibfnamefont {F.~J.}\
  \bibnamefont {Decker}}, \bibinfo {author} {\bibfnamefont {Y.}~\bibnamefont
  {Ding}}, \bibinfo {author} {\bibfnamefont {D.}~\bibnamefont {Dowell}},
  \bibinfo {author} {\bibfnamefont {S.}~\bibnamefont {Edstrom}}, \bibinfo
  {author} {\bibfnamefont {A.}~\bibnamefont {Fisher}}, \bibinfo {author}
  {\bibfnamefont {J.}~\bibnamefont {Frisch}}, \bibinfo {author} {\bibfnamefont
  {S.}~\bibnamefont {Gilevich}}, \bibinfo {author} {\bibfnamefont
  {J.}~\bibnamefont {Hastings}}, \bibinfo {author} {\bibfnamefont
  {G.}~\bibnamefont {Hays}}, \bibinfo {author} {\bibfnamefont {P.}~\bibnamefont
  {Hering}}, \bibinfo {author} {\bibfnamefont {Z.}~\bibnamefont {Huang}},
  \bibinfo {author} {\bibfnamefont {R.}~\bibnamefont {Iverson}}, \bibinfo
  {author} {\bibfnamefont {H.}~\bibnamefont {Loos}}, \bibinfo {author}
  {\bibfnamefont {M.}~\bibnamefont {Messerschmidt}}, \bibinfo {author}
  {\bibfnamefont {A.}~\bibnamefont {Miahnahri}}, \bibinfo {author}
  {\bibfnamefont {S.}~\bibnamefont {Moeller}}, \bibinfo {author} {\bibfnamefont
  {H.~D.}\ \bibnamefont {Nuhn}}, \bibinfo {author} {\bibfnamefont
  {G.}~\bibnamefont {Pile}}, \bibinfo {author} {\bibfnamefont {D.}~\bibnamefont
  {Ratner}}, \bibinfo {author} {\bibfnamefont {J.}~\bibnamefont {Rzepiela}},
  \bibinfo {author} {\bibfnamefont {D.}~\bibnamefont {Schultz}}, \bibinfo
  {author} {\bibfnamefont {T.}~\bibnamefont {Smith}}, \bibinfo {author}
  {\bibfnamefont {P.}~\bibnamefont {Stefan}}, \bibinfo {author} {\bibfnamefont
  {H.}~\bibnamefont {Tompkins}}, \bibinfo {author} {\bibfnamefont
  {J.}~\bibnamefont {Turner}}, \bibinfo {author} {\bibfnamefont
  {J.}~\bibnamefont {Welch}}, \bibinfo {author} {\bibfnamefont
  {W.}~\bibnamefont {White}}, \bibinfo {author} {\bibfnamefont
  {J.}~\bibnamefont {Wu}}, \bibinfo {author} {\bibfnamefont {G.}~\bibnamefont
  {Yocky}}, and\ \bibinfo {author} {\bibfnamefont {J.}~\bibnamefont
  {Galayda}},\ }\bibfield  {title} {\bibinfo {title} {First lasing and
  operation of an angstrom-wavelength free-electron laser},\ }\href
  {https://doi.org/10.1038/nphoton.2010.176} {\bibfield  {journal} {\bibinfo
  {journal} {Nat. Photon.}\ }\textbf {\bibinfo {volume} {4}},\ \bibinfo {pages}
  {641} (\bibinfo {year} {2010})}\BibitemShut {NoStop}%
\bibitem [{\citenamefont {Decking}\ \emph {et~al.}(2020)\citenamefont
  {Decking}, \citenamefont {Abeghyan}, \citenamefont {Abramian}, \citenamefont
  {Abramsky}, \citenamefont {Aguirre}, \citenamefont {Albrecht}, \citenamefont
  {Alou}, \citenamefont {Altarelli}, \citenamefont {Altmann}, \citenamefont
  {Amyan}, \citenamefont {Anashin}, \citenamefont {Apostolov}, \citenamefont
  {Appel}, \citenamefont {Auguste}, \citenamefont {Ayvazyan}, \citenamefont
  {Baark}, \citenamefont {Babies}, \citenamefont {Baboi}, \citenamefont {Bak},
  \citenamefont {Balandin}, \citenamefont {Baldinger}, \citenamefont
  {Baranasic}, \citenamefont {Barbanotti}, \citenamefont {Belikov},
  \citenamefont {Belokurov}, \citenamefont {Belova}, \citenamefont {Belyakov},
  \citenamefont {Berry}, \citenamefont {Bertucci}, \citenamefont {Beutner},
  \citenamefont {Block}, \citenamefont {Bl{\"o}cher}, \citenamefont
  {B{\"o}ckmann}, \citenamefont {Bohm}, \citenamefont {B{\"o}hnert},
  \citenamefont {Bondar}, \citenamefont {Bondarchuk}, \citenamefont {Bonezzi},
  \citenamefont {Borowiec}, \citenamefont {B{\"o}sch}, \citenamefont
  {B{\"o}senberg}, \citenamefont {Bosotti}, \citenamefont {B{\"o}spflug},
  \citenamefont {Bousonville}, \citenamefont {Boyd}, \citenamefont {Bozhko},
  \citenamefont {Brand}, \citenamefont {Branlard}, \citenamefont {Briechle},
  \citenamefont {Brinker}, \citenamefont {Brinker}, \citenamefont {Brinkmann},
  \citenamefont {Brockhauser}, \citenamefont {Brovko}, \citenamefont
  {Br{\"u}ck}, \citenamefont {Br{\"u}dgam}, \citenamefont {Butkowski},
  \citenamefont {B{\"u}ttner}, \citenamefont {Calero}, \citenamefont
  {Castro-Carballo}, \citenamefont {Cattalanotto}, \citenamefont {Charrier},
  \citenamefont {Chen}, \citenamefont {Cherepenko}, \citenamefont {Cheskidov},
  \citenamefont {Chiodini}, \citenamefont {Chong}, \citenamefont {Choroba},
  \citenamefont {Chorowski}, \citenamefont {Churanov}, \citenamefont
  {Cichalewski}, \citenamefont {Clausen}, \citenamefont {Clement},
  \citenamefont {Clou{\'e}}, \citenamefont {Cobos}, \citenamefont {Coppola},
  \citenamefont {Cunis}, \citenamefont {Czuba}, \citenamefont {Czwalinna},
  \citenamefont {D'Almagne}, \citenamefont {Dammann}, \citenamefont {Danared},
  \citenamefont {de~Zubiaurre~Wagner}, \citenamefont {Delfs}, \citenamefont
  {Delfs}, \citenamefont {Dietrich}, \citenamefont {Dietrich}, \citenamefont
  {Dohlus}, \citenamefont {Dommach}, \citenamefont {Donat}, \citenamefont
  {Dong}, \citenamefont {Doynikov}, \citenamefont {Dressel}, \citenamefont
  {Duda}, \citenamefont {Duda}, \citenamefont {Eckoldt}, \citenamefont {Ehsan},
  \citenamefont {Eidam}, \citenamefont {Eints}, \citenamefont {Engling},
  \citenamefont {Englisch}, \citenamefont {Ermakov}, \citenamefont {Escherich},
  \citenamefont {Eschke}, \citenamefont {Saldin}, \citenamefont {Faesing},
  \citenamefont {Fallou}, \citenamefont {Felber}, \citenamefont {Fenner},
  \citenamefont {Fernandes}, \citenamefont {Fern{\'a}ndez}, \citenamefont
  {Feuker}, \citenamefont {Filippakopoulos}, \citenamefont {Floettmann},
  \citenamefont {Fogel}, \citenamefont {Fontaine}, \citenamefont {Franc{\'e}s},
  \citenamefont {Martin}, \citenamefont {Freund}, \citenamefont {Freyermuth},
  \citenamefont {Friedland}, \citenamefont {Fr{\"o}hlich}, \citenamefont
  {Fusetti}, \citenamefont {Fydrych}, \citenamefont {Gallas}, \citenamefont
  {Garc{\'i}a}, \citenamefont {Garcia-Tabares}, \citenamefont {Geloni},
  \citenamefont {Gerasimova}, \citenamefont {Gerth}, \citenamefont
  {Ge{\ss}ler}, \citenamefont {Gharibyan}, \citenamefont {Gloor}, \citenamefont
  {G{\l}owinkowski}, \citenamefont {Goessel}, \citenamefont
  {Go{\l}{\k{e}}biewski}, \citenamefont {Golubeva}, \citenamefont {Grabowski},
  \citenamefont {Graeff}, \citenamefont {Grebentsov}, \citenamefont {Grecki},
  \citenamefont {Grevsmuehl}, \citenamefont {Gross}, \citenamefont
  {Grosse-Wortmann}, \citenamefont {Gr{\"u}nert}, \citenamefont {Grunewald},
  \citenamefont {Grzegory}, \citenamefont {Feng}, \citenamefont {Guler},
  \citenamefont {Gusev}, \citenamefont {Gutierrez}, \citenamefont {Hagge},
  \citenamefont {Hamberg}, \citenamefont {Hanneken}, \citenamefont {Harms},
  \citenamefont {Hartl}, \citenamefont {Hauberg}, \citenamefont {Hauf},
  \citenamefont {Hauschildt}, \citenamefont {Hauser}, \citenamefont {Havlicek},
  \citenamefont {Hedqvist}, \citenamefont {Heidbrook}, \citenamefont
  {Hellberg}, \citenamefont {Henning}, \citenamefont {Hensler}, \citenamefont
  {Hermann}, \citenamefont {Hidv{\'e}gi}, \citenamefont {Hierholzer},
  \citenamefont {Hintz}, \citenamefont {Hoffmann}, \citenamefont {Hoffmann},
  \citenamefont {Hoffmann}, \citenamefont {Holler}, \citenamefont {H{\"u}ning},
  \citenamefont {Ignatenko}, \citenamefont {Ilchen}, \citenamefont {Iluk},
  \citenamefont {Iversen}, \citenamefont {Izquierdo}, \citenamefont {Jachmann},
  \citenamefont {Jardon}, \citenamefont {Jastrow}, \citenamefont {Jensch},
  \citenamefont {Jensen}, \citenamefont {Je{\.{z}}abek}, \citenamefont {Jidda},
  \citenamefont {Jin}, \citenamefont {Johansson}, \citenamefont {Jonas},
  \citenamefont {Kaabi}, \citenamefont {Kaefer}, \citenamefont {Kammering},
  \citenamefont {Kapitza}, \citenamefont {Karabekyan}, \citenamefont
  {Karstensen}, \citenamefont {Kasprzak}, \citenamefont {Katalev},
  \citenamefont {Keese}, \citenamefont {Keil}, \citenamefont {Kholopov},
  \citenamefont {Killenberger}, \citenamefont {Kitaev}, \citenamefont
  {Klimchenko}, \citenamefont {Klos}, \citenamefont {Knebel}, \citenamefont
  {Koch}, \citenamefont {Koepke}, \citenamefont {K{\"o}hler}, \citenamefont
  {K{\"o}hler}, \citenamefont {Kohlstrunk}, \citenamefont {Konopkova},
  \citenamefont {Konstantinov}, \citenamefont {Kook}, \citenamefont {Koprek},
  \citenamefont {K{\"o}rfer}, \citenamefont {Korth}, \citenamefont {Kosarev},
  \citenamefont {Kosi{\'{n}}ski}, \citenamefont {Kostin}, \citenamefont {Kot},
  \citenamefont {Kotarba}, \citenamefont {Kozak}, \citenamefont {Kozak},
  \citenamefont {Kramert}, \citenamefont {Krasilnikov}, \citenamefont
  {Krasnov}, \citenamefont {Krause}, \citenamefont {Kravchuk}, \citenamefont
  {Krebs}, \citenamefont {Kretschmer}, \citenamefont {Kreutzkamp},
  \citenamefont {Kr{\"o}plin}, \citenamefont {Krzysik}, \citenamefont {Kube},
  \citenamefont {Kuehn}, \citenamefont {Kujala}, \citenamefont {Kulikov},
  \citenamefont {Kuzminych}, \citenamefont {La~Civita}, \citenamefont
  {Lacroix}, \citenamefont {Lamb}, \citenamefont {Lancetov}, \citenamefont
  {Larsson}, \citenamefont {Le~Pinvidic}, \citenamefont {Lederer},
  \citenamefont {Lensch}, \citenamefont {Lenz}, \citenamefont {Leuschner},
  \citenamefont {Levenhagen}, \citenamefont {Li}, \citenamefont {Liebing},
  \citenamefont {Lilje}, \citenamefont {Limberg}, \citenamefont {Lipka},
  \citenamefont {List}, \citenamefont {Liu}, \citenamefont {Liu}, \citenamefont
  {Lorbeer}, \citenamefont {Lorkiewicz}, \citenamefont {Lu}, \citenamefont
  {Ludwig}, \citenamefont {Machau}, \citenamefont {Maciocha}, \citenamefont
  {Madec}, \citenamefont {Magueur}, \citenamefont {Maiano}, \citenamefont
  {Maksimova}, \citenamefont {Malcher}, \citenamefont {Maltezopoulos},
  \citenamefont {Mamoshkina}, \citenamefont {Manschwetus}, \citenamefont
  {Marcellini}, \citenamefont {Marinkovic}, \citenamefont {Martinez},
  \citenamefont {Martirosyan}, \citenamefont {Maschmann}, \citenamefont
  {Maslov}, \citenamefont {Matheisen}, \citenamefont {Mavric}, \citenamefont
  {Mei{\ss}ner}, \citenamefont {Meissner}, \citenamefont {Messerschmidt},
  \citenamefont {Meyners}, \citenamefont {Michalski}, \citenamefont
  {Michelato}, \citenamefont {Mildner}, \citenamefont {Moe}, \citenamefont
  {Moglia}, \citenamefont {Mohr}, \citenamefont {Mohr}, \citenamefont
  {M{\"o}ller}, \citenamefont {Mommerz}, \citenamefont {Monaco}, \citenamefont
  {Montiel}, \citenamefont {Moretti}, \citenamefont {Morozov}, \citenamefont
  {Morozov}, and\ \citenamefont {Mross}}]{Decking:NatPhot14:391}%
  \BibitemOpen
  \bibfield  {author} {\bibinfo {author} {\bibfnamefont {W.}~\bibnamefont
  {Decking}}, \bibinfo {author} {\bibfnamefont {S.}~\bibnamefont {Abeghyan}},
  \bibinfo {author} {\bibfnamefont {P.}~\bibnamefont {Abramian}}, \bibinfo
  {author} {\bibfnamefont {A.}~\bibnamefont {Abramsky}}, \bibinfo {author}
  {\bibfnamefont {A.}~\bibnamefont {Aguirre}}, \bibinfo {author} {\bibfnamefont
  {C.}~\bibnamefont {Albrecht}}, \bibinfo {author} {\bibfnamefont
  {P.}~\bibnamefont {Alou}}, \bibinfo {author} {\bibfnamefont {M.}~\bibnamefont
  {Altarelli}}, \bibinfo {author} {\bibfnamefont {P.}~\bibnamefont {Altmann}},
  \bibinfo {author} {\bibfnamefont {K.}~\bibnamefont {Amyan}}, \bibinfo
  {author} {\bibfnamefont {V.}~\bibnamefont {Anashin}}, \bibinfo {author}
  {\bibfnamefont {E.}~\bibnamefont {Apostolov}}, \bibinfo {author}
  {\bibfnamefont {K.}~\bibnamefont {Appel}}, \bibinfo {author} {\bibfnamefont
  {D.}~\bibnamefont {Auguste}}, \bibinfo {author} {\bibfnamefont
  {V.}~\bibnamefont {Ayvazyan}}, \bibinfo {author} {\bibfnamefont
  {S.}~\bibnamefont {Baark}}, \bibinfo {author} {\bibfnamefont
  {F.}~\bibnamefont {Babies}}, \bibinfo {author} {\bibfnamefont
  {N.}~\bibnamefont {Baboi}}, \bibinfo {author} {\bibfnamefont
  {P.}~\bibnamefont {Bak}}, \bibinfo {author} {\bibfnamefont {V.}~\bibnamefont
  {Balandin}}, \bibinfo {author} {\bibfnamefont {R.}~\bibnamefont {Baldinger}},
  \bibinfo {author} {\bibfnamefont {B.}~\bibnamefont {Baranasic}}, \bibinfo
  {author} {\bibfnamefont {S.}~\bibnamefont {Barbanotti}}, \bibinfo {author}
  {\bibfnamefont {O.}~\bibnamefont {Belikov}}, \bibinfo {author} {\bibfnamefont
  {V.}~\bibnamefont {Belokurov}}, \bibinfo {author} {\bibfnamefont
  {L.}~\bibnamefont {Belova}}, \bibinfo {author} {\bibfnamefont
  {V.}~\bibnamefont {Belyakov}}, \bibinfo {author} {\bibfnamefont
  {S.}~\bibnamefont {Berry}}, \bibinfo {author} {\bibfnamefont
  {M.}~\bibnamefont {Bertucci}}, \bibinfo {author} {\bibfnamefont
  {B.}~\bibnamefont {Beutner}}, \bibinfo {author} {\bibfnamefont
  {A.}~\bibnamefont {Block}}, \bibinfo {author} {\bibfnamefont
  {M.}~\bibnamefont {Bl{\"o}cher}}, \bibinfo {author} {\bibfnamefont
  {T.}~\bibnamefont {B{\"o}ckmann}}, \bibinfo {author} {\bibfnamefont
  {C.}~\bibnamefont {Bohm}}, \bibinfo {author} {\bibfnamefont {M.}~\bibnamefont
  {B{\"o}hnert}}, \bibinfo {author} {\bibfnamefont {V.}~\bibnamefont {Bondar}},
  \bibinfo {author} {\bibfnamefont {E.}~\bibnamefont {Bondarchuk}}, \bibinfo
  {author} {\bibfnamefont {M.}~\bibnamefont {Bonezzi}}, \bibinfo {author}
  {\bibfnamefont {P.}~\bibnamefont {Borowiec}}, \bibinfo {author}
  {\bibfnamefont {C.}~\bibnamefont {B{\"o}sch}}, \bibinfo {author}
  {\bibfnamefont {U.}~\bibnamefont {B{\"o}senberg}}, \bibinfo {author}
  {\bibfnamefont {A.}~\bibnamefont {Bosotti}}, \bibinfo {author} {\bibfnamefont
  {R.}~\bibnamefont {B{\"o}spflug}}, \bibinfo {author} {\bibfnamefont
  {M.}~\bibnamefont {Bousonville}}, \bibinfo {author} {\bibfnamefont
  {E.}~\bibnamefont {Boyd}}, \bibinfo {author} {\bibfnamefont {Y.}~\bibnamefont
  {Bozhko}}, \bibinfo {author} {\bibfnamefont {A.}~\bibnamefont {Brand}},
  \bibinfo {author} {\bibfnamefont {J.}~\bibnamefont {Branlard}}, \bibinfo
  {author} {\bibfnamefont {S.}~\bibnamefont {Briechle}}, \bibinfo {author}
  {\bibfnamefont {F.}~\bibnamefont {Brinker}}, \bibinfo {author} {\bibfnamefont
  {S.}~\bibnamefont {Brinker}}, \bibinfo {author} {\bibfnamefont
  {R.}~\bibnamefont {Brinkmann}}, \bibinfo {author} {\bibfnamefont
  {S.}~\bibnamefont {Brockhauser}}, \bibinfo {author} {\bibfnamefont
  {O.}~\bibnamefont {Brovko}}, \bibinfo {author} {\bibfnamefont
  {H.}~\bibnamefont {Br{\"u}ck}}, \bibinfo {author} {\bibfnamefont
  {A.}~\bibnamefont {Br{\"u}dgam}}, \bibinfo {author} {\bibfnamefont
  {L.}~\bibnamefont {Butkowski}}, \bibinfo {author} {\bibfnamefont
  {T.}~\bibnamefont {B{\"u}ttner}}, \bibinfo {author} {\bibfnamefont
  {J.}~\bibnamefont {Calero}}, \bibinfo {author} {\bibfnamefont
  {E.}~\bibnamefont {Castro-Carballo}}, \bibinfo {author} {\bibfnamefont
  {G.}~\bibnamefont {Cattalanotto}}, \bibinfo {author} {\bibfnamefont
  {J.}~\bibnamefont {Charrier}}, \bibinfo {author} {\bibfnamefont
  {J.}~\bibnamefont {Chen}}, \bibinfo {author} {\bibfnamefont {A.}~\bibnamefont
  {Cherepenko}}, \bibinfo {author} {\bibfnamefont {V.}~\bibnamefont
  {Cheskidov}}, \bibinfo {author} {\bibfnamefont {M.}~\bibnamefont {Chiodini}},
  \bibinfo {author} {\bibfnamefont {A.}~\bibnamefont {Chong}}, \bibinfo
  {author} {\bibfnamefont {S.}~\bibnamefont {Choroba}}, \bibinfo {author}
  {\bibfnamefont {M.}~\bibnamefont {Chorowski}}, \bibinfo {author}
  {\bibfnamefont {D.}~\bibnamefont {Churanov}}, \bibinfo {author}
  {\bibfnamefont {W.}~\bibnamefont {Cichalewski}}, \bibinfo {author}
  {\bibfnamefont {M.}~\bibnamefont {Clausen}}, \bibinfo {author} {\bibfnamefont
  {W.}~\bibnamefont {Clement}}, \bibinfo {author} {\bibfnamefont
  {C.}~\bibnamefont {Clou{\'e}}}, \bibinfo {author} {\bibfnamefont {J.~A.}\
  \bibnamefont {Cobos}}, \bibinfo {author} {\bibfnamefont {N.}~\bibnamefont
  {Coppola}}, \bibinfo {author} {\bibfnamefont {S.}~\bibnamefont {Cunis}},
  \bibinfo {author} {\bibfnamefont {K.}~\bibnamefont {Czuba}}, \bibinfo
  {author} {\bibfnamefont {M.}~\bibnamefont {Czwalinna}}, \bibinfo {author}
  {\bibfnamefont {B.}~\bibnamefont {D'Almagne}}, \bibinfo {author}
  {\bibfnamefont {J.}~\bibnamefont {Dammann}}, \bibinfo {author} {\bibfnamefont
  {H.}~\bibnamefont {Danared}}, \bibinfo {author} {\bibfnamefont
  {A.}~\bibnamefont {de~Zubiaurre~Wagner}}, \bibinfo {author} {\bibfnamefont
  {A.}~\bibnamefont {Delfs}}, \bibinfo {author} {\bibfnamefont
  {T.}~\bibnamefont {Delfs}}, \bibinfo {author} {\bibfnamefont
  {F.}~\bibnamefont {Dietrich}}, \bibinfo {author} {\bibfnamefont
  {T.}~\bibnamefont {Dietrich}}, \bibinfo {author} {\bibfnamefont
  {M.}~\bibnamefont {Dohlus}}, \bibinfo {author} {\bibfnamefont
  {M.}~\bibnamefont {Dommach}}, \bibinfo {author} {\bibfnamefont
  {A.}~\bibnamefont {Donat}}, \bibinfo {author} {\bibfnamefont
  {X.}~\bibnamefont {Dong}}, \bibinfo {author} {\bibfnamefont {N.}~\bibnamefont
  {Doynikov}}, \bibinfo {author} {\bibfnamefont {M.}~\bibnamefont {Dressel}},
  \bibinfo {author} {\bibfnamefont {M.}~\bibnamefont {Duda}}, \bibinfo {author}
  {\bibfnamefont {P.}~\bibnamefont {Duda}}, \bibinfo {author} {\bibfnamefont
  {H.}~\bibnamefont {Eckoldt}}, \bibinfo {author} {\bibfnamefont
  {W.}~\bibnamefont {Ehsan}}, \bibinfo {author} {\bibfnamefont
  {J.}~\bibnamefont {Eidam}}, \bibinfo {author} {\bibfnamefont
  {F.}~\bibnamefont {Eints}}, \bibinfo {author} {\bibfnamefont
  {C.}~\bibnamefont {Engling}}, \bibinfo {author} {\bibfnamefont
  {U.}~\bibnamefont {Englisch}}, \bibinfo {author} {\bibfnamefont
  {A.}~\bibnamefont {Ermakov}}, \bibinfo {author} {\bibfnamefont
  {K.}~\bibnamefont {Escherich}}, \bibinfo {author} {\bibfnamefont
  {J.}~\bibnamefont {Eschke}}, \bibinfo {author} {\bibfnamefont
  {E.}~\bibnamefont {Saldin}}, \bibinfo {author} {\bibfnamefont
  {M.}~\bibnamefont {Faesing}}, \bibinfo {author} {\bibfnamefont
  {A.}~\bibnamefont {Fallou}}, \bibinfo {author} {\bibfnamefont
  {M.}~\bibnamefont {Felber}}, \bibinfo {author} {\bibfnamefont
  {M.}~\bibnamefont {Fenner}}, \bibinfo {author} {\bibfnamefont
  {B.}~\bibnamefont {Fernandes}}, \bibinfo {author} {\bibfnamefont {J.~M.}\
  \bibnamefont {Fern{\'a}ndez}}, \bibinfo {author} {\bibfnamefont
  {S.}~\bibnamefont {Feuker}}, \bibinfo {author} {\bibfnamefont
  {K.}~\bibnamefont {Filippakopoulos}}, \bibinfo {author} {\bibfnamefont
  {K.}~\bibnamefont {Floettmann}}, \bibinfo {author} {\bibfnamefont
  {V.}~\bibnamefont {Fogel}}, \bibinfo {author} {\bibfnamefont
  {M.}~\bibnamefont {Fontaine}}, \bibinfo {author} {\bibfnamefont
  {A.}~\bibnamefont {Franc{\'e}s}}, \bibinfo {author} {\bibfnamefont {I.~F.}\
  \bibnamefont {Martin}}, \bibinfo {author} {\bibfnamefont {W.}~\bibnamefont
  {Freund}}, \bibinfo {author} {\bibfnamefont {T.}~\bibnamefont {Freyermuth}},
  \bibinfo {author} {\bibfnamefont {M.}~\bibnamefont {Friedland}}, \bibinfo
  {author} {\bibfnamefont {L.}~\bibnamefont {Fr{\"o}hlich}}, \bibinfo {author}
  {\bibfnamefont {M.}~\bibnamefont {Fusetti}}, \bibinfo {author} {\bibfnamefont
  {J.}~\bibnamefont {Fydrych}}, \bibinfo {author} {\bibfnamefont
  {A.}~\bibnamefont {Gallas}}, \bibinfo {author} {\bibfnamefont
  {O.}~\bibnamefont {Garc{\'i}a}}, \bibinfo {author} {\bibfnamefont
  {L.}~\bibnamefont {Garcia-Tabares}}, \bibinfo {author} {\bibfnamefont
  {G.}~\bibnamefont {Geloni}}, \bibinfo {author} {\bibfnamefont
  {N.}~\bibnamefont {Gerasimova}}, \bibinfo {author} {\bibfnamefont
  {C.}~\bibnamefont {Gerth}}, \bibinfo {author} {\bibfnamefont
  {P.}~\bibnamefont {Ge{\ss}ler}}, \bibinfo {author} {\bibfnamefont
  {V.}~\bibnamefont {Gharibyan}}, \bibinfo {author} {\bibfnamefont
  {M.}~\bibnamefont {Gloor}}, \bibinfo {author} {\bibfnamefont
  {J.}~\bibnamefont {G{\l}owinkowski}}, \bibinfo {author} {\bibfnamefont
  {A.}~\bibnamefont {Goessel}}, \bibinfo {author} {\bibfnamefont
  {Z.}~\bibnamefont {Go{\l}{\k{e}}biewski}}, \bibinfo {author} {\bibfnamefont
  {N.}~\bibnamefont {Golubeva}}, \bibinfo {author} {\bibfnamefont
  {W.}~\bibnamefont {Grabowski}}, \bibinfo {author} {\bibfnamefont
  {W.}~\bibnamefont {Graeff}}, \bibinfo {author} {\bibfnamefont
  {A.}~\bibnamefont {Grebentsov}}, \bibinfo {author} {\bibfnamefont
  {M.}~\bibnamefont {Grecki}}, \bibinfo {author} {\bibfnamefont
  {T.}~\bibnamefont {Grevsmuehl}}, \bibinfo {author} {\bibfnamefont
  {M.}~\bibnamefont {Gross}}, \bibinfo {author} {\bibfnamefont
  {U.}~\bibnamefont {Grosse-Wortmann}}, \bibinfo {author} {\bibfnamefont
  {J.}~\bibnamefont {Gr{\"u}nert}}, \bibinfo {author} {\bibfnamefont
  {S.}~\bibnamefont {Grunewald}}, \bibinfo {author} {\bibfnamefont
  {P.}~\bibnamefont {Grzegory}}, \bibinfo {author} {\bibfnamefont
  {G.}~\bibnamefont {Feng}}, \bibinfo {author} {\bibfnamefont {H.}~\bibnamefont
  {Guler}}, \bibinfo {author} {\bibfnamefont {G.}~\bibnamefont {Gusev}},
  \bibinfo {author} {\bibfnamefont {J.~L.}\ \bibnamefont {Gutierrez}}, \bibinfo
  {author} {\bibfnamefont {L.}~\bibnamefont {Hagge}}, \bibinfo {author}
  {\bibfnamefont {M.}~\bibnamefont {Hamberg}}, \bibinfo {author} {\bibfnamefont
  {R.}~\bibnamefont {Hanneken}}, \bibinfo {author} {\bibfnamefont
  {E.}~\bibnamefont {Harms}}, \bibinfo {author} {\bibfnamefont
  {I.}~\bibnamefont {Hartl}}, \bibinfo {author} {\bibfnamefont
  {A.}~\bibnamefont {Hauberg}}, \bibinfo {author} {\bibfnamefont
  {S.}~\bibnamefont {Hauf}}, \bibinfo {author} {\bibfnamefont {J.}~\bibnamefont
  {Hauschildt}}, \bibinfo {author} {\bibfnamefont {J.}~\bibnamefont {Hauser}},
  \bibinfo {author} {\bibfnamefont {J.}~\bibnamefont {Havlicek}}, \bibinfo
  {author} {\bibfnamefont {A.}~\bibnamefont {Hedqvist}}, \bibinfo {author}
  {\bibfnamefont {N.}~\bibnamefont {Heidbrook}}, \bibinfo {author}
  {\bibfnamefont {F.}~\bibnamefont {Hellberg}}, \bibinfo {author}
  {\bibfnamefont {D.}~\bibnamefont {Henning}}, \bibinfo {author} {\bibfnamefont
  {O.}~\bibnamefont {Hensler}}, \bibinfo {author} {\bibfnamefont
  {T.}~\bibnamefont {Hermann}}, \bibinfo {author} {\bibfnamefont
  {A.}~\bibnamefont {Hidv{\'e}gi}}, \bibinfo {author} {\bibfnamefont
  {M.}~\bibnamefont {Hierholzer}}, \bibinfo {author} {\bibfnamefont
  {H.}~\bibnamefont {Hintz}}, \bibinfo {author} {\bibfnamefont
  {F.}~\bibnamefont {Hoffmann}}, \bibinfo {author} {\bibfnamefont
  {M.}~\bibnamefont {Hoffmann}}, \bibinfo {author} {\bibfnamefont
  {M.}~\bibnamefont {Hoffmann}}, \bibinfo {author} {\bibfnamefont
  {Y.}~\bibnamefont {Holler}}, \bibinfo {author} {\bibfnamefont
  {M.}~\bibnamefont {H{\"u}ning}}, \bibinfo {author} {\bibfnamefont
  {A.}~\bibnamefont {Ignatenko}}, \bibinfo {author} {\bibfnamefont
  {M.}~\bibnamefont {Ilchen}}, \bibinfo {author} {\bibfnamefont
  {A.}~\bibnamefont {Iluk}}, \bibinfo {author} {\bibfnamefont {J.}~\bibnamefont
  {Iversen}}, \bibinfo {author} {\bibfnamefont {M.}~\bibnamefont {Izquierdo}},
  \bibinfo {author} {\bibfnamefont {L.}~\bibnamefont {Jachmann}}, \bibinfo
  {author} {\bibfnamefont {N.}~\bibnamefont {Jardon}}, \bibinfo {author}
  {\bibfnamefont {U.}~\bibnamefont {Jastrow}}, \bibinfo {author} {\bibfnamefont
  {K.}~\bibnamefont {Jensch}}, \bibinfo {author} {\bibfnamefont
  {J.}~\bibnamefont {Jensen}}, \bibinfo {author} {\bibfnamefont
  {M.}~\bibnamefont {Je{\.{z}}abek}}, \bibinfo {author} {\bibfnamefont
  {M.}~\bibnamefont {Jidda}}, \bibinfo {author} {\bibfnamefont
  {H.}~\bibnamefont {Jin}}, \bibinfo {author} {\bibfnamefont {N.}~\bibnamefont
  {Johansson}}, \bibinfo {author} {\bibfnamefont {R.}~\bibnamefont {Jonas}},
  \bibinfo {author} {\bibfnamefont {W.}~\bibnamefont {Kaabi}}, \bibinfo
  {author} {\bibfnamefont {D.}~\bibnamefont {Kaefer}}, \bibinfo {author}
  {\bibfnamefont {R.}~\bibnamefont {Kammering}}, \bibinfo {author}
  {\bibfnamefont {H.}~\bibnamefont {Kapitza}}, \bibinfo {author} {\bibfnamefont
  {S.}~\bibnamefont {Karabekyan}}, \bibinfo {author} {\bibfnamefont
  {S.}~\bibnamefont {Karstensen}}, \bibinfo {author} {\bibfnamefont
  {K.}~\bibnamefont {Kasprzak}}, \bibinfo {author} {\bibfnamefont
  {V.}~\bibnamefont {Katalev}}, \bibinfo {author} {\bibfnamefont
  {D.}~\bibnamefont {Keese}}, \bibinfo {author} {\bibfnamefont
  {B.}~\bibnamefont {Keil}}, \bibinfo {author} {\bibfnamefont {M.}~\bibnamefont
  {Kholopov}}, \bibinfo {author} {\bibfnamefont {M.}~\bibnamefont
  {Killenberger}}, \bibinfo {author} {\bibfnamefont {B.}~\bibnamefont
  {Kitaev}}, \bibinfo {author} {\bibfnamefont {Y.}~\bibnamefont {Klimchenko}},
  \bibinfo {author} {\bibfnamefont {R.}~\bibnamefont {Klos}}, \bibinfo {author}
  {\bibfnamefont {L.}~\bibnamefont {Knebel}}, \bibinfo {author} {\bibfnamefont
  {A.}~\bibnamefont {Koch}}, \bibinfo {author} {\bibfnamefont {M.}~\bibnamefont
  {Koepke}}, \bibinfo {author} {\bibfnamefont {S.}~\bibnamefont {K{\"o}hler}},
  \bibinfo {author} {\bibfnamefont {W.}~\bibnamefont {K{\"o}hler}}, \bibinfo
  {author} {\bibfnamefont {N.}~\bibnamefont {Kohlstrunk}}, \bibinfo {author}
  {\bibfnamefont {Z.}~\bibnamefont {Konopkova}}, \bibinfo {author}
  {\bibfnamefont {A.}~\bibnamefont {Konstantinov}}, \bibinfo {author}
  {\bibfnamefont {W.}~\bibnamefont {Kook}}, \bibinfo {author} {\bibfnamefont
  {W.}~\bibnamefont {Koprek}}, \bibinfo {author} {\bibfnamefont
  {M.}~\bibnamefont {K{\"o}rfer}}, \bibinfo {author} {\bibfnamefont
  {O.}~\bibnamefont {Korth}}, \bibinfo {author} {\bibfnamefont
  {A.}~\bibnamefont {Kosarev}}, \bibinfo {author} {\bibfnamefont
  {K.}~\bibnamefont {Kosi{\'{n}}ski}}, \bibinfo {author} {\bibfnamefont
  {D.}~\bibnamefont {Kostin}}, \bibinfo {author} {\bibfnamefont
  {Y.}~\bibnamefont {Kot}}, \bibinfo {author} {\bibfnamefont {A.}~\bibnamefont
  {Kotarba}}, \bibinfo {author} {\bibfnamefont {T.}~\bibnamefont {Kozak}},
  \bibinfo {author} {\bibfnamefont {V.}~\bibnamefont {Kozak}}, \bibinfo
  {author} {\bibfnamefont {R.}~\bibnamefont {Kramert}}, \bibinfo {author}
  {\bibfnamefont {M.}~\bibnamefont {Krasilnikov}}, \bibinfo {author}
  {\bibfnamefont {A.}~\bibnamefont {Krasnov}}, \bibinfo {author} {\bibfnamefont
  {B.}~\bibnamefont {Krause}}, \bibinfo {author} {\bibfnamefont
  {L.}~\bibnamefont {Kravchuk}}, \bibinfo {author} {\bibfnamefont
  {O.}~\bibnamefont {Krebs}}, \bibinfo {author} {\bibfnamefont
  {R.}~\bibnamefont {Kretschmer}}, \bibinfo {author} {\bibfnamefont
  {J.}~\bibnamefont {Kreutzkamp}}, \bibinfo {author} {\bibfnamefont
  {O.}~\bibnamefont {Kr{\"o}plin}}, \bibinfo {author} {\bibfnamefont
  {K.}~\bibnamefont {Krzysik}}, \bibinfo {author} {\bibfnamefont
  {G.}~\bibnamefont {Kube}}, \bibinfo {author} {\bibfnamefont {H.}~\bibnamefont
  {Kuehn}}, \bibinfo {author} {\bibfnamefont {N.}~\bibnamefont {Kujala}},
  \bibinfo {author} {\bibfnamefont {V.}~\bibnamefont {Kulikov}}, \bibinfo
  {author} {\bibfnamefont {V.}~\bibnamefont {Kuzminych}}, \bibinfo {author}
  {\bibfnamefont {D.}~\bibnamefont {La~Civita}}, \bibinfo {author}
  {\bibfnamefont {M.}~\bibnamefont {Lacroix}}, \bibinfo {author} {\bibfnamefont
  {T.}~\bibnamefont {Lamb}}, \bibinfo {author} {\bibfnamefont {A.}~\bibnamefont
  {Lancetov}}, \bibinfo {author} {\bibfnamefont {M.}~\bibnamefont {Larsson}},
  \bibinfo {author} {\bibfnamefont {D.}~\bibnamefont {Le~Pinvidic}}, \bibinfo
  {author} {\bibfnamefont {S.}~\bibnamefont {Lederer}}, \bibinfo {author}
  {\bibfnamefont {T.}~\bibnamefont {Lensch}}, \bibinfo {author} {\bibfnamefont
  {D.}~\bibnamefont {Lenz}}, \bibinfo {author} {\bibfnamefont {A.}~\bibnamefont
  {Leuschner}}, \bibinfo {author} {\bibfnamefont {F.}~\bibnamefont
  {Levenhagen}}, \bibinfo {author} {\bibfnamefont {Y.}~\bibnamefont {Li}},
  \bibinfo {author} {\bibfnamefont {J.}~\bibnamefont {Liebing}}, \bibinfo
  {author} {\bibfnamefont {L.}~\bibnamefont {Lilje}}, \bibinfo {author}
  {\bibfnamefont {T.}~\bibnamefont {Limberg}}, \bibinfo {author} {\bibfnamefont
  {D.}~\bibnamefont {Lipka}}, \bibinfo {author} {\bibfnamefont
  {B.}~\bibnamefont {List}}, \bibinfo {author} {\bibfnamefont {J.}~\bibnamefont
  {Liu}}, \bibinfo {author} {\bibfnamefont {S.}~\bibnamefont {Liu}}, \bibinfo
  {author} {\bibfnamefont {B.}~\bibnamefont {Lorbeer}}, \bibinfo {author}
  {\bibfnamefont {J.}~\bibnamefont {Lorkiewicz}}, \bibinfo {author}
  {\bibfnamefont {H.~H.}\ \bibnamefont {Lu}}, \bibinfo {author} {\bibfnamefont
  {F.}~\bibnamefont {Ludwig}}, \bibinfo {author} {\bibfnamefont
  {K.}~\bibnamefont {Machau}}, \bibinfo {author} {\bibfnamefont
  {W.}~\bibnamefont {Maciocha}}, \bibinfo {author} {\bibfnamefont
  {C.}~\bibnamefont {Madec}}, \bibinfo {author} {\bibfnamefont
  {C.}~\bibnamefont {Magueur}}, \bibinfo {author} {\bibfnamefont
  {C.}~\bibnamefont {Maiano}}, \bibinfo {author} {\bibfnamefont
  {I.}~\bibnamefont {Maksimova}}, \bibinfo {author} {\bibfnamefont
  {K.}~\bibnamefont {Malcher}}, \bibinfo {author} {\bibfnamefont
  {T.}~\bibnamefont {Maltezopoulos}}, \bibinfo {author} {\bibfnamefont
  {E.}~\bibnamefont {Mamoshkina}}, \bibinfo {author} {\bibfnamefont
  {B.}~\bibnamefont {Manschwetus}}, \bibinfo {author} {\bibfnamefont
  {F.}~\bibnamefont {Marcellini}}, \bibinfo {author} {\bibfnamefont
  {G.}~\bibnamefont {Marinkovic}}, \bibinfo {author} {\bibfnamefont
  {T.}~\bibnamefont {Martinez}}, \bibinfo {author} {\bibfnamefont
  {H.}~\bibnamefont {Martirosyan}}, \bibinfo {author} {\bibfnamefont
  {W.}~\bibnamefont {Maschmann}}, \bibinfo {author} {\bibfnamefont
  {M.}~\bibnamefont {Maslov}}, \bibinfo {author} {\bibfnamefont
  {A.}~\bibnamefont {Matheisen}}, \bibinfo {author} {\bibfnamefont
  {U.}~\bibnamefont {Mavric}}, \bibinfo {author} {\bibfnamefont
  {J.}~\bibnamefont {Mei{\ss}ner}}, \bibinfo {author} {\bibfnamefont
  {K.}~\bibnamefont {Meissner}}, \bibinfo {author} {\bibfnamefont
  {M.}~\bibnamefont {Messerschmidt}}, \bibinfo {author} {\bibfnamefont
  {N.}~\bibnamefont {Meyners}}, \bibinfo {author} {\bibfnamefont
  {G.}~\bibnamefont {Michalski}}, \bibinfo {author} {\bibfnamefont
  {P.}~\bibnamefont {Michelato}}, \bibinfo {author} {\bibfnamefont
  {N.}~\bibnamefont {Mildner}}, \bibinfo {author} {\bibfnamefont
  {M.}~\bibnamefont {Moe}}, \bibinfo {author} {\bibfnamefont {F.}~\bibnamefont
  {Moglia}}, \bibinfo {author} {\bibfnamefont {C.}~\bibnamefont {Mohr}},
  \bibinfo {author} {\bibfnamefont {S.}~\bibnamefont {Mohr}}, \bibinfo {author}
  {\bibfnamefont {W.}~\bibnamefont {M{\"o}ller}}, \bibinfo {author}
  {\bibfnamefont {M.}~\bibnamefont {Mommerz}}, \bibinfo {author} {\bibfnamefont
  {L.}~\bibnamefont {Monaco}}, \bibinfo {author} {\bibfnamefont
  {C.}~\bibnamefont {Montiel}}, \bibinfo {author} {\bibfnamefont
  {M.}~\bibnamefont {Moretti}}, \bibinfo {author} {\bibfnamefont
  {I.}~\bibnamefont {Morozov}}, \bibinfo {author} {\bibfnamefont
  {P.}~\bibnamefont {Morozov}}, and\ \bibinfo {author} {\bibfnamefont
  {D.}~\bibnamefont {Mross}},\ }\bibfield  {title} {\bibinfo {title} {A
  {MHz}-repetition-rate hard x-ray free-electron laser driven by a
  superconducting linear accelerator},\ }\href
  {https://doi.org/10.1038/s41566-020-0607-z} {\bibfield  {journal} {\bibinfo
  {journal} {Nat. Photon.}\ }\textbf {\bibinfo {volume} {14}},\ \bibinfo
  {pages} {391} (\bibinfo {year} {2020})}\BibitemShut {NoStop}%
\bibitem [{\citenamefont {Lorenz}\ \emph {et~al.}(2012)\citenamefont {Lorenz},
  \citenamefont {Kabachnik}, \citenamefont {Weckert}, and\ \citenamefont
  {Vartanyants}}]{Lorenz:PRE86:051911}%
  \BibitemOpen
  \bibfield  {author} {\bibinfo {author} {\bibfnamefont {U.}~\bibnamefont
  {Lorenz}}, \bibinfo {author} {\bibfnamefont {N.~M.}\ \bibnamefont
  {Kabachnik}}, \bibinfo {author} {\bibfnamefont {E.}~\bibnamefont {Weckert}},\
  and\ \bibinfo {author} {\bibfnamefont {I.~A.}\ \bibnamefont {Vartanyants}},\
  }\bibfield  {title} {\bibinfo {title} {Impact of ultrafast electronic damage
  in single-particle x-ray imaging experiments},\ }\href
  {https://doi.org/10.1103/PhysRevE.86.051911} {\bibfield  {journal} {\bibinfo
  {journal} {Phys. Rev. E}\ }\textbf {\bibinfo {volume} {86}},\ \bibinfo
  {pages} {051911} (\bibinfo {year} {2012})},\ \Eprint
  {https://arxiv.org/abs/1206.6960} {arXiv:1206.6960 [physics]}\BibitemShut
  {NoStop}%
\bibitem [{\citenamefont {Barty}\ \emph {et~al.}(2013)\citenamefont {Barty},
  \citenamefont {K{\"u}pper}, and\ \citenamefont
  {Chapman}}]{Barty:ARPC64:415}%
  \BibitemOpen
  \bibfield  {author} {\bibinfo {author} {\bibfnamefont {A.}~\bibnamefont
  {Barty}}, \bibinfo {author} {\bibfnamefont {J.}~\bibnamefont {K{\"u}pper}},\
  and\ \bibinfo {author} {\bibfnamefont {H.~N.}\ \bibnamefont {Chapman}},\
  }\bibfield  {title} {\bibinfo {title} {Molecular imaging using x-ray
  free-electron lasers},\ }\href
  {https://doi.org/10.1146/annurev-physchem-032511-143708} {\bibfield
  {journal} {\bibinfo  {journal} {Annu. Rev. Phys. Chem.}\ }\textbf {\bibinfo
  {volume} {64}},\ \bibinfo {pages} {415} (\bibinfo {year} {2013})}\BibitemShut
  {NoStop}%
\bibitem [{\citenamefont {Pande}\ \emph {et~al.}(2016)\citenamefont {Pande},
  \citenamefont {Hutchison}, \citenamefont {Groenhof}, \citenamefont {Aquila},
  \citenamefont {Robinson}, \citenamefont {Tenboer}, \citenamefont {Basu},
  \citenamefont {Boutet}, \citenamefont {DePonte}, \citenamefont {Liang},
  \citenamefont {White}, \citenamefont {Zatsepin}, \citenamefont {Yefanov},
  \citenamefont {Morozov}, \citenamefont {Oberthuer}, \citenamefont {Gati},
  \citenamefont {Subramanian}, \citenamefont {James}, \citenamefont {Zhao},
  \citenamefont {Koralek}, \citenamefont {Brayshaw}, \citenamefont {Kupitz},
  \citenamefont {Conrad}, \citenamefont {Roy-Chowdhury}, \citenamefont {Coe},
  \citenamefont {Metz}, \citenamefont {Xavier}, \citenamefont {Grant},
  \citenamefont {Koglin}, \citenamefont {Ketawala}, \citenamefont {Fromme},
  \citenamefont {{\v S}rajer}, \citenamefont {Henning}, \citenamefont {Spence},
  \citenamefont {Ourmazd}, \citenamefont {Schwander}, \citenamefont
  {Weierstall}, \citenamefont {Frank}, \citenamefont {Fromme}, \citenamefont
  {Barty}, \citenamefont {Chapman}, \citenamefont {Moffat}, \citenamefont {van
  Thor}, and\ \citenamefont {Schmidt}}]{Pande:Science352:725}%
  \BibitemOpen
  \bibfield  {author} {\bibinfo {author} {\bibfnamefont {K.}~\bibnamefont
  {Pande}}, \bibinfo {author} {\bibfnamefont {C.~D.~M.}\ \bibnamefont
  {Hutchison}}, \bibinfo {author} {\bibfnamefont {G.}~\bibnamefont {Groenhof}},
  \bibinfo {author} {\bibfnamefont {A.}~\bibnamefont {Aquila}}, \bibinfo
  {author} {\bibfnamefont {J.~S.}\ \bibnamefont {Robinson}}, \bibinfo {author}
  {\bibfnamefont {J.}~\bibnamefont {Tenboer}}, \bibinfo {author} {\bibfnamefont
  {S.}~\bibnamefont {Basu}}, \bibinfo {author} {\bibfnamefont {S.}~\bibnamefont
  {Boutet}}, \bibinfo {author} {\bibfnamefont {D.~P.}\ \bibnamefont {DePonte}},
  \bibinfo {author} {\bibfnamefont {M.}~\bibnamefont {Liang}}, \bibinfo
  {author} {\bibfnamefont {T.~A.}\ \bibnamefont {White}}, \bibinfo {author}
  {\bibfnamefont {N.~A.}\ \bibnamefont {Zatsepin}}, \bibinfo {author}
  {\bibfnamefont {O.}~\bibnamefont {Yefanov}}, \bibinfo {author} {\bibfnamefont
  {D.}~\bibnamefont {Morozov}}, \bibinfo {author} {\bibfnamefont
  {D.}~\bibnamefont {Oberthuer}}, \bibinfo {author} {\bibfnamefont
  {C.}~\bibnamefont {Gati}}, \bibinfo {author} {\bibfnamefont {G.}~\bibnamefont
  {Subramanian}}, \bibinfo {author} {\bibfnamefont {D.}~\bibnamefont {James}},
  \bibinfo {author} {\bibfnamefont {Y.}~\bibnamefont {Zhao}}, \bibinfo {author}
  {\bibfnamefont {J.}~\bibnamefont {Koralek}}, \bibinfo {author} {\bibfnamefont
  {J.}~\bibnamefont {Brayshaw}}, \bibinfo {author} {\bibfnamefont
  {C.}~\bibnamefont {Kupitz}}, \bibinfo {author} {\bibfnamefont
  {C.}~\bibnamefont {Conrad}}, \bibinfo {author} {\bibfnamefont
  {S.}~\bibnamefont {Roy-Chowdhury}}, \bibinfo {author} {\bibfnamefont {J.~D.}\
  \bibnamefont {Coe}}, \bibinfo {author} {\bibfnamefont {M.}~\bibnamefont
  {Metz}}, \bibinfo {author} {\bibfnamefont {P.~L.}\ \bibnamefont {Xavier}},
  \bibinfo {author} {\bibfnamefont {T.~D.}\ \bibnamefont {Grant}}, \bibinfo
  {author} {\bibfnamefont {J.~E.}\ \bibnamefont {Koglin}}, \bibinfo {author}
  {\bibfnamefont {G.}~\bibnamefont {Ketawala}}, \bibinfo {author}
  {\bibfnamefont {R.}~\bibnamefont {Fromme}}, \bibinfo {author} {\bibfnamefont
  {V.}~\bibnamefont {{\v S}rajer}}, \bibinfo {author} {\bibfnamefont
  {R.}~\bibnamefont {Henning}}, \bibinfo {author} {\bibfnamefont {J.~C.~H.}\
  \bibnamefont {Spence}}, \bibinfo {author} {\bibfnamefont {A.}~\bibnamefont
  {Ourmazd}}, \bibinfo {author} {\bibfnamefont {P.}~\bibnamefont {Schwander}},
  \bibinfo {author} {\bibfnamefont {U.}~\bibnamefont {Weierstall}}, \bibinfo
  {author} {\bibfnamefont {M.}~\bibnamefont {Frank}}, \bibinfo {author}
  {\bibfnamefont {P.}~\bibnamefont {Fromme}}, \bibinfo {author} {\bibfnamefont
  {A.}~\bibnamefont {Barty}}, \bibinfo {author} {\bibfnamefont {H.~N.}\
  \bibnamefont {Chapman}}, \bibinfo {author} {\bibfnamefont {K.}~\bibnamefont
  {Moffat}}, \bibinfo {author} {\bibfnamefont {J.~J.}\ \bibnamefont {van
  Thor}}, and\ \bibinfo {author} {\bibfnamefont {M.}~\bibnamefont {Schmidt}},\
  }\bibfield  {title} {\bibinfo {title} {Femtosecond structural dynamics drives
  the trans/cis isomerization in photoactive yellow protein},\ }\href
  {https://doi.org/10.1126/science.aad5081} {\bibfield  {journal} {\bibinfo
  {journal} {Science}\ }\textbf {\bibinfo {volume} {352}},\ \bibinfo {pages}
  {725} (\bibinfo {year} {2016})}\BibitemShut {NoStop}%
\bibitem [{\citenamefont {Ourmazd}(2019)}]{Ourmazd:NatMeth16:941}%
  \BibitemOpen
  \bibfield  {author} {\bibinfo {author} {\bibfnamefont {A.}~\bibnamefont
  {Ourmazd}},\ }\bibfield  {title} {\bibinfo {title} {Cryo-{EM}, {XFEL}s and
  the structure conundrum in structural biology},\ }\href
  {https://doi.org/10.1038/s41592-019-0587-4} {\bibfield  {journal} {\bibinfo
  {journal} {Nat. Meth.}\ }\textbf {\bibinfo {volume} {16}},\ \bibinfo {pages}
  {941} (\bibinfo {year} {2019})}\BibitemShut {NoStop}%
\bibitem [{\citenamefont {Ayyer}\ \emph {et~al.}(2019)\citenamefont {Ayyer},
  \citenamefont {Morgan}, \citenamefont {Aquila}, \citenamefont {DeMirci},
  \citenamefont {Hogue}, \citenamefont {Kirian}, \citenamefont {Xavier},
  \citenamefont {Yoon}, \citenamefont {Chapman}, and\ \citenamefont
  {Barty}}]{Ayyer:OptExp27:37816}%
  \BibitemOpen
  \bibfield  {author} {\bibinfo {author} {\bibfnamefont {K.}~\bibnamefont
  {Ayyer}}, \bibinfo {author} {\bibfnamefont {A.~J.}\ \bibnamefont {Morgan}},
  \bibinfo {author} {\bibfnamefont {A.}~\bibnamefont {Aquila}}, \bibinfo
  {author} {\bibfnamefont {H.}~\bibnamefont {DeMirci}}, \bibinfo {author}
  {\bibfnamefont {B.~G.}\ \bibnamefont {Hogue}}, \bibinfo {author}
  {\bibfnamefont {R.~A.}\ \bibnamefont {Kirian}}, \bibinfo {author}
  {\bibfnamefont {P.~L.}\ \bibnamefont {Xavier}}, \bibinfo {author}
  {\bibfnamefont {C.~H.}\ \bibnamefont {Yoon}}, \bibinfo {author}
  {\bibfnamefont {H.~N.}\ \bibnamefont {Chapman}}, and\ \bibinfo {author}
  {\bibfnamefont {A.}~\bibnamefont {Barty}},\ }\bibfield  {title} {\bibinfo
  {title} {Low-signal limit of x-ray single particle diffractive imaging},\
  }\href {https://doi.org/10.1364/oe.27.037816} {\bibfield  {journal} {\bibinfo
   {journal} {Opt. Exp.}\ }\textbf {\bibinfo {volume} {27}},\ \bibinfo {pages}
  {37816} (\bibinfo {year} {2019})}\BibitemShut {NoStop}%
\bibitem [{\citenamefont {Hantke}\ \emph {et~al.}(2018)\citenamefont {Hantke},
  \citenamefont {Bielecki}, \citenamefont {Kulyk}, \citenamefont {Westphal},
  \citenamefont {Larsson}, \citenamefont {Svenda}, \citenamefont {Reddy},
  \citenamefont {Kirian}, \citenamefont {Andreasson}, \citenamefont {Hajdu},\
  and\ \citenamefont {Maia}}]{Hantke:IUCr5:673}%
  \BibitemOpen
  \bibfield  {author} {\bibinfo {author} {\bibfnamefont {M.~F.}\ \bibnamefont
  {Hantke}}, \bibinfo {author} {\bibfnamefont {J.}~\bibnamefont {Bielecki}},
  \bibinfo {author} {\bibfnamefont {O.}~\bibnamefont {Kulyk}}, \bibinfo
  {author} {\bibfnamefont {D.}~\bibnamefont {Westphal}}, \bibinfo {author}
  {\bibfnamefont {D.~S.~D.}\ \bibnamefont {Larsson}}, \bibinfo {author}
  {\bibfnamefont {M.}~\bibnamefont {Svenda}}, \bibinfo {author} {\bibfnamefont
  {H.~K.~N.}\ \bibnamefont {Reddy}}, \bibinfo {author} {\bibfnamefont {R.~A.}\
  \bibnamefont {Kirian}}, \bibinfo {author} {\bibfnamefont {J.}~\bibnamefont
  {Andreasson}}, \bibinfo {author} {\bibfnamefont {J.}~\bibnamefont {Hajdu}},\
  and\ \bibinfo {author} {\bibfnamefont {F.~R. N.~C.}\ \bibnamefont {Maia}},\
  }\bibfield  {title} {\bibinfo {title} {Rayleigh-scattering microscopy for
  tracking and sizing nanoparticles in focused aerosol beams},\ }\href
  {https://doi.org/10.1107/S2052252518010837} {\bibfield  {journal} {\bibinfo
  {journal} {IUCrJ}\ }\textbf {\bibinfo {volume} {5}},\ \bibinfo {pages} {673}
  (\bibinfo {year} {2018})}\BibitemShut {NoStop}%
\bibitem [{\citenamefont {Ayyer}\ \emph {et~al.}(2021)\citenamefont {Ayyer},
  \citenamefont {Xavier}, \citenamefont {Bielecki}, \citenamefont {Shen},
  \citenamefont {Daurer}, \citenamefont {Samanta}, \citenamefont {Awel},
  \citenamefont {Bean}, \citenamefont {Barty}, \citenamefont {Bergemann},
  \citenamefont {Ekeberg}, \citenamefont {Estillore}, \citenamefont {Fangohr},
  \citenamefont {Giewekemeyer}, \citenamefont {Hunter}, \citenamefont
  {Karnevskiy}, \citenamefont {Kirian}, \citenamefont {Kirkwood}, \citenamefont
  {Kim}, \citenamefont {Koliyadu}, \citenamefont {Lange}, \citenamefont
  {Letrun}, \citenamefont {L\"{u}bke}, \citenamefont {Michelat}, \citenamefont
  {Morgan}, \citenamefont {Roth}, \citenamefont {Sato}, \citenamefont
  {Sikorski}, \citenamefont {Schulz}, \citenamefont {Spence}, \citenamefont
  {Vagovic}, \citenamefont {Wollweber}, \citenamefont {Worbs}, \citenamefont
  {Yefanov}, \citenamefont {Zhuang}, \citenamefont {Maia}, \citenamefont
  {Horke}, \citenamefont {K\"{u}pper}, \citenamefont {Loh}, \citenamefont
  {Mancuso}, and\ \citenamefont {Chapman}}]{Ayyer:Optica8:15}%
  \BibitemOpen
  \bibfield  {author} {\bibinfo {author} {\bibfnamefont {K.}~\bibnamefont
  {Ayyer}}, \bibinfo {author} {\bibfnamefont {P.~L.}\ \bibnamefont {Xavier}},
  \bibinfo {author} {\bibfnamefont {J.}~\bibnamefont {Bielecki}}, \bibinfo
  {author} {\bibfnamefont {Z.}~\bibnamefont {Shen}}, \bibinfo {author}
  {\bibfnamefont {B.~J.}\ \bibnamefont {Daurer}}, \bibinfo {author}
  {\bibfnamefont {A.~K.}\ \bibnamefont {Samanta}}, \bibinfo {author}
  {\bibfnamefont {S.}~\bibnamefont {Awel}}, \bibinfo {author} {\bibfnamefont
  {R.}~\bibnamefont {Bean}}, \bibinfo {author} {\bibfnamefont {A.}~\bibnamefont
  {Barty}}, \bibinfo {author} {\bibfnamefont {M.}~\bibnamefont {Bergemann}},
  \bibinfo {author} {\bibfnamefont {T.}~\bibnamefont {Ekeberg}}, \bibinfo
  {author} {\bibfnamefont {A.~D.}\ \bibnamefont {Estillore}}, \bibinfo {author}
  {\bibfnamefont {H.}~\bibnamefont {Fangohr}}, \bibinfo {author} {\bibfnamefont
  {K.}~\bibnamefont {Giewekemeyer}}, \bibinfo {author} {\bibfnamefont {M.~S.}\
  \bibnamefont {Hunter}}, \bibinfo {author} {\bibfnamefont {M.}~\bibnamefont
  {Karnevskiy}}, \bibinfo {author} {\bibfnamefont {R.~A.}\ \bibnamefont
  {Kirian}}, \bibinfo {author} {\bibfnamefont {H.}~\bibnamefont {Kirkwood}},
  \bibinfo {author} {\bibfnamefont {Y.}~\bibnamefont {Kim}}, \bibinfo {author}
  {\bibfnamefont {J.}~\bibnamefont {Koliyadu}}, \bibinfo {author}
  {\bibfnamefont {H.}~\bibnamefont {Lange}}, \bibinfo {author} {\bibfnamefont
  {R.}~\bibnamefont {Letrun}}, \bibinfo {author} {\bibfnamefont
  {J.}~\bibnamefont {L\"{u}bke}}, \bibinfo {author} {\bibfnamefont
  {T.}~\bibnamefont {Michelat}}, \bibinfo {author} {\bibfnamefont {A.~J.}\
  \bibnamefont {Morgan}}, \bibinfo {author} {\bibfnamefont {N.}~\bibnamefont
  {Roth}}, \bibinfo {author} {\bibfnamefont {T.}~\bibnamefont {Sato}}, \bibinfo
  {author} {\bibfnamefont {M.}~\bibnamefont {Sikorski}}, \bibinfo {author}
  {\bibfnamefont {F.}~\bibnamefont {Schulz}}, \bibinfo {author} {\bibfnamefont
  {J.~C.~H.}\ \bibnamefont {Spence}}, \bibinfo {author} {\bibfnamefont
  {P.}~\bibnamefont {Vagovic}}, \bibinfo {author} {\bibfnamefont
  {T.}~\bibnamefont {Wollweber}}, \bibinfo {author} {\bibfnamefont
  {L.}~\bibnamefont {Worbs}}, \bibinfo {author} {\bibfnamefont
  {O.}~\bibnamefont {Yefanov}}, \bibinfo {author} {\bibfnamefont
  {Y.}~\bibnamefont {Zhuang}}, \bibinfo {author} {\bibfnamefont {F.~R. N.~C.}\
  \bibnamefont {Maia}}, \bibinfo {author} {\bibfnamefont {D.~A.}\ \bibnamefont
  {Horke}}, \bibinfo {author} {\bibfnamefont {J.}~\bibnamefont {K\"{u}pper}},
  \bibinfo {author} {\bibfnamefont {N.~D.}\ \bibnamefont {Loh}}, \bibinfo
  {author} {\bibfnamefont {A.~P.}\ \bibnamefont {Mancuso}}, and\ \bibinfo
  {author} {\bibfnamefont {H.~N.}\ \bibnamefont {Chapman}},\ }\bibfield
  {title} {\bibinfo {title} {{3D} diffractive imaging of nanoparticle ensembles
  using an x-ray laser},\ }\href {https://doi.org/10.1364/OPTICA.410851}
  {\bibfield  {journal} {\bibinfo  {journal} {Optica}\ }\textbf {\bibinfo
  {volume} {8}},\ \bibinfo {pages} {15} (\bibinfo {year} {2021})},\ \Eprint
  {https://arxiv.org/abs/2007.13597} {arXiv:2007.13597 [physics]}\BibitemShut
  {NoStop}%
\bibitem [{\citenamefont {Fung}\ \emph {et~al.}(2009)\citenamefont {Fung},
  \citenamefont {Shneerson}, \citenamefont {Saldin}, and\ \citenamefont
  {Ourmazd}}]{Fung:NatPhys5:64}%
  \BibitemOpen
  \bibfield  {author} {\bibinfo {author} {\bibfnamefont {R.}~\bibnamefont
  {Fung}}, \bibinfo {author} {\bibfnamefont {V.}~\bibnamefont {Shneerson}},
  \bibinfo {author} {\bibfnamefont {D.}~\bibnamefont {Saldin}}, and\ \bibinfo
  {author} {\bibfnamefont {A.}~\bibnamefont {Ourmazd}},\ }\bibfield  {title}
  {\bibinfo {title} {Structure from fleeting illumination of faint spinning
  objects in flight},\ }\href {https://doi.org/10.1038/nphys1129} {\bibfield
  {journal} {\bibinfo  {journal} {Nat. Phys.}\ }\textbf {\bibinfo {volume}
  {5}},\ \bibinfo {pages} {64} (\bibinfo {year} {2009})}\BibitemShut {NoStop}%
\bibitem [{\citenamefont {Bortel} and\ \citenamefont
  {Faigel}(2007)}]{Bortel:JStructBiol:158}%
  \BibitemOpen
  \bibfield  {author} {\bibinfo {author} {\bibfnamefont {G.}~\bibnamefont
  {Bortel}} and\ \bibinfo {author} {\bibfnamefont {G.}~\bibnamefont
  {Faigel}},\ }\bibfield  {title} {\bibinfo {title} {Classification of
  continuous diffraction patterns: a numerical study},\ }\href
  {https://doi.org/10.1016/j.jsb.2006.10.018} {\bibfield  {journal} {\bibinfo
  {journal} {J. Struct. Biol.}\ }\textbf {\bibinfo {volume} {158}},\ \bibinfo
  {pages} {10} (\bibinfo {year} {2007})}\BibitemShut {NoStop}%
\bibitem [{\citenamefont {Samanta}\ \emph {et~al.}(2020)\citenamefont
  {Samanta}, \citenamefont {Amin}, \citenamefont {Estillore}, \citenamefont
  {Roth}, \citenamefont {Worbs}, \citenamefont {Horke}, and\ \citenamefont
  {Küpper}}]{Samanta:StructDyn7:024304}%
  \BibitemOpen
  \bibfield  {author} {\bibinfo {author} {\bibfnamefont {A.~K.}\ \bibnamefont
  {Samanta}}, \bibinfo {author} {\bibfnamefont {M.}~\bibnamefont {Amin}},
  \bibinfo {author} {\bibfnamefont {A.~D.}\ \bibnamefont {Estillore}}, \bibinfo
  {author} {\bibfnamefont {N.}~\bibnamefont {Roth}}, \bibinfo {author}
  {\bibfnamefont {L.}~\bibnamefont {Worbs}}, \bibinfo {author} {\bibfnamefont
  {D.~A.}\ \bibnamefont {Horke}}, and\ \bibinfo {author} {\bibfnamefont
  {J.}~\bibnamefont {Küpper}},\ }\bibfield  {title} {\bibinfo {title}
  {Controlled beams of shockfrozen, isolated, biological and artificial
  nanoparticles},\ }\href {https://doi.org/10.1063/4.0000004} {\bibfield
  {journal} {\bibinfo  {journal} {Struct. Dyn.}\ }\textbf {\bibinfo {volume}
  {7}},\ \bibinfo {pages} {024304} (\bibinfo {year} {2020})},\ \Eprint
  {https://arxiv.org/abs/1910.12606} {arXiv:1910.12606 [physics]}\BibitemShut
  {NoStop}%
\bibitem [{\citenamefont {Roth}\ \emph {et~al.}(2018)\citenamefont {Roth},
  \citenamefont {Awel}, \citenamefont {Horke}, and\ \citenamefont
  {Küpper}}]{Roth:JAS124:17}%
  \BibitemOpen
  \bibfield  {author} {\bibinfo {author} {\bibfnamefont {N.}~\bibnamefont
  {Roth}}, \bibinfo {author} {\bibfnamefont {S.}~\bibnamefont {Awel}}, \bibinfo
  {author} {\bibfnamefont {D.~A.}\ \bibnamefont {Horke}}, and\ \bibinfo
  {author} {\bibfnamefont {J.}~\bibnamefont {Küpper}},\ }\bibfield  {title}
  {\bibinfo {title} {Optimizing aerodynamic lenses for single-particle
  imaging},\ }\href {https://doi.org/10.1016/j.jaerosci.2018.06.010} {\bibfield
   {journal} {\bibinfo  {journal} {J. Aerosol. Sci.}\ }\textbf {\bibinfo
  {volume} {124}},\ \bibinfo {pages} {17} (\bibinfo {year} {2018})},\ \Eprint
  {https://arxiv.org/abs/1712.01795} {arXiv:1712.01795 [physics]}\BibitemShut
  {NoStop}%
\bibitem [{\citenamefont {Chang}\ \emph {et~al.}(2015)\citenamefont {Chang},
  \citenamefont {Horke}, \citenamefont {Trippel}, and\ \citenamefont
  {Küpper}}]{Chang:IRPC34:557}%
  \BibitemOpen
  \bibfield  {author} {\bibinfo {author} {\bibfnamefont {Y.-P.}\ \bibnamefont
  {Chang}}, \bibinfo {author} {\bibfnamefont {D.~A.}\ \bibnamefont {Horke}},
  \bibinfo {author} {\bibfnamefont {S.}~\bibnamefont {Trippel}}, and\ \bibinfo
  {author} {\bibfnamefont {J.}~\bibnamefont {Küpper}},\ }\bibfield  {title}
  {\bibinfo {title} {Spatially-controlled complex molecules and their
  applications},\ }\href {https://doi.org/10.1080/0144235X.2015.1077838}
  {\bibfield  {journal} {\bibinfo  {journal} {Int. Rev. Phys. Chem.}\ }\textbf
  {\bibinfo {volume} {34}},\ \bibinfo {pages} {557} (\bibinfo {year} {2015})},\
  \Eprint {https://arxiv.org/abs/1505.05632} {arXiv:1505.05632 [physics]}\BibitemShut {NoStop}%
\bibitem [{\citenamefont {Liu}\ \emph {et~al.}(1995)\citenamefont {Liu},
  \citenamefont {Ziemann}, \citenamefont {Kittelson}, and\ \citenamefont
  {McMurry}}]{Liu:AST22:293}%
  \BibitemOpen
  \bibfield  {author} {\bibinfo {author} {\bibfnamefont {P.}~\bibnamefont
  {Liu}}, \bibinfo {author} {\bibfnamefont {P.~J.}\ \bibnamefont {Ziemann}},
  \bibinfo {author} {\bibfnamefont {D.~B.}\ \bibnamefont {Kittelson}}, and\
  \bibinfo {author} {\bibfnamefont {P.~H.}\ \bibnamefont {McMurry}},\
  }\bibfield  {title} {\bibinfo {title} {Generating particle beams of
  controlled dimensions and divergence: {I.} theory of particle motion in
  aerodynamic lenses and nozzle expansions},\ }\href
  {https://doi.org/10.1080/02786829408959748} {\bibfield  {journal} {\bibinfo
  {journal} {Aerosol Sci. Techn.}\ }\textbf {\bibinfo {volume} {22}},\ \bibinfo
  {pages} {293} (\bibinfo {year} {1995})}\BibitemShut {NoStop}%
\bibitem [{\citenamefont {Wang} and\ \citenamefont
  {McMurry}(2006{\natexlab{a}})}]{Wang:IJMS258:30}%
  \BibitemOpen
  \bibfield  {author} {\bibinfo {author} {\bibfnamefont {X.}~\bibnamefont
  {Wang}} and\ \bibinfo {author} {\bibfnamefont {P.~H.}\ \bibnamefont
  {McMurry}},\ }\bibfield  {title} {\bibinfo {title} {An experimental study of
  nanoparticle focusing with aerodynamic lenses},\ }\href
  {https://doi.org/https://doi.org/10.1016/j.ijms.2006.06.008} {\bibfield
  {journal} {\bibinfo  {journal} {Int. J. Mass Spectrom.}\ }\textbf {\bibinfo
  {volume} {258}},\ \bibinfo {pages} {30} (\bibinfo {year}
  {2006}{\natexlab{a}})}\BibitemShut {NoStop}%
\bibitem [{\citenamefont {Wang}\ \emph {et~al.}(2005)\citenamefont {Wang},
  \citenamefont {Gidwani}, \citenamefont {Girshick}, and\ \citenamefont
  {McMurry}}]{Wang:AST39:624}%
  \BibitemOpen
  \bibfield  {author} {\bibinfo {author} {\bibfnamefont {X.}~\bibnamefont
  {Wang}}, \bibinfo {author} {\bibfnamefont {A.}~\bibnamefont {Gidwani}},
  \bibinfo {author} {\bibfnamefont {S.~L.}\ \bibnamefont {Girshick}}, and\
  \bibinfo {author} {\bibfnamefont {P.~H.}\ \bibnamefont {McMurry}},\
  }\bibfield  {title} {\bibinfo {title} {Aerodynamic focusing of nanoparticles:
  {II.} numerical simulation of particle motion through aerodynamic lenses},\
  }\href {https://doi.org/10.1080/02786820500181950} {\bibfield  {journal}
  {\bibinfo  {journal} {Aerosol Sci. Techn.}\ }\textbf {\bibinfo {volume}
  {39}},\ \bibinfo {pages} {624} (\bibinfo {year} {2005})}\BibitemShut
  {NoStop}%
\bibitem [{\citenamefont {Wang} and\ \citenamefont
  {McMurry}(2006{\natexlab{b}})}]{Wang:AST40:320}%
  \BibitemOpen
  \bibfield  {author} {\bibinfo {author} {\bibfnamefont {X.}~\bibnamefont
  {Wang}} and\ \bibinfo {author} {\bibfnamefont {P.~H.}\ \bibnamefont
  {McMurry}},\ }\bibfield  {title} {\bibinfo {title} {A design tool for
  aerodynamic lens systems},\ }\href
  {https://doi.org/10.1080/02786820600615063} {\bibfield  {journal} {\bibinfo
  {journal} {Aerosol Sci. Technol.}\ }\textbf {\bibinfo {volume} {40}},\
  \bibinfo {pages} {320} (\bibinfo {year} {2006}{\natexlab{b}})}\BibitemShut
  {NoStop}%
\bibitem [{\citenamefont {Worbs}\ \emph {et~al.}(2020)\citenamefont {Worbs},
  \citenamefont {Roth}, \citenamefont {Lübke}, \citenamefont {Estillore},
  \citenamefont {Xavier}, \citenamefont {Samanta}, and\ \citenamefont
  {Küpper}}]{Worbs:geomopt:inprep}%
  \BibitemOpen
  \bibfield  {author} {\bibinfo {author} {\bibfnamefont {L.}~\bibnamefont
  {Worbs}}, \bibinfo {author} {\bibfnamefont {N.}~\bibnamefont {Roth}},
  \bibinfo {author} {\bibfnamefont {J.}~\bibnamefont {Lübke}}, \bibinfo
  {author} {\bibfnamefont {A.}~\bibnamefont {Estillore}}, \bibinfo {author}
  {\bibfnamefont {P.~L.}\ \bibnamefont {Xavier}}, \bibinfo {author}
  {\bibfnamefont {A.~K.}\ \bibnamefont {Samanta}}, and\ \bibinfo {author}
  {\bibfnamefont {J.}~\bibnamefont {Küpper}},\ }\bibfield  {title} {\bibinfo
  {title} {Optimizing the geometry of nanoparticle injectors}} (\bibinfo {year}
  {2020}),\ \bibinfo {note} {in preparation}\BibitemShut {NoStop}%
\bibitem [{\citenamefont {Sobolev}\ \emph {et~al.}(2020)\citenamefont
  {Sobolev}, \citenamefont {Zolotarev}, \citenamefont {Giewekemeyer},
  \citenamefont {Bielecki}, \citenamefont {Okamoto}, \citenamefont {Reddy},
  \citenamefont {Andreasson}, \citenamefont {Ayyer}, \citenamefont {Barak},
  \citenamefont {Bari}, \citenamefont {Barty}, \citenamefont {Bean},
  \citenamefont {Bobkov}, \citenamefont {Chapman}, \citenamefont {Chojnowski},
  \citenamefont {Daurer}, \citenamefont {Dörner}, \citenamefont {Ekeberg},
  \citenamefont {Flückiger}, \citenamefont {Galzitskaya}, \citenamefont
  {Gelisio}, \citenamefont {Hauf}, \citenamefont {Hogue}, \citenamefont
  {Horke}, \citenamefont {Hosseinizadeh}, \citenamefont {Ilyin}, \citenamefont
  {Jung}, \citenamefont {Kim}, \citenamefont {Kim}, \citenamefont {Kirian},
  \citenamefont {Kirkwood}, \citenamefont {Kulyk}, \citenamefont {Letrun},
  \citenamefont {Loh}, \citenamefont {Messerschmidt}, \citenamefont {Mühlig},
  \citenamefont {Ourmazd}, \citenamefont {Raab}, \citenamefont {Rode},
  \citenamefont {Rose}, \citenamefont {Round}, \citenamefont {Sato},
  \citenamefont {Schubert}, \citenamefont {Schwander}, \citenamefont
  {Sellberg}, \citenamefont {Sikorski}, \citenamefont {Silenzi}, \citenamefont
  {Song}, \citenamefont {Spence}, \citenamefont {Stern}, \citenamefont
  {Sztuk-Dambietz}, \citenamefont {Teslyuk}, \citenamefont {Timneanu},
  \citenamefont {Trebbin}, \citenamefont {Uetrecht}, \citenamefont
  {Weinhausen}, \citenamefont {Williams}, \citenamefont {Xavier}, \citenamefont
  {Xu}, \citenamefont {Vartanyants}, \citenamefont {Lamzin}, \citenamefont
  {Mancuso}, and\ \citenamefont {Maia}}]{Sobolev:CommPhys3:97}%
  \BibitemOpen
  \bibfield  {author} {\bibinfo {author} {\bibfnamefont {E.}~\bibnamefont
  {Sobolev}}, \bibinfo {author} {\bibfnamefont {S.}~\bibnamefont {Zolotarev}},
  \bibinfo {author} {\bibfnamefont {K.}~\bibnamefont {Giewekemeyer}}, \bibinfo
  {author} {\bibfnamefont {J.}~\bibnamefont {Bielecki}}, \bibinfo {author}
  {\bibfnamefont {K.}~\bibnamefont {Okamoto}}, \bibinfo {author} {\bibfnamefont
  {H.~K.~N.}\ \bibnamefont {Reddy}}, \bibinfo {author} {\bibfnamefont
  {J.}~\bibnamefont {Andreasson}}, \bibinfo {author} {\bibfnamefont
  {K.}~\bibnamefont {Ayyer}}, \bibinfo {author} {\bibfnamefont
  {I.}~\bibnamefont {Barak}}, \bibinfo {author} {\bibfnamefont
  {S.}~\bibnamefont {Bari}}, \bibinfo {author} {\bibfnamefont {A.}~\bibnamefont
  {Barty}}, \bibinfo {author} {\bibfnamefont {R.}~\bibnamefont {Bean}},
  \bibinfo {author} {\bibfnamefont {S.}~\bibnamefont {Bobkov}}, \bibinfo
  {author} {\bibfnamefont {H.~N.}\ \bibnamefont {Chapman}}, \bibinfo {author}
  {\bibfnamefont {G.}~\bibnamefont {Chojnowski}}, \bibinfo {author}
  {\bibfnamefont {B.~J.}\ \bibnamefont {Daurer}}, \bibinfo {author}
  {\bibfnamefont {K.}~\bibnamefont {Dörner}}, \bibinfo {author} {\bibfnamefont
  {T.}~\bibnamefont {Ekeberg}}, \bibinfo {author} {\bibfnamefont
  {L.}~\bibnamefont {Flückiger}}, \bibinfo {author} {\bibfnamefont
  {O.}~\bibnamefont {Galzitskaya}}, \bibinfo {author} {\bibfnamefont
  {L.}~\bibnamefont {Gelisio}}, \bibinfo {author} {\bibfnamefont
  {S.}~\bibnamefont {Hauf}}, \bibinfo {author} {\bibfnamefont {B.~G.}\
  \bibnamefont {Hogue}}, \bibinfo {author} {\bibfnamefont {D.~A.}\ \bibnamefont
  {Horke}}, \bibinfo {author} {\bibfnamefont {A.}~\bibnamefont
  {Hosseinizadeh}}, \bibinfo {author} {\bibfnamefont {V.}~\bibnamefont
  {Ilyin}}, \bibinfo {author} {\bibfnamefont {C.}~\bibnamefont {Jung}},
  \bibinfo {author} {\bibfnamefont {C.}~\bibnamefont {Kim}}, \bibinfo {author}
  {\bibfnamefont {Y.}~\bibnamefont {Kim}}, \bibinfo {author} {\bibfnamefont
  {R.~A.}\ \bibnamefont {Kirian}}, \bibinfo {author} {\bibfnamefont
  {H.}~\bibnamefont {Kirkwood}}, \bibinfo {author} {\bibfnamefont
  {O.}~\bibnamefont {Kulyk}}, \bibinfo {author} {\bibfnamefont
  {R.}~\bibnamefont {Letrun}}, \bibinfo {author} {\bibfnamefont
  {D.}~\bibnamefont {Loh}}, \bibinfo {author} {\bibfnamefont {M.}~\bibnamefont
  {Messerschmidt}}, \bibinfo {author} {\bibfnamefont {K.}~\bibnamefont
  {Mühlig}}, \bibinfo {author} {\bibfnamefont {A.}~\bibnamefont {Ourmazd}},
  \bibinfo {author} {\bibfnamefont {N.}~\bibnamefont {Raab}}, \bibinfo {author}
  {\bibfnamefont {A.~V.}\ \bibnamefont {Rode}}, \bibinfo {author}
  {\bibfnamefont {M.}~\bibnamefont {Rose}}, \bibinfo {author} {\bibfnamefont
  {A.}~\bibnamefont {Round}}, \bibinfo {author} {\bibfnamefont
  {T.}~\bibnamefont {Sato}}, \bibinfo {author} {\bibfnamefont {R.}~\bibnamefont
  {Schubert}}, \bibinfo {author} {\bibfnamefont {P.}~\bibnamefont {Schwander}},
  \bibinfo {author} {\bibfnamefont {J.~A.}\ \bibnamefont {Sellberg}}, \bibinfo
  {author} {\bibfnamefont {M.}~\bibnamefont {Sikorski}}, \bibinfo {author}
  {\bibfnamefont {A.}~\bibnamefont {Silenzi}}, \bibinfo {author} {\bibfnamefont
  {C.}~\bibnamefont {Song}}, \bibinfo {author} {\bibfnamefont {J.~C.~H.}\
  \bibnamefont {Spence}}, \bibinfo {author} {\bibfnamefont {S.}~\bibnamefont
  {Stern}}, \bibinfo {author} {\bibfnamefont {J.}~\bibnamefont
  {Sztuk-Dambietz}}, \bibinfo {author} {\bibfnamefont {A.}~\bibnamefont
  {Teslyuk}}, \bibinfo {author} {\bibfnamefont {N.}~\bibnamefont {Timneanu}},
  \bibinfo {author} {\bibfnamefont {M.}~\bibnamefont {Trebbin}}, \bibinfo
  {author} {\bibfnamefont {C.}~\bibnamefont {Uetrecht}}, \bibinfo {author}
  {\bibfnamefont {B.}~\bibnamefont {Weinhausen}}, \bibinfo {author}
  {\bibfnamefont {G.~J.}\ \bibnamefont {Williams}}, \bibinfo {author}
  {\bibfnamefont {P.~L.}\ \bibnamefont {Xavier}}, \bibinfo {author}
  {\bibfnamefont {C.}~\bibnamefont {Xu}}, \bibinfo {author} {\bibfnamefont
  {I.}~\bibnamefont {Vartanyants}}, \bibinfo {author} {\bibfnamefont
  {V.}~\bibnamefont {Lamzin}}, \bibinfo {author} {\bibfnamefont
  {A.}~\bibnamefont {Mancuso}}, and\ \bibinfo {author} {\bibfnamefont {F.~R.
  N.~C.}\ \bibnamefont {Maia}},\ }\bibfield  {title} {\bibinfo {title}
  {Megahertz single-particle imaging at the {E}uropean {XFEL}},\ }\href
  {https://doi.org/10.1038/s42005-020-0362-y} {\bibfield  {journal} {\bibinfo
  {journal} {Comm. Phys}\ }\textbf {\bibinfo {volume} {3}},\ \bibinfo {pages}
  {97} (\bibinfo {year} {2020})},\ \Eprint {https://arxiv.org/abs/1912.10796}
  {arXiv:1912.10796 [physics]}\BibitemShut {NoStop}%
\bibitem [{\citenamefont {Bielecki}\ \emph {et~al.}(2019)\citenamefont
  {Bielecki}, \citenamefont {Hantke}, \citenamefont {Daurer}, \citenamefont
  {Reddy}, \citenamefont {Hasse}, \citenamefont {Larsson}, \citenamefont
  {Gunn}, \citenamefont {Svenda}, \citenamefont {Munke}, \citenamefont
  {Sellberg}, \citenamefont {Flueckiger}, \citenamefont {Pietrini},
  \citenamefont {Nettelblad}, \citenamefont {Lundholm}, \citenamefont
  {Carlsson}, \citenamefont {Okamoto}, \citenamefont {Timneanu}, \citenamefont
  {Westphal}, \citenamefont {Kulyk}, \citenamefont {Higashiura}, \citenamefont
  {van~der Schot}, \citenamefont {Loh}, \citenamefont {Wysong}, \citenamefont
  {Bostedt}, \citenamefont {Gorkhover}, \citenamefont {Iwan}, \citenamefont
  {Seibert}, \citenamefont {Osipov}, \citenamefont {Walter}, \citenamefont
  {Hart}, \citenamefont {Bucher}, \citenamefont {Ulmer}, \citenamefont {Ray},
  \citenamefont {Carini}, \citenamefont {Ferguson}, \citenamefont {Andersson},
  \citenamefont {Andreasson}, \citenamefont {Hajdu}, and\ \citenamefont
  {Maia}}]{Bielecki:SciAdv5:eaav8801}%
  \BibitemOpen
  \bibfield  {author} {\bibinfo {author} {\bibfnamefont {J.}~\bibnamefont
  {Bielecki}}, \bibinfo {author} {\bibfnamefont {M.~F.}\ \bibnamefont
  {Hantke}}, \bibinfo {author} {\bibfnamefont {B.~J.}\ \bibnamefont {Daurer}},
  \bibinfo {author} {\bibfnamefont {H.~K.~N.}\ \bibnamefont {Reddy}}, \bibinfo
  {author} {\bibfnamefont {D.}~\bibnamefont {Hasse}}, \bibinfo {author}
  {\bibfnamefont {D.~S.~D.}\ \bibnamefont {Larsson}}, \bibinfo {author}
  {\bibfnamefont {L.~H.}\ \bibnamefont {Gunn}}, \bibinfo {author}
  {\bibfnamefont {M.}~\bibnamefont {Svenda}}, \bibinfo {author} {\bibfnamefont
  {A.}~\bibnamefont {Munke}}, \bibinfo {author} {\bibfnamefont {J.~A.}\
  \bibnamefont {Sellberg}}, \bibinfo {author} {\bibfnamefont {L.}~\bibnamefont
  {Flueckiger}}, \bibinfo {author} {\bibfnamefont {A.}~\bibnamefont
  {Pietrini}}, \bibinfo {author} {\bibfnamefont {C.}~\bibnamefont
  {Nettelblad}}, \bibinfo {author} {\bibfnamefont {I.}~\bibnamefont
  {Lundholm}}, \bibinfo {author} {\bibfnamefont {G.}~\bibnamefont {Carlsson}},
  \bibinfo {author} {\bibfnamefont {K.}~\bibnamefont {Okamoto}}, \bibinfo
  {author} {\bibfnamefont {N.}~\bibnamefont {Timneanu}}, \bibinfo {author}
  {\bibfnamefont {D.}~\bibnamefont {Westphal}}, \bibinfo {author}
  {\bibfnamefont {O.}~\bibnamefont {Kulyk}}, \bibinfo {author} {\bibfnamefont
  {A.}~\bibnamefont {Higashiura}}, \bibinfo {author} {\bibfnamefont
  {G.}~\bibnamefont {van~der Schot}}, \bibinfo {author} {\bibfnamefont
  {N.-T.~D.}\ \bibnamefont {Loh}}, \bibinfo {author} {\bibfnamefont {T.~E.}\
  \bibnamefont {Wysong}}, \bibinfo {author} {\bibfnamefont {C.}~\bibnamefont
  {Bostedt}}, \bibinfo {author} {\bibfnamefont {T.}~\bibnamefont {Gorkhover}},
  \bibinfo {author} {\bibfnamefont {B.}~\bibnamefont {Iwan}}, \bibinfo {author}
  {\bibfnamefont {M.~M.}\ \bibnamefont {Seibert}}, \bibinfo {author}
  {\bibfnamefont {T.}~\bibnamefont {Osipov}}, \bibinfo {author} {\bibfnamefont
  {P.}~\bibnamefont {Walter}}, \bibinfo {author} {\bibfnamefont
  {P.}~\bibnamefont {Hart}}, \bibinfo {author} {\bibfnamefont {M.}~\bibnamefont
  {Bucher}}, \bibinfo {author} {\bibfnamefont {A.}~\bibnamefont {Ulmer}},
  \bibinfo {author} {\bibfnamefont {D.}~\bibnamefont {Ray}}, \bibinfo {author}
  {\bibfnamefont {G.}~\bibnamefont {Carini}}, \bibinfo {author} {\bibfnamefont
  {K.~R.}\ \bibnamefont {Ferguson}}, \bibinfo {author} {\bibfnamefont
  {I.}~\bibnamefont {Andersson}}, \bibinfo {author} {\bibfnamefont
  {J.}~\bibnamefont {Andreasson}}, \bibinfo {author} {\bibfnamefont
  {J.}~\bibnamefont {Hajdu}}, and\ \bibinfo {author} {\bibfnamefont {F.~R.
  N.~C.}\ \bibnamefont {Maia}},\ }\bibfield  {title} {\bibinfo {title}
  {Electrospray sample injection for single-particle imaging with x-ray
  lasers},\ }\href {https://doi.org/10.1126/sciadv.aav8801} {\bibfield
  {journal} {\bibinfo  {journal} {Science Advances}\ }\textbf {\bibinfo
  {volume} {5}},\ \bibinfo {pages} {eaav8801} (\bibinfo {year}
  {2019})}\BibitemShut {NoStop}%
\bibitem [{\citenamefont {Roth}\ \emph {et~al.}(2020)\citenamefont {Roth},
  \citenamefont {Amin}, \citenamefont {Samanta}, and\ \citenamefont
  {Küpper}}]{Roth:microscopic-drag-force:inprep}%
  \BibitemOpen
  \bibfield  {author} {\bibinfo {author} {\bibfnamefont {N.}~\bibnamefont
  {Roth}}, \bibinfo {author} {\bibfnamefont {M.}~\bibnamefont {Amin}}, \bibinfo
  {author} {\bibfnamefont {A.~K.}\ \bibnamefont {Samanta}}, and\ \bibinfo
  {author} {\bibfnamefont {J.}~\bibnamefont {Küpper}},\ }\bibfield  {title}
  {\bibinfo {title} {Microscopic force for aerosol transport},\ }\Eprint
  {https://arxiv.org/abs/2006.10652} {arXiv:2006.10652 [physics]}  (\bibinfo
  {year} {2020}),\ \bibinfo {note} {in preparation}\BibitemShut {NoStop}%
\bibitem [{\citenamefont {Eckerskorn}\ \emph {et~al.}(2015)\citenamefont
  {Eckerskorn}, \citenamefont {Bowman}, \citenamefont {Kirian}, \citenamefont
  {Awel}, \citenamefont {Wiedorn}, \citenamefont {K{\"u}pper}, \citenamefont
  {Padgett}, \citenamefont {Chapman}, and\ \citenamefont
  {Rode}}]{Eckerskorn:PRAppl4:064001}%
  \BibitemOpen
  \bibfield  {author} {\bibinfo {author} {\bibfnamefont {N.}~\bibnamefont
  {Eckerskorn}}, \bibinfo {author} {\bibfnamefont {R.}~\bibnamefont {Bowman}},
  \bibinfo {author} {\bibfnamefont {R.~A.}\ \bibnamefont {Kirian}}, \bibinfo
  {author} {\bibfnamefont {S.}~\bibnamefont {Awel}}, \bibinfo {author}
  {\bibfnamefont {M.}~\bibnamefont {Wiedorn}}, \bibinfo {author} {\bibfnamefont
  {J.}~\bibnamefont {K{\"u}pper}}, \bibinfo {author} {\bibfnamefont {M.~J.}\
  \bibnamefont {Padgett}}, \bibinfo {author} {\bibfnamefont {H.~N.}\
  \bibnamefont {Chapman}}, and\ \bibinfo {author} {\bibfnamefont {A.~V.}\
  \bibnamefont {Rode}},\ }\bibfield  {title} {\bibinfo {title} {Optically
  induced forces imposed in an optical funnel on a stream of particles in air
  or vacuum},\ }\href {https://doi.org/10.1103/PhysRevApplied.4.064001}
  {\bibfield  {journal} {\bibinfo  {journal} {Phys. Rev. Appl.}\ }\textbf
  {\bibinfo {volume} {4}},\ \bibinfo {pages} {064001} (\bibinfo {year}
  {2015})}\BibitemShut {NoStop}%
\bibitem [{\citenamefont {Desyatnikov}\ \emph {et~al.}(2009)\citenamefont
  {Desyatnikov}, \citenamefont {Shvedov}, \citenamefont {Rode}, \citenamefont
  {Krolikowski}, and\ \citenamefont {Kivshar}}]{Desyatnikov:OptExp17:8201}%
  \BibitemOpen
  \bibfield  {author} {\bibinfo {author} {\bibfnamefont {A.~S.}\ \bibnamefont
  {Desyatnikov}}, \bibinfo {author} {\bibfnamefont {V.~G.}\ \bibnamefont
  {Shvedov}}, \bibinfo {author} {\bibfnamefont {A.~V.}\ \bibnamefont {Rode}},
  \bibinfo {author} {\bibfnamefont {W.}~\bibnamefont {Krolikowski}}, and\
  \bibinfo {author} {\bibfnamefont {Y.~S.}\ \bibnamefont {Kivshar}},\
  }\bibfield  {title} {\bibinfo {title} {Photophoretic manipulation of
  absorbing aerosol particles with vortex beams: theory versus experiment},\
  }\href {https://doi.org/10.1364/OE.17.008201} {\bibfield  {journal} {\bibinfo
   {journal} {Opt. Exp.}\ }\textbf {\bibinfo {volume} {17}},\ \bibinfo {pages}
  {8201} (\bibinfo {year} {2009})}\BibitemShut {NoStop}%
\bibitem [{\citenamefont {Daurer}\ \emph {et~al.}(2017)\citenamefont {Daurer},
  \citenamefont {Okamoto}, \citenamefont {Bielecki}, \citenamefont {Maia},
  \citenamefont {M{\"{u}}hlig}, \citenamefont {Seibert}, \citenamefont
  {Hantke}, \citenamefont {Nettelblad}, \citenamefont {Benner}, \citenamefont
  {Svenda}, \citenamefont {Timneanu}, \citenamefont {Ekeberg}, \citenamefont
  {Loh}, \citenamefont {Pietrini}, \citenamefont {Zani}, \citenamefont {Rath},
  \citenamefont {Westphal}, \citenamefont {Kirian}, \citenamefont {Awel},
  \citenamefont {Wiedorn}, \citenamefont {van~der Schot}, \citenamefont
  {Carlsson}, \citenamefont {Hasse}, \citenamefont {Sellberg}, \citenamefont
  {Barty}, \citenamefont {Andreasson}, \citenamefont {Boutet}, \citenamefont
  {Williams}, \citenamefont {Koglin}, \citenamefont {Andersson}, \citenamefont
  {Hajdu}, and\ \citenamefont {Larsson}}]{Daurer:IUCrJ4:3}%
  \BibitemOpen
  \bibfield  {author} {\bibinfo {author} {\bibfnamefont {B.~J.}\ \bibnamefont
  {Daurer}}, \bibinfo {author} {\bibfnamefont {K.}~\bibnamefont {Okamoto}},
  \bibinfo {author} {\bibfnamefont {J.}~\bibnamefont {Bielecki}}, \bibinfo
  {author} {\bibfnamefont {F.~R. N.~C.}\ \bibnamefont {Maia}}, \bibinfo
  {author} {\bibfnamefont {K.}~\bibnamefont {M{\"{u}}hlig}}, \bibinfo {author}
  {\bibfnamefont {M.~M.}\ \bibnamefont {Seibert}}, \bibinfo {author}
  {\bibfnamefont {M.~F.}\ \bibnamefont {Hantke}}, \bibinfo {author}
  {\bibfnamefont {C.}~\bibnamefont {Nettelblad}}, \bibinfo {author}
  {\bibfnamefont {W.~H.}\ \bibnamefont {Benner}}, \bibinfo {author}
  {\bibfnamefont {M.}~\bibnamefont {Svenda}}, \bibinfo {author} {\bibfnamefont
  {N.}~\bibnamefont {Timneanu}}, \bibinfo {author} {\bibfnamefont
  {T.}~\bibnamefont {Ekeberg}}, \bibinfo {author} {\bibfnamefont {N.~D.}\
  \bibnamefont {Loh}}, \bibinfo {author} {\bibfnamefont {A.}~\bibnamefont
  {Pietrini}}, \bibinfo {author} {\bibfnamefont {A.}~\bibnamefont {Zani}},
  \bibinfo {author} {\bibfnamefont {A.~D.}\ \bibnamefont {Rath}}, \bibinfo
  {author} {\bibfnamefont {D.}~\bibnamefont {Westphal}}, \bibinfo {author}
  {\bibfnamefont {R.~A.}\ \bibnamefont {Kirian}}, \bibinfo {author}
  {\bibfnamefont {S.}~\bibnamefont {Awel}}, \bibinfo {author} {\bibfnamefont
  {M.~O.}\ \bibnamefont {Wiedorn}}, \bibinfo {author} {\bibfnamefont
  {G.}~\bibnamefont {van~der Schot}}, \bibinfo {author} {\bibfnamefont {G.~H.}\
  \bibnamefont {Carlsson}}, \bibinfo {author} {\bibfnamefont {D.}~\bibnamefont
  {Hasse}}, \bibinfo {author} {\bibfnamefont {J.~A.}\ \bibnamefont {Sellberg}},
  \bibinfo {author} {\bibfnamefont {A.}~\bibnamefont {Barty}}, \bibinfo
  {author} {\bibfnamefont {J.}~\bibnamefont {Andreasson}}, \bibinfo {author}
  {\bibfnamefont {S.}~\bibnamefont {Boutet}}, \bibinfo {author} {\bibfnamefont
  {G.}~\bibnamefont {Williams}}, \bibinfo {author} {\bibfnamefont
  {J.}~\bibnamefont {Koglin}}, \bibinfo {author} {\bibfnamefont
  {I.}~\bibnamefont {Andersson}}, \bibinfo {author} {\bibfnamefont
  {J.}~\bibnamefont {Hajdu}}, and\ \bibinfo {author} {\bibfnamefont
  {D.~S.~D.}\ \bibnamefont {Larsson}},\ }\bibfield  {title} {\bibinfo {title}
  {Experimental strategies for imaging bioparticles with femtosecond hard x-ray
  pulses},\ }\href {https://doi.org/10.1107/S2052252517003591} {\bibfield
  {journal} {\bibinfo  {journal} {IUCrJ}\ }\textbf {\bibinfo {volume} {4}},\
  \bibinfo {pages} {251} (\bibinfo {year} {2017})}\BibitemShut {NoStop}%
\bibitem [{\citenamefont {Yachmenev}\ \emph {et~al.}(2019)\citenamefont
  {Yachmenev}, \citenamefont {Onvlee}, \citenamefont {Zak}, \citenamefont
  {Owens}, and\ \citenamefont {Küpper}}]{Yachmenev:PRL123:243202}%
  \BibitemOpen
  \bibfield  {author} {\bibinfo {author} {\bibfnamefont {A.}~\bibnamefont
  {Yachmenev}}, \bibinfo {author} {\bibfnamefont {J.}~\bibnamefont {Onvlee}},
  \bibinfo {author} {\bibfnamefont {E.}~\bibnamefont {Zak}}, \bibinfo {author}
  {\bibfnamefont {A.}~\bibnamefont {Owens}}, and\ \bibinfo {author}
  {\bibfnamefont {J.}~\bibnamefont {Küpper}},\ }\bibfield  {title} {\bibinfo
  {title} {Field-induced diastereomers for chiral separation},\ }\href
  {https://doi.org/10.1103/PhysRevLett.123.243202} {\bibfield  {journal}
  {\bibinfo  {journal} {Phys. Rev. Lett.}\ }\textbf {\bibinfo {volume} {123}},\
  \bibinfo {pages} {243202} (\bibinfo {year} {2019})},\ \Eprint
  {https://arxiv.org/abs/1905.07166} {arXiv:1905.07166 [physics]}\BibitemShut
  {NoStop}%
\bibitem [{\citenamefont {Stapelfeldt} and\ \citenamefont
  {Seideman}(2003)}]{Stapelfeldt:RMP75:543}%
  \BibitemOpen
  \bibfield  {author} {\bibinfo {author} {\bibfnamefont {H.}~\bibnamefont
  {Stapelfeldt}} and\ \bibinfo {author} {\bibfnamefont {T.}~\bibnamefont
  {Seideman}},\ }\bibfield  {title} {\bibinfo {title} {Colloquium: Aligning
  molecules with strong laser pulses},\ }\href
  {https://doi.org/10.1103/RevModPhys.75.543} {\bibfield  {journal} {\bibinfo
  {journal} {Rev. Mod. Phys.}\ }\textbf {\bibinfo {volume} {75}},\ \bibinfo
  {pages} {543} (\bibinfo {year} {2003})}\BibitemShut {NoStop}%
\bibitem [{\citenamefont {Spence} and\ \citenamefont
  {Doak}(2004)}]{Spence:PRL92:198102}%
  \BibitemOpen
  \bibfield  {author} {\bibinfo {author} {\bibfnamefont {J.~C.~H.}\
  \bibnamefont {Spence}} and\ \bibinfo {author} {\bibfnamefont {R.~B.}\
  \bibnamefont {Doak}},\ }\bibfield  {title} {\bibinfo {title} {Single molecule
  diffraction},\ }\href {https://doi.org/10.1103/PhysRevLett.92.198102}
  {\bibfield  {journal} {\bibinfo  {journal} {Phys. Rev. Lett.}\ }\textbf
  {\bibinfo {volume} {92}},\ \bibinfo {pages} {198102} (\bibinfo {year}
  {2004})}\BibitemShut {NoStop}%
\bibitem [{\citenamefont {Holmegaard}\ \emph {et~al.}(2009)\citenamefont
  {Holmegaard}, \citenamefont {Nielsen}, \citenamefont {Nevo}, \citenamefont
  {Stapelfeldt}, \citenamefont {Filsinger}, \citenamefont {K{\"u}pper}, and\
  \citenamefont {Meijer}}]{Holmegaard:PRL102:023001}%
  \BibitemOpen
  \bibfield  {author} {\bibinfo {author} {\bibfnamefont {L.}~\bibnamefont
  {Holmegaard}}, \bibinfo {author} {\bibfnamefont {J.~H.}\ \bibnamefont
  {Nielsen}}, \bibinfo {author} {\bibfnamefont {I.}~\bibnamefont {Nevo}},
  \bibinfo {author} {\bibfnamefont {H.}~\bibnamefont {Stapelfeldt}}, \bibinfo
  {author} {\bibfnamefont {F.}~\bibnamefont {Filsinger}}, \bibinfo {author}
  {\bibfnamefont {J.}~\bibnamefont {K{\"u}pper}}, and\ \bibinfo {author}
  {\bibfnamefont {G.}~\bibnamefont {Meijer}},\ }\bibfield  {title} {\bibinfo
  {title} {Laser-induced alignment and orientation of quantum-state-selected
  large molecules},\ }\href {https://doi.org/10.1103/PhysRevLett.102.023001}
  {\bibfield  {journal} {\bibinfo  {journal} {Phys. Rev. Lett.}\ }\textbf
  {\bibinfo {volume} {102}},\ \bibinfo {pages} {023001} (\bibinfo {year}
  {2009})},\ \Eprint {https://arxiv.org/abs/0810.2307} {arXiv:0810.2307
  [physics]}\BibitemShut {NoStop}%
\bibitem [{\citenamefont {K{\"u}pper}\ \emph {et~al.}(2014)\citenamefont
  {K{\"u}pper}, \citenamefont {Stern}, \citenamefont {Holmegaard},
  \citenamefont {Filsinger}, \citenamefont {Rouz\'{e}e}, \citenamefont
  {Rudenko}, \citenamefont {Johnsson}, \citenamefont {Martin}, \citenamefont
  {Adolph}, \citenamefont {Aquila}, \citenamefont {Bajt}, \citenamefont
  {Barty}, \citenamefont {Bostedt}, \citenamefont {Bozek}, \citenamefont
  {Caleman}, \citenamefont {Coffee}, \citenamefont {Coppola}, \citenamefont
  {Delmas}, \citenamefont {Epp}, \citenamefont {Erk}, \citenamefont {Foucar},
  \citenamefont {Gorkhover}, \citenamefont {Gumprecht}, \citenamefont
  {Hartmann}, \citenamefont {Hartmann}, \citenamefont {Hauser}, \citenamefont
  {Holl}, \citenamefont {H{\"o}mke}, \citenamefont {Kimmel}, \citenamefont
  {Krasniqi}, \citenamefont {K{\"u}hnel}, \citenamefont {Maurer}, \citenamefont
  {Messerschmidt}, \citenamefont {Moshammer}, \citenamefont {Reich},
  \citenamefont {Rudek}, \citenamefont {Santra}, \citenamefont {Schlichting},
  \citenamefont {Schmidt}, \citenamefont {Schorb}, \citenamefont {Schulz},
  \citenamefont {Soltau}, \citenamefont {Spence}, \citenamefont {Starodub},
  \citenamefont {Str{\"u}der}, \citenamefont {Th{\o}gersen}, \citenamefont
  {Vrakking}, \citenamefont {Weidenspointner}, \citenamefont {White},
  \citenamefont {Wunderer}, \citenamefont {Meijer}, \citenamefont {Ullrich},
  \citenamefont {Stapelfeldt}, \citenamefont {Rolles}, and\ \citenamefont
  {Chapman}}]{Kuepper:PRL112:083002}%
  \BibitemOpen
  \bibfield  {author} {\bibinfo {author} {\bibfnamefont {J.}~\bibnamefont
  {K{\"u}pper}}, \bibinfo {author} {\bibfnamefont {S.}~\bibnamefont {Stern}},
  \bibinfo {author} {\bibfnamefont {L.}~\bibnamefont {Holmegaard}}, \bibinfo
  {author} {\bibfnamefont {F.}~\bibnamefont {Filsinger}}, \bibinfo {author}
  {\bibfnamefont {A.}~\bibnamefont {Rouz\'{e}e}}, \bibinfo {author}
  {\bibfnamefont {A.}~\bibnamefont {Rudenko}}, \bibinfo {author} {\bibfnamefont
  {P.}~\bibnamefont {Johnsson}}, \bibinfo {author} {\bibfnamefont {A.~V.}\
  \bibnamefont {Martin}}, \bibinfo {author} {\bibfnamefont {M.}~\bibnamefont
  {Adolph}}, \bibinfo {author} {\bibfnamefont {A.}~\bibnamefont {Aquila}},
  \bibinfo {author} {\bibfnamefont {S.}~\bibnamefont {Bajt}}, \bibinfo {author}
  {\bibfnamefont {A.}~\bibnamefont {Barty}}, \bibinfo {author} {\bibfnamefont
  {C.}~\bibnamefont {Bostedt}}, \bibinfo {author} {\bibfnamefont
  {J.}~\bibnamefont {Bozek}}, \bibinfo {author} {\bibfnamefont
  {C.}~\bibnamefont {Caleman}}, \bibinfo {author} {\bibfnamefont
  {R.}~\bibnamefont {Coffee}}, \bibinfo {author} {\bibfnamefont
  {N.}~\bibnamefont {Coppola}}, \bibinfo {author} {\bibfnamefont
  {T.}~\bibnamefont {Delmas}}, \bibinfo {author} {\bibfnamefont
  {S.}~\bibnamefont {Epp}}, \bibinfo {author} {\bibfnamefont {B.}~\bibnamefont
  {Erk}}, \bibinfo {author} {\bibfnamefont {L.}~\bibnamefont {Foucar}},
  \bibinfo {author} {\bibfnamefont {T.}~\bibnamefont {Gorkhover}}, \bibinfo
  {author} {\bibfnamefont {L.}~\bibnamefont {Gumprecht}}, \bibinfo {author}
  {\bibfnamefont {A.}~\bibnamefont {Hartmann}}, \bibinfo {author}
  {\bibfnamefont {R.}~\bibnamefont {Hartmann}}, \bibinfo {author}
  {\bibfnamefont {G.}~\bibnamefont {Hauser}}, \bibinfo {author} {\bibfnamefont
  {P.}~\bibnamefont {Holl}}, \bibinfo {author} {\bibfnamefont {A.}~\bibnamefont
  {H{\"o}mke}}, \bibinfo {author} {\bibfnamefont {N.}~\bibnamefont {Kimmel}},
  \bibinfo {author} {\bibfnamefont {F.}~\bibnamefont {Krasniqi}}, \bibinfo
  {author} {\bibfnamefont {K.-U.}\ \bibnamefont {K{\"u}hnel}}, \bibinfo
  {author} {\bibfnamefont {J.}~\bibnamefont {Maurer}}, \bibinfo {author}
  {\bibfnamefont {M.}~\bibnamefont {Messerschmidt}}, \bibinfo {author}
  {\bibfnamefont {R.}~\bibnamefont {Moshammer}}, \bibinfo {author}
  {\bibfnamefont {C.}~\bibnamefont {Reich}}, \bibinfo {author} {\bibfnamefont
  {B.}~\bibnamefont {Rudek}}, \bibinfo {author} {\bibfnamefont
  {R.}~\bibnamefont {Santra}}, \bibinfo {author} {\bibfnamefont
  {I.}~\bibnamefont {Schlichting}}, \bibinfo {author} {\bibfnamefont
  {C.}~\bibnamefont {Schmidt}}, \bibinfo {author} {\bibfnamefont
  {S.}~\bibnamefont {Schorb}}, \bibinfo {author} {\bibfnamefont
  {J.}~\bibnamefont {Schulz}}, \bibinfo {author} {\bibfnamefont
  {H.}~\bibnamefont {Soltau}}, \bibinfo {author} {\bibfnamefont {J.~C.~H.}\
  \bibnamefont {Spence}}, \bibinfo {author} {\bibfnamefont {D.}~\bibnamefont
  {Starodub}}, \bibinfo {author} {\bibfnamefont {L.}~\bibnamefont
  {Str{\"u}der}}, \bibinfo {author} {\bibfnamefont {J.}~\bibnamefont
  {Th{\o}gersen}}, \bibinfo {author} {\bibfnamefont {M.~J.~J.}\ \bibnamefont
  {Vrakking}}, \bibinfo {author} {\bibfnamefont {G.}~\bibnamefont
  {Weidenspointner}}, \bibinfo {author} {\bibfnamefont {T.~A.}\ \bibnamefont
  {White}}, \bibinfo {author} {\bibfnamefont {C.}~\bibnamefont {Wunderer}},
  \bibinfo {author} {\bibfnamefont {G.}~\bibnamefont {Meijer}}, \bibinfo
  {author} {\bibfnamefont {J.}~\bibnamefont {Ullrich}}, \bibinfo {author}
  {\bibfnamefont {H.}~\bibnamefont {Stapelfeldt}}, \bibinfo {author}
  {\bibfnamefont {D.}~\bibnamefont {Rolles}}, and\ \bibinfo {author}
  {\bibfnamefont {H.~N.}\ \bibnamefont {Chapman}},\ }\bibfield  {title}
  {\bibinfo {title} {X-ray diffraction from isolated and strongly aligned
  gas-phase molecules with a free-electron laser},\ }\href
  {https://doi.org/10.1103/PhysRevLett.112.083002} {\bibfield  {journal}
  {\bibinfo  {journal} {Phys. Rev. Lett.}\ }\textbf {\bibinfo {volume} {112}},\
  \bibinfo {pages} {083002} (\bibinfo {year} {2014})},\ \Eprint
  {https://arxiv.org/abs/1307.4577} {arXiv:1307.4577 [physics]}\BibitemShut
  {NoStop}%
\bibitem [{\citenamefont {Karamatskos}\ \emph {et~al.}(2019)\citenamefont
  {Karamatskos}, \citenamefont {Raabe}, \citenamefont {Mullins}, \citenamefont
  {Trabattoni}, \citenamefont {Stammer}, \citenamefont {Goldsztejn},
  \citenamefont {Johansen}, \citenamefont {D{\l}ugo{\l}\k{e}cki}, \citenamefont
  {Stapelfeldt}, \citenamefont {Vrakking}, \citenamefont {Trippel},
  \citenamefont {Rouzée}, and\ \citenamefont
  {Küpper}}]{Karamatskos:NatComm10:3364}%
  \BibitemOpen
  \bibfield  {author} {\bibinfo {author} {\bibfnamefont {E.~T.}\ \bibnamefont
  {Karamatskos}}, \bibinfo {author} {\bibfnamefont {S.}~\bibnamefont {Raabe}},
  \bibinfo {author} {\bibfnamefont {T.}~\bibnamefont {Mullins}}, \bibinfo
  {author} {\bibfnamefont {A.}~\bibnamefont {Trabattoni}}, \bibinfo {author}
  {\bibfnamefont {P.}~\bibnamefont {Stammer}}, \bibinfo {author} {\bibfnamefont
  {G.}~\bibnamefont {Goldsztejn}}, \bibinfo {author} {\bibfnamefont {R.~R.}\
  \bibnamefont {Johansen}}, \bibinfo {author} {\bibfnamefont {K.}~\bibnamefont
  {D{\l}ugo{\l}\k{e}cki}}, \bibinfo {author} {\bibfnamefont {H.}~\bibnamefont
  {Stapelfeldt}}, \bibinfo {author} {\bibfnamefont {M.~J.~J.}\ \bibnamefont
  {Vrakking}}, \bibinfo {author} {\bibfnamefont {S.}~\bibnamefont {Trippel}},
  \bibinfo {author} {\bibfnamefont {A.}~\bibnamefont {Rouzée}}, and\ \bibinfo
  {author} {\bibfnamefont {J.}~\bibnamefont {Küpper}},\ }\bibfield  {title}
  {\bibinfo {title} {Molecular movie of ultrafast coherent rotational dynamics
  of {OCS}},\ }\href {https://doi.org/10.1038/s41467-019-11122-y} {\bibfield
  {journal} {\bibinfo  {journal} {Nat. Commun.}\ }\textbf {\bibinfo {volume}
  {10}},\ \bibinfo {pages} {3364} (\bibinfo {year} {2019})},\ \Eprint
  {https://arxiv.org/abs/1807.01034} {arXiv:1807.01034 [physics]}\BibitemShut
  {NoStop}%
\bibitem [{\citenamefont {Miao}\ \emph {et~al.}(2007)\citenamefont {Miao},
  \citenamefont {Ishikawa}, \citenamefont {Shen}, and\ \citenamefont
  {Earnest}}]{Miao:ARPC59:387}%
  \BibitemOpen
  \bibfield  {author} {\bibinfo {author} {\bibfnamefont {J.}~\bibnamefont
  {Miao}}, \bibinfo {author} {\bibfnamefont {T.}~\bibnamefont {Ishikawa}},
  \bibinfo {author} {\bibfnamefont {Q.}~\bibnamefont {Shen}}, and\ \bibinfo
  {author} {\bibfnamefont {T.}~\bibnamefont {Earnest}},\ }\bibfield  {title}
  {\bibinfo {title} {Extending x-ray crystallography to allow the imaging of
  noncrystalline materials, cells, and single protein complexes},\ }\href
  {https://doi.org/10.1146/annurev.physchem.59.032607.093642} {\bibfield
  {journal} {\bibinfo  {journal} {Annu. Rev. Phys. Chem.}\ }\textbf {\bibinfo
  {volume} {59}},\ \bibinfo {pages} {387} (\bibinfo {year} {2007})}\BibitemShut
  {NoStop}%
\bibitem [{\citenamefont {Miao}\ \emph {et~al.}(1999)\citenamefont {Miao},
  \citenamefont {Charalambous}, \citenamefont {Kirz}, and\ \citenamefont
  {Sayre}}]{Miao:Nature400:342}%
  \BibitemOpen
  \bibfield  {author} {\bibinfo {author} {\bibfnamefont {J.}~\bibnamefont
  {Miao}}, \bibinfo {author} {\bibfnamefont {P.}~\bibnamefont {Charalambous}},
  \bibinfo {author} {\bibfnamefont {J.}~\bibnamefont {Kirz}}, and\ \bibinfo
  {author} {\bibfnamefont {D.}~\bibnamefont {Sayre}},\ }\bibfield  {title}
  {\bibinfo {title} {Extending the methodology of x-ray crystallography to
  allow imaging of micrometre-sized non-crystalline specimens},\ }\href
  {https://doi.org/10.1038/22498} {\bibfield  {journal} {\bibinfo  {journal}
  {Nature}\ }\textbf {\bibinfo {volume} {400}},\ \bibinfo {pages} {342}
  (\bibinfo {year} {1999})}\BibitemShut {NoStop}%
\bibitem [{\citenamefont {Welker}(2019)}]{Welker:thesis:2019}%
  \BibitemOpen
  \bibfield  {author} {\bibinfo {author} {\bibfnamefont {S.}~\bibnamefont
  {Welker}},\ }\emph {\bibinfo {title} {Simulation of artificial and biological
  nanoparticles’ trajectories in hybrid force-fields}},\ \href
  {https://bib-pubdb1.desy.de/record/426150} {\bibinfo {type} {Bachelor
  thesis}},\ \bibinfo  {school} {Universität Hamburg} (\bibinfo {year}
  {2019})\BibitemShut {NoStop}%
\bibitem [{\citenamefont {Filsinger}\ \emph {et~al.}(2009)\citenamefont
  {Filsinger}, \citenamefont {K{\"u}pper}, \citenamefont {Meijer},
  \citenamefont {Holmegaard}, \citenamefont {Nielsen}, \citenamefont {Nevo},
  \citenamefont {Hansen}, and\ \citenamefont
  {Stapelfeldt}}]{Filsinger:JCP131:064309}%
  \BibitemOpen
  \bibfield  {author} {\bibinfo {author} {\bibfnamefont {F.}~\bibnamefont
  {Filsinger}}, \bibinfo {author} {\bibfnamefont {J.}~\bibnamefont
  {K{\"u}pper}}, \bibinfo {author} {\bibfnamefont {G.}~\bibnamefont {Meijer}},
  \bibinfo {author} {\bibfnamefont {L.}~\bibnamefont {Holmegaard}}, \bibinfo
  {author} {\bibfnamefont {J.~H.}\ \bibnamefont {Nielsen}}, \bibinfo {author}
  {\bibfnamefont {I.}~\bibnamefont {Nevo}}, \bibinfo {author} {\bibfnamefont
  {J.~L.}\ \bibnamefont {Hansen}}, and\ \bibinfo {author} {\bibfnamefont
  {H.}~\bibnamefont {Stapelfeldt}},\ }\bibfield  {title} {\bibinfo {title}
  {Quantum-state selection, alignment, and orientation of large molecules using
  static electric and laser fields},\ }\href
  {https://doi.org/10.1063/1.3194287} {\bibfield  {journal} {\bibinfo
  {journal} {J. Chem. Phys.}\ }\textbf {\bibinfo {volume} {131}},\ \bibinfo
  {pages} {064309} (\bibinfo {year} {2009})},\ \Eprint
  {https://arxiv.org/abs/0903.5413} {arXiv:0903.5413 [physics]}\BibitemShut
  {NoStop}%
\bibitem [{\citenamefont {van Rossum} and\ \citenamefont
  {Talin}(2007)}]{python:website:PEP3119}%
  \BibitemOpen
  \bibfield  {author} {\bibinfo {author} {\bibfnamefont {G.}~\bibnamefont {van
  Rossum}} and\ \bibinfo {author} {\bibnamefont {Talin}},\ }\href
  {https://www.python.org/dev/peps/pep-3119/} {\bibinfo {title} {{PEP~3119}:
  {I}ntroducing abstract base classes}},\ \bibinfo {howpublished} {Website,
  URL: \url{https://www.python.org/dev/peps/pep-3119}} (\bibinfo {year}
  {2007}),\ \bibinfo {note} {accessed on 2019-10-04}\BibitemShut {NoStop}%
\bibitem [{\citenamefont {Petzold}(1983)}]{Petzold:SIAM:4:136}%
  \BibitemOpen
  \bibfield  {author} {\bibinfo {author} {\bibfnamefont {L.}~\bibnamefont
  {Petzold}},\ }\bibfield  {title} {\bibinfo {title} {Automatic selection of
  methods for solving stiff and nonstiff systems of ordinary differential
  equations},\ }\href {https://doi.org/10.1137/0904010} {\bibfield  {journal}
  {\bibinfo  {journal} {SIAM J. Sci. \& Stat. Comp.}\ }\textbf {\bibinfo
  {volume} {4}},\ \bibinfo {pages} {136} (\bibinfo {year} {1983})}\BibitemShut
  {NoStop}%
\bibitem [{\citenamefont {{Virtanen}}\ \emph {et~al.}(2020)\citenamefont
  {{Virtanen}}, \citenamefont {{Gommers}}, \citenamefont {{Oliphant}},
  \citenamefont {{Haberland}}, \citenamefont {{Reddy}}, \citenamefont
  {{Cournapeau}}, \citenamefont {{Burovski}}, \citenamefont {{Peterson}},
  \citenamefont {{Weckesser}}, \citenamefont {{Bright}}, \citenamefont {{van
  der Walt}}, \citenamefont {{Brett}}, \citenamefont {{Wilson}}, \citenamefont
  {{Jarrod Millman}}, \citenamefont {{Mayorov}}, \citenamefont {{Nelson}},
  \citenamefont {{Jones}}, \citenamefont {{Kern}}, \citenamefont {{Larson}},
  \citenamefont {{Carey}}, \citenamefont {{Polat}}, \citenamefont {{Feng}},
  \citenamefont {{Moore}}, \citenamefont {{VanderPlas}}, \citenamefont
  {{Laxalde}}, \citenamefont {{Perktold}}, \citenamefont {{Cimrman}},
  \citenamefont {{Henriksen}}, \citenamefont {{Quintero}}, \citenamefont
  {{Harris}}, \citenamefont {{Archibald}}, \citenamefont {{Ribeiro}},
  \citenamefont {{Pedregosa}}, \citenamefont {{van Mulbregt}}, and\
  \citenamefont {{SciPy 1.0 Contributors}}}]{Virtanen:NatMeth17:261}%
  \BibitemOpen
  \bibfield  {author} {\bibinfo {author} {\bibfnamefont {P.}~\bibnamefont
  {{Virtanen}}}, \bibinfo {author} {\bibfnamefont {R.}~\bibnamefont
  {{Gommers}}}, \bibinfo {author} {\bibfnamefont {T.~E.}\ \bibnamefont
  {{Oliphant}}}, \bibinfo {author} {\bibfnamefont {M.}~\bibnamefont
  {{Haberland}}}, \bibinfo {author} {\bibfnamefont {T.}~\bibnamefont
  {{Reddy}}}, \bibinfo {author} {\bibfnamefont {D.}~\bibnamefont
  {{Cournapeau}}}, \bibinfo {author} {\bibfnamefont {E.}~\bibnamefont
  {{Burovski}}}, \bibinfo {author} {\bibfnamefont {P.}~\bibnamefont
  {{Peterson}}}, \bibinfo {author} {\bibfnamefont {W.}~\bibnamefont
  {{Weckesser}}}, \bibinfo {author} {\bibfnamefont {J.}~\bibnamefont
  {{Bright}}}, \bibinfo {author} {\bibfnamefont {S.~J.}\ \bibnamefont {{van der
  Walt}}}, \bibinfo {author} {\bibfnamefont {M.}~\bibnamefont {{Brett}}},
  \bibinfo {author} {\bibfnamefont {J.}~\bibnamefont {{Wilson}}}, \bibinfo
  {author} {\bibfnamefont {K.}~\bibnamefont {{Jarrod Millman}}}, \bibinfo
  {author} {\bibfnamefont {N.}~\bibnamefont {{Mayorov}}}, \bibinfo {author}
  {\bibfnamefont {A.~R.~J.}\ \bibnamefont {{Nelson}}}, \bibinfo {author}
  {\bibfnamefont {E.}~\bibnamefont {{Jones}}}, \bibinfo {author} {\bibfnamefont
  {R.}~\bibnamefont {{Kern}}}, \bibinfo {author} {\bibfnamefont
  {E.}~\bibnamefont {{Larson}}}, \bibinfo {author} {\bibfnamefont
  {C.}~\bibnamefont {{Carey}}}, \bibinfo {author} {\bibfnamefont
  {{\.I}.}~\bibnamefont {{Polat}}}, \bibinfo {author} {\bibfnamefont
  {Y.}~\bibnamefont {{Feng}}}, \bibinfo {author} {\bibfnamefont {E.~W.}\
  \bibnamefont {{Moore}}}, \bibinfo {author} {\bibfnamefont {J.}~\bibnamefont
  {{VanderPlas}}}, \bibinfo {author} {\bibfnamefont {D.}~\bibnamefont
  {{Laxalde}}}, \bibinfo {author} {\bibfnamefont {J.}~\bibnamefont
  {{Perktold}}}, \bibinfo {author} {\bibfnamefont {R.}~\bibnamefont
  {{Cimrman}}}, \bibinfo {author} {\bibfnamefont {I.}~\bibnamefont
  {{Henriksen}}}, \bibinfo {author} {\bibfnamefont {E.~A.}\ \bibnamefont
  {{Quintero}}}, \bibinfo {author} {\bibfnamefont {C.~R.}\ \bibnamefont
  {{Harris}}}, \bibinfo {author} {\bibfnamefont {A.~M.}\ \bibnamefont
  {{Archibald}}}, \bibinfo {author} {\bibfnamefont {A.~H.}\ \bibnamefont
  {{Ribeiro}}}, \bibinfo {author} {\bibfnamefont {F.}~\bibnamefont
  {{Pedregosa}}}, \bibinfo {author} {\bibfnamefont {P.}~\bibnamefont {{van
  Mulbregt}}}, and\ \bibinfo {author} {\bibnamefont {{SciPy 1.0
  Contributors}}},\ }\bibfield  {title} {\bibinfo {title} {{SciPy}~1.0:
  fundamental algorithms for scientific computing in {Python}},\ }\href
  {https://doi.org/10.1038/s41592-019-0686-2} {\bibfield  {journal} {\bibinfo
  {journal} {Nat. Meth.}\ }\textbf {\bibinfo {volume} {17}},\ \bibinfo {pages}
  {261} (\bibinfo {year} {2020})}\BibitemShut {NoStop}%
\bibitem [{\citenamefont {{van der Walt}}\ \emph {et~al.}(2011)\citenamefont
  {{van der Walt}}, \citenamefont {{Colbert}}, and\ \citenamefont
  {{Varoquaux}}}]{vanderWalt:CSE13:22}%
  \BibitemOpen
  \bibfield  {author} {\bibinfo {author} {\bibfnamefont {S.}~\bibnamefont {{van
  der Walt}}}, \bibinfo {author} {\bibfnamefont {S.~C.}\ \bibnamefont
  {{Colbert}}}, and\ \bibinfo {author} {\bibfnamefont {G.}~\bibnamefont
  {{Varoquaux}}},\ }\bibfield  {title} {\bibinfo {title} {{The} {NumPy} array:
  {A} structure for efficient numerical computation},\ }\href
  {https://doi.org/10.1109/MCSE.2011.37} {\bibfield  {journal} {\bibinfo
  {journal} {Comp. in Sci. \& Eng.}\ }\textbf {\bibinfo {volume} {13}},\
  \bibinfo {pages} {22} (\bibinfo {year} {2011})}\BibitemShut {NoStop}%
\bibitem [{\citenamefont {Lam}\ \emph {et~al.}(2015)\citenamefont {Lam},
  \citenamefont {Pitrou}, and\ \citenamefont {Seibert}}]{Lam:LLVM15:7}%
  \BibitemOpen
  \bibfield  {author} {\bibinfo {author} {\bibfnamefont {S.~K.}\ \bibnamefont
  {Lam}}, \bibinfo {author} {\bibfnamefont {A.}~\bibnamefont {Pitrou}}, and\
  \bibinfo {author} {\bibfnamefont {S.}~\bibnamefont {Seibert}},\ }\bibfield
  {title} {\bibinfo {title} {Numba: A {LLVM}-based {Python} {JIT} compiler},\
  }in\ \href {https://doi.org/10.1145/2833157.2833162} {\emph {\bibinfo
  {booktitle} {Proceedings of the Second Workshop on the LLVM Compiler
  Infrastructure in HPC}}},\ \bibinfo {series and number} {LLVM ’15}\
  (\bibinfo  {publisher} {Association for Computing Machinery},\ \bibinfo
  {address} {New York, NY, USA},\ \bibinfo {year} {2015})\BibitemShut {NoStop}%
\bibitem [{\citenamefont {Behnel}\ \emph {et~al.}(2011)\citenamefont {Behnel},
  \citenamefont {Bradshaw}, \citenamefont {Citro}, \citenamefont {Dalcin},
  \citenamefont {Seljebotn}, and\ \citenamefont
  {Smith}}]{Behnel:CompSciEng13:31}%
  \BibitemOpen
  \bibfield  {author} {\bibinfo {author} {\bibfnamefont {S.}~\bibnamefont
  {Behnel}}, \bibinfo {author} {\bibfnamefont {R.}~\bibnamefont {Bradshaw}},
  \bibinfo {author} {\bibfnamefont {C.}~\bibnamefont {Citro}}, \bibinfo
  {author} {\bibfnamefont {L.}~\bibnamefont {Dalcin}}, \bibinfo {author}
  {\bibfnamefont {D.~S.}\ \bibnamefont {Seljebotn}}, and\ \bibinfo {author}
  {\bibfnamefont {K.}~\bibnamefont {Smith}},\ }\bibfield  {title} {\bibinfo
  {title} {{Cython}: The best of both worlds},\ }\href
  {https://doi.org/10.1109/MCSE.2010.118} {\bibfield  {journal} {\bibinfo
  {journal} {Comp. Sci. \& Eng.}\ }\textbf {\bibinfo {volume} {13}},\ \bibinfo
  {pages} {31} (\bibinfo {year} {2011})}\BibitemShut {NoStop}%
\bibitem [{\citenamefont {{Stokes}}(1851)}]{Stokes:TCaPS9:8}%
  \BibitemOpen
  \bibfield  {author} {\bibinfo {author} {\bibfnamefont {G.~G.}\ \bibnamefont
  {{Stokes}}},\ }\bibfield  {title} {\bibinfo {title} {On the effect of the
  internal friction of fluids on the motion of pendulums},\ }\href
  {https://ui.adsabs.harvard.edu/abs/1851TCaPS...9....8S} {\bibfield  {journal}
  {\bibinfo  {journal} {Trans. Cambridge Phil. Soc.}\ }\textbf {\bibinfo
  {volume} {9}},\ \bibinfo {pages} {8} (\bibinfo {year} {1851})}\BibitemShut
  {NoStop}%
\bibitem [{\citenamefont {Cunningham} and\ \citenamefont
  {Larmor}(1910)}]{Cunningham:PRSA83:357}%
  \BibitemOpen
  \bibfield  {author} {\bibinfo {author} {\bibfnamefont {E.}~\bibnamefont
  {Cunningham}} and\ \bibinfo {author} {\bibfnamefont {J.}~\bibnamefont
  {Larmor}},\ }\bibfield  {title} {\bibinfo {title} {On the velocity of steady
  fall of spherical particles through fluid medium},\ }\href
  {https://doi.org/10.1098/rspa.1910.0024} {\bibfield  {journal} {\bibinfo
  {journal} {Proc. Royal Soc. London A}\ }\textbf {\bibinfo {volume} {83}},\
  \bibinfo {pages} {357} (\bibinfo {year} {1910})}\BibitemShut {NoStop}%
\bibitem [{\citenamefont {Li} and\ \citenamefont
  {Ahmadi}(1992)}]{Li:AST16:209}%
  \BibitemOpen
  \bibfield  {author} {\bibinfo {author} {\bibfnamefont {A.}~\bibnamefont
  {Li}} and\ \bibinfo {author} {\bibfnamefont {G.}~\bibnamefont {Ahmadi}},\
  }\bibfield  {title} {\bibinfo {title} {Dispersion and deposition of spherical
  particles from point sources in a turbulent channel flow},\ }\href
  {https://doi.org/10.1080/02786829208959550} {\bibfield  {journal} {\bibinfo
  {journal} {Aerosol Sci. Techn.}\ }\textbf {\bibinfo {volume} {16}},\ \bibinfo
  {pages} {209} (\bibinfo {year} {1992})}\BibitemShut {NoStop}%
\bibitem [{\citenamefont {Epstein}(1924)}]{Epstein:PR23:710}%
  \BibitemOpen
  \bibfield  {author} {\bibinfo {author} {\bibfnamefont {P.~S.}\ \bibnamefont
  {Epstein}},\ }\bibfield  {title} {\bibinfo {title} {On the resistance
  experienced by spheres in their motion through gases},\ }\href
  {https://doi.org/10.1103/PhysRev.23.710} {\bibfield  {journal} {\bibinfo
  {journal} {Phys. Rev.}\ }\textbf {\bibinfo {volume} {23}},\ \bibinfo {pages}
  {710} (\bibinfo {year} {1924})}\BibitemShut {NoStop}%
\bibitem [{\citenamefont {Bowman} and\ \citenamefont
  {Padgett}(2013)}]{Bowman:RPP76:026401}%
  \BibitemOpen
  \bibfield  {author} {\bibinfo {author} {\bibfnamefont {R.~W.}\ \bibnamefont
  {Bowman}} and\ \bibinfo {author} {\bibfnamefont {M.~J.}\ \bibnamefont
  {Padgett}},\ }\bibfield  {title} {\bibinfo {title} {Optical trapping and
  binding},\ }\href {http://stacks.iop.org/0034-4885/76/i=2/a=026401}
  {\bibfield  {journal} {\bibinfo  {journal} {Rep. Prog. Phys.}\ }\textbf
  {\bibinfo {volume} {76}},\ \bibinfo {pages} {026401} (\bibinfo {year}
  {2013})}\BibitemShut {NoStop}%
\bibitem [{\citenamefont {Eckerskorn}(2016)}]{Eckerskorn:thesis:2016}%
  \BibitemOpen
  \bibfield  {author} {\bibinfo {author} {\bibfnamefont {N.~O.}\ \bibnamefont
  {Eckerskorn}},\ }\emph {\bibinfo {title} {Trapping and guiding microscopic
  particles with light-induced forces}},\ \href
  {https://doi.org/10.25911/5d76352c8c03a} {\bibinfo {type} {Dissertation}},\
  \bibinfo  {school} {College of Physical {\&} Mathematical Sciences, Research
  School of Physics and Engineering, Laser Physics Centre}, \bibinfo {address}
  {Australia} (\bibinfo {year} {2016})\BibitemShut {NoStop}%
\bibitem [{\citenamefont {Shvedov}\ \emph {et~al.}(2009)\citenamefont
  {Shvedov}, \citenamefont {Desyatnikov}, \citenamefont {Rode}, \citenamefont
  {Krolikowski}, and\ \citenamefont {Kivshar}}]{Shvedov:OptExp17:5743}%
  \BibitemOpen
  \bibfield  {author} {\bibinfo {author} {\bibfnamefont {V.~G.}\ \bibnamefont
  {Shvedov}}, \bibinfo {author} {\bibfnamefont {A.~S.}\ \bibnamefont
  {Desyatnikov}}, \bibinfo {author} {\bibfnamefont {A.~V.}\ \bibnamefont
  {Rode}}, \bibinfo {author} {\bibfnamefont {W.}~\bibnamefont {Krolikowski}},\
  and\ \bibinfo {author} {\bibfnamefont {Y.~S.}\ \bibnamefont {Kivshar}},\
  }\bibfield  {title} {\bibinfo {title} {Optical guiding of absorbing
  nanoclusters in air},\ }\href {https://doi.org/10.1364/OE.17.005743}
  {\bibfield  {journal} {\bibinfo  {journal} {Opt. Exp.}\ }\textbf {\bibinfo
  {volume} {17}},\ \bibinfo {pages} {5743} (\bibinfo {year}
  {2009})}\BibitemShut {NoStop}%
\bibitem [{\citenamefont {Shvedov}\ \emph {et~al.}(2011)\citenamefont
  {Shvedov}, \citenamefont {Hnatovsky}, \citenamefont {Rode}, and\
  \citenamefont {Krolikowski}}]{Shvedov:OptExp19:17350}%
  \BibitemOpen
  \bibfield  {author} {\bibinfo {author} {\bibfnamefont {V.~G.}\ \bibnamefont
  {Shvedov}}, \bibinfo {author} {\bibfnamefont {C.}~\bibnamefont {Hnatovsky}},
  \bibinfo {author} {\bibfnamefont {A.~V.}\ \bibnamefont {Rode}}, and\
  \bibinfo {author} {\bibfnamefont {W.}~\bibnamefont {Krolikowski}},\
  }\bibfield  {title} {\bibinfo {title} {Robust trapping and manipulation of
  airborne particles with a bottle beam},\ }\href
  {https://doi.org/10.1364/OE.19.017350} {\bibfield  {journal} {\bibinfo
  {journal} {Opt. Exp.}\ }\textbf {\bibinfo {volume} {19}},\ \bibinfo {pages}
  {17350} (\bibinfo {year} {2011})}\BibitemShut {NoStop}%
\bibitem [{\citenamefont {Awel}\ \emph {et~al.}(2020)\citenamefont {Awel},
  \citenamefont {Lavin-Varela}, \citenamefont {Roth}, \citenamefont {Horke},
  \citenamefont {Rode}, \citenamefont {Kirian}, \citenamefont {K{\"u}pper},\
  and\ \citenamefont {Chapman}}]{Awel:optical-funnel:inprep}%
  \BibitemOpen
  \bibfield  {author} {\bibinfo {author} {\bibfnamefont {S.}~\bibnamefont
  {Awel}}, \bibinfo {author} {\bibfnamefont {S.}~\bibnamefont {Lavin-Varela}},
  \bibinfo {author} {\bibfnamefont {N.}~\bibnamefont {Roth}}, \bibinfo {author}
  {\bibfnamefont {D.~A.}\ \bibnamefont {Horke}}, \bibinfo {author}
  {\bibfnamefont {A.~V.}\ \bibnamefont {Rode}}, \bibinfo {author}
  {\bibfnamefont {R.~A.}\ \bibnamefont {Kirian}}, \bibinfo {author}
  {\bibfnamefont {J.}~\bibnamefont {K{\"u}pper}}, and\ \bibinfo {author}
  {\bibfnamefont {H.~N.}\ \bibnamefont {Chapman}},\ }\bibfield  {title}
  {\bibinfo {title} {Optical funnel to guide and focus virus particles for
  x-ray laser imaging}} (\bibinfo {year} {2020}),\ \bibinfo {note}
  {submitted}\BibitemShut {NoStop}%
\bibitem [{Com()}]{Comsol:Multiphysics:5.5}%
  \BibitemOpen
  \href@noop {} {}\bibinfo {note} {COMSOL Multiphysics v.\ 5.5.\
  \url{http://www.comsol.com}. COMSOL AB, Stockholm, Sweden}\BibitemShut
  {NoStop}%
\bibitem [{\citenamefont {{Saini}}\ \emph {et~al.}(2011)\citenamefont
  {{Saini}}, \citenamefont {{Jin}}, \citenamefont {{Hood}}, \citenamefont
  {{Barker}}, \citenamefont {{Mehrotra}}, and\ \citenamefont
  {{Biswas}}}]{Saini:ICHPC2011:1}%
  \BibitemOpen
  \bibfield  {author} {\bibinfo {author} {\bibfnamefont {S.}~\bibnamefont
  {{Saini}}}, \bibinfo {author} {\bibfnamefont {H.}~\bibnamefont {{Jin}}},
  \bibinfo {author} {\bibfnamefont {R.}~\bibnamefont {{Hood}}}, \bibinfo
  {author} {\bibfnamefont {D.}~\bibnamefont {{Barker}}}, \bibinfo {author}
  {\bibfnamefont {P.}~\bibnamefont {{Mehrotra}}}, and\ \bibinfo {author}
  {\bibfnamefont {R.}~\bibnamefont {{Biswas}}},\ }\bibfield  {title} {\bibinfo
  {title} {The impact of hyper-threading on processor resource utilization in
  production applications},\ }in\ \href
  {https://doi.org/10.1109/HiPC.2011.6152743} {\emph {\bibinfo {booktitle}
  {2011 18th International Conference on High Performance Computing}}}\
  (\bibinfo {year} {2011})\ pp.\ \bibinfo {pages} {1--10}\BibitemShut {NoStop}%
\bibitem [{\citenamefont {{NVIDIA
  Corporation}}(2020)}]{NVIDIA:website:CUDAcppbestpractices}%
  \BibitemOpen
  \bibfield  {author} {\bibinfo {author} {\bibnamefont {{NVIDIA
  Corporation}}},\ }\href
  {https://docs.nvidia.com/cuda/cuda-c-best-practices-guide/index.html#data-transfer-between-host-and-device}
  {\bibinfo {title} {{CUDA} {C}++ best practices guide}} (\bibinfo {year}
  {2020}),\ \bibinfo {note} {accessed on 2020-12-09}\BibitemShut {NoStop}%
\bibitem [{\citenamefont {Seiffert}\ \emph {et~al.}(2017)\citenamefont
  {Seiffert}, \citenamefont {Liu}, \citenamefont {Zherebtsov}, \citenamefont
  {Trabattoni}, \citenamefont {Rupp}, \citenamefont {Castrovilli},
  \citenamefont {Galli}, \citenamefont {Submann}, \citenamefont
  {Wintersperger}, \citenamefont {Stierle}, \citenamefont {Sansone},
  \citenamefont {Poletto}, \citenamefont {Frassetto}, \citenamefont {Halfpap},
  \citenamefont {Mondes}, \citenamefont {Graf}, \citenamefont {Rühl},
  \citenamefont {Krausz}, \citenamefont {Nisoli}, \citenamefont {Fennel},
  \citenamefont {Calegari}, and\ \citenamefont
  {Kling}}]{Seiffert:NatPhys13:766}%
  \BibitemOpen
  \bibfield  {author} {\bibinfo {author} {\bibfnamefont {L.}~\bibnamefont
  {Seiffert}}, \bibinfo {author} {\bibfnamefont {Q.}~\bibnamefont {Liu}},
  \bibinfo {author} {\bibfnamefont {S.}~\bibnamefont {Zherebtsov}}, \bibinfo
  {author} {\bibfnamefont {A.}~\bibnamefont {Trabattoni}}, \bibinfo {author}
  {\bibfnamefont {P.}~\bibnamefont {Rupp}}, \bibinfo {author} {\bibfnamefont
  {M.~C.}\ \bibnamefont {Castrovilli}}, \bibinfo {author} {\bibfnamefont
  {M.}~\bibnamefont {Galli}}, \bibinfo {author} {\bibfnamefont
  {F.}~\bibnamefont {Submann}}, \bibinfo {author} {\bibfnamefont
  {K.}~\bibnamefont {Wintersperger}}, \bibinfo {author} {\bibfnamefont
  {J.}~\bibnamefont {Stierle}}, \bibinfo {author} {\bibfnamefont
  {G.}~\bibnamefont {Sansone}}, \bibinfo {author} {\bibfnamefont
  {L.}~\bibnamefont {Poletto}}, \bibinfo {author} {\bibfnamefont
  {F.}~\bibnamefont {Frassetto}}, \bibinfo {author} {\bibfnamefont
  {I.}~\bibnamefont {Halfpap}}, \bibinfo {author} {\bibfnamefont
  {V.}~\bibnamefont {Mondes}}, \bibinfo {author} {\bibfnamefont
  {C.}~\bibnamefont {Graf}}, \bibinfo {author} {\bibfnamefont {E.}~\bibnamefont
  {Rühl}}, \bibinfo {author} {\bibfnamefont {F.}~\bibnamefont {Krausz}},
  \bibinfo {author} {\bibfnamefont {M.}~\bibnamefont {Nisoli}}, \bibinfo
  {author} {\bibfnamefont {T.}~\bibnamefont {Fennel}}, \bibinfo {author}
  {\bibfnamefont {F.}~\bibnamefont {Calegari}}, and\ \bibinfo {author}
  {\bibfnamefont {M.~F.}\ \bibnamefont {Kling}},\ }\bibfield  {title} {\bibinfo
  {title} {Attosecond chronoscopy of electron scattering in dielectric
  nanoparticles},\ }\href {https://doi.org/10.1038/nphys4129} {\bibfield
  {journal} {\bibinfo  {journal} {Nat. Phys.}\ }\textbf {\bibinfo {volume}
  {13}},\ \bibinfo {pages} {766 } (\bibinfo {year} {2017})}\BibitemShut
  {NoStop}%
\bibitem [{\citenamefont {Aquila}\ \emph {et~al.}(2015)\citenamefont {Aquila},
  \citenamefont {Barty}, \citenamefont {Bostedt}, \citenamefont {Boutet},
  \citenamefont {Carini}, \citenamefont {DePonte}, \citenamefont {Drell},
  \citenamefont {Doniach}, \citenamefont {Downing}, \citenamefont {Earnest},
  \citenamefont {Elmlund}, \citenamefont {Elser}, \citenamefont {G{\"u}hr},
  \citenamefont {Hajdu}, \citenamefont {Hastings}, \citenamefont {Hau-Riege},
  \citenamefont {Huang}, \citenamefont {Lattman}, \citenamefont {Maia},
  \citenamefont {Marchesini}, \citenamefont {Ourmazd}, \citenamefont
  {Pellegrini}, \citenamefont {Santra}, \citenamefont {Schlichting},
  \citenamefont {Schroer}, \citenamefont {Spence}, \citenamefont {Vartanyants},
  \citenamefont {Wakatsuki}, \citenamefont {Weis}, and\ \citenamefont
  {Williams}}]{Aquila:StructDyn2:041701}%
  \BibitemOpen
  \bibfield  {author} {\bibinfo {author} {\bibfnamefont {A.}~\bibnamefont
  {Aquila}}, \bibinfo {author} {\bibfnamefont {A.}~\bibnamefont {Barty}},
  \bibinfo {author} {\bibfnamefont {C.}~\bibnamefont {Bostedt}}, \bibinfo
  {author} {\bibfnamefont {S.}~\bibnamefont {Boutet}}, \bibinfo {author}
  {\bibfnamefont {G.}~\bibnamefont {Carini}}, \bibinfo {author} {\bibfnamefont
  {D.}~\bibnamefont {DePonte}}, \bibinfo {author} {\bibfnamefont
  {P.}~\bibnamefont {Drell}}, \bibinfo {author} {\bibfnamefont
  {S.}~\bibnamefont {Doniach}}, \bibinfo {author} {\bibfnamefont {K.~H.}\
  \bibnamefont {Downing}}, \bibinfo {author} {\bibfnamefont {T.}~\bibnamefont
  {Earnest}}, \bibinfo {author} {\bibfnamefont {H.}~\bibnamefont {Elmlund}},
  \bibinfo {author} {\bibfnamefont {V.}~\bibnamefont {Elser}}, \bibinfo
  {author} {\bibfnamefont {M.}~\bibnamefont {G{\"u}hr}}, \bibinfo {author}
  {\bibfnamefont {J.}~\bibnamefont {Hajdu}}, \bibinfo {author} {\bibfnamefont
  {J.}~\bibnamefont {Hastings}}, \bibinfo {author} {\bibfnamefont {S.~P.}\
  \bibnamefont {Hau-Riege}}, \bibinfo {author} {\bibfnamefont {Z.}~\bibnamefont
  {Huang}}, \bibinfo {author} {\bibfnamefont {E.~E.}\ \bibnamefont {Lattman}},
  \bibinfo {author} {\bibfnamefont {F.~R. N.~C.}\ \bibnamefont {Maia}},
  \bibinfo {author} {\bibfnamefont {S.}~\bibnamefont {Marchesini}}, \bibinfo
  {author} {\bibfnamefont {A.}~\bibnamefont {Ourmazd}}, \bibinfo {author}
  {\bibfnamefont {C.}~\bibnamefont {Pellegrini}}, \bibinfo {author}
  {\bibfnamefont {R.}~\bibnamefont {Santra}}, \bibinfo {author} {\bibfnamefont
  {I.}~\bibnamefont {Schlichting}}, \bibinfo {author} {\bibfnamefont
  {C.}~\bibnamefont {Schroer}}, \bibinfo {author} {\bibfnamefont {J.~C.~H.}\
  \bibnamefont {Spence}}, \bibinfo {author} {\bibfnamefont {I.~A.}\
  \bibnamefont {Vartanyants}}, \bibinfo {author} {\bibfnamefont
  {S.}~\bibnamefont {Wakatsuki}}, \bibinfo {author} {\bibfnamefont {W.~I.}\
  \bibnamefont {Weis}}, and\ \bibinfo {author} {\bibfnamefont {G.~J.}\
  \bibnamefont {Williams}},\ }\bibfield  {title} {\bibinfo {title} {The linac
  coherent light source single particle imaging road map},\ }\href
  {https://doi.org/10.1063/1.4918726} {\bibfield  {journal} {\bibinfo
  {journal} {Struct. Dyn.}\ }\textbf {\bibinfo {volume} {2}},\ \bibinfo {pages}
  {041701} (\bibinfo {year} {2015})}\BibitemShut {NoStop}%
\bibitem [{\citenamefont {Li}\ \emph {et~al.}(2019)\citenamefont {Li},
  \citenamefont {Shi}, \citenamefont {Cao}, \citenamefont {Liu}, and\
  \citenamefont {Küpper}}]{Li:PRAppl11:064036}%
  \BibitemOpen
  \bibfield  {author} {\bibinfo {author} {\bibfnamefont {Z.}~\bibnamefont
  {Li}}, \bibinfo {author} {\bibfnamefont {L.}~\bibnamefont {Shi}}, \bibinfo
  {author} {\bibfnamefont {L.}~\bibnamefont {Cao}}, \bibinfo {author}
  {\bibfnamefont {Z.}~\bibnamefont {Liu}}, and\ \bibinfo {author}
  {\bibfnamefont {J.}~\bibnamefont {Küpper}},\ }\bibfield  {title} {\bibinfo
  {title} {Acoustic funnel and buncher for nanoparticle injection},\ }\href
  {https://doi.org/10.1103/PhysRevApplied.11.064036} {\bibfield  {journal}
  {\bibinfo  {journal} {Phys. Rev. Appl.}\ }\textbf {\bibinfo {volume} {11}},\
  \bibinfo {pages} {064036} (\bibinfo {year} {2019})},\ \Eprint
  {https://arxiv.org/abs/1803.07472} {arXiv:1803.07472 [physics]}\BibitemShut
  {NoStop}%
\bibitem [{\citenamefont {Dalcín}\ \emph {et~al.}(2005)\citenamefont
  {Dalcín}, \citenamefont {Paz}, and\ \citenamefont
  {Storti}}]{Dalcin:JParDistComp:2005}%
  \BibitemOpen
  \bibfield  {author} {\bibinfo {author} {\bibfnamefont {L.}~\bibnamefont
  {Dalcín}}, \bibinfo {author} {\bibfnamefont {R.}~\bibnamefont {Paz}}, and\
  \bibinfo {author} {\bibfnamefont {M.}~\bibnamefont {Storti}},\ }\bibfield
  {title} {\bibinfo {title} {{MPI} for {Python}},\ }\href
  {https://doi.org/10.1016/j.jpdc.2005.03.010} {\bibfield  {journal} {\bibinfo
  {journal} {J. Parallel Distr. Comp.}\ }\textbf {\bibinfo {volume} {65}},\
  \bibinfo {pages} {1108} (\bibinfo {year} {2005})}\BibitemShut {NoStop}%
\end{thebibliography}%


\begin{thebibliography}{2}%
\makeatletter
\providecommand \@ifxundefined [1]{%
 \@ifx{#1\undefined}
}%
\providecommand \@ifnum [1]{%
 \ifnum #1\expandafter \@firstoftwo
 \else \expandafter \@secondoftwo
 \fi
}%
\providecommand \@ifx [1]{%
 \ifx #1\expandafter \@firstoftwo
 \else \expandafter \@secondoftwo
 \fi
}%
\providecommand \natexlab [1]{#1}%
\providecommand \enquote  [1]{``#1''}%
\providecommand \bibnamefont  [1]{#1}%
\providecommand \bibfnamefont [1]{#1}%
\providecommand \citenamefont [1]{#1}%
\providecommand \href@noop [0]{\@secondoftwo}%
\providecommand \href [0]{\begingroup \@sanitize@url \@href}%
\providecommand \@href[1]{\@@startlink{#1}\@@href}%
\providecommand \@@href[1]{\endgroup#1\@@endlink}%
\providecommand \@sanitize@url [0]{\catcode `\\12\catcode `\$12\catcode
  `\&12\catcode `\#12\catcode `\^12\catcode `\_12\catcode `\%12\relax}%
\providecommand \@@startlink[1]{}%
\providecommand \@@endlink[0]{}%
\providecommand \url  [0]{\begingroup\@sanitize@url \@url }%
\providecommand \@url [1]{\endgroup\@href {#1}{\urlprefix }}%
\providecommand \urlprefix  [0]{URL }%
\providecommand \Eprint [0]{\href }%
\providecommand \doibase [0]{https://doi.org/}%
\providecommand \selectlanguage [0]{\@gobble}%
\providecommand \bibinfo  [0]{\@secondoftwo}%
\providecommand \bibfield  [0]{\@secondoftwo}%
\providecommand \translation [1]{[#1]}%
\providecommand \BibitemOpen [0]{}%
\providecommand \bibitemStop [0]{}%
\providecommand \bibitemNoStop [0]{.\EOS\space}%
\providecommand \EOS [0]{\spacefactor3000\relax}%
\providecommand \BibitemShut  [1]{\csname bibitem#1\endcsname}%
\let\auto@bib@innerbib\@empty
\bibitem [{\citenamefont {Desyatnikov}\ \emph {et~al.}(2009)\citenamefont
  {Desyatnikov}, \citenamefont {Shvedov}, \citenamefont {Rode}, \citenamefont
  {Krolikowski}, and\ \citenamefont {Kivshar}}]{Desyatnikov:OptExp17:8201}%
  \BibitemOpen
  \bibfield  {author} {\bibinfo {author} {\bibfnamefont {A.~S.}\ \bibnamefont
  {Desyatnikov}}, \bibinfo {author} {\bibfnamefont {V.~G.}\ \bibnamefont
  {Shvedov}}, \bibinfo {author} {\bibfnamefont {A.~V.}\ \bibnamefont {Rode}},
  \bibinfo {author} {\bibfnamefont {W.}~\bibnamefont {Krolikowski}}, and\
  \bibinfo {author} {\bibfnamefont {Y.~S.}\ \bibnamefont {Kivshar}},\
  }\bibfield  {title} {\bibinfo {title} {Photophoretic manipulation of
  absorbing aerosol particles with vortex beams: theory versus experiment},\
  }\href {https://doi.org/10.1364/OE.17.008201} {\bibfield  {journal} {\bibinfo
   {journal} {Opt. Exp.}\ }\textbf {\bibinfo {volume} {17}},\ \bibinfo {pages}
  {8201} (\bibinfo {year} {2009})}\BibitemShut {NoStop}%
\bibitem [{\citenamefont {Welker}(2019)}]{Welker:thesis:2019}%
  \BibitemOpen
  \bibfield  {author} {\bibinfo {author} {\bibfnamefont {S.}~\bibnamefont
  {Welker}},\ }\emph {\bibinfo {title} {Simulation of artificial and biological
  nanoparticles’ trajectories in hybrid force-fields}},\ \href
  {https://bib-pubdb1.desy.de/record/426150} {\bibinfo {type} {Bachelor
  thesis}},\ \bibinfo  {school} {Universität Hamburg} (\bibinfo {year}
  {2019})\BibitemShut {NoStop}%
\end{thebibliography}%
\onecolumngrid
\listofnotes
\end{document}


\hyphenation{Brownian-Motion-Property-Updater}
\hyphenation{Simple-Z-Detector}\
\hyphenation{Stokes-Drag-Force-Field}
\hyphenation{Property-Updater}
\hyphenation{Result-Storage}
\hyphenation{CMInject}

\title{Supplementary information: \cminject: Python framework for the numerical simulation of
  nanoparticle injection pipelines}%
\author{Simon~Welker}\cfeldesy%
\author{Muhamed~Amin}\cfeldesy\groco\groinst%
\author{Jochen~Küpper}\jkemail\cmiweb\cfeldesy\uhhcui\uhhphys%
\date{\today}%
\maketitle

\section{Example definition of a Setup in CMInject}
The code in \autoref{code:example-setup} is an example definition of a CMInject setup, \ie, it
defines a subclass of the abstract base class \texttt{cminject.base.Setup}. This setup can be
simulated as a concrete virtual experiment with the \texttt{cminject} command provided in
\autoref{code:example-setup-run}. The required file
``\texttt{examples/example\textunderscore{}field.h5}'' is provided in the
supplementary materials, and is also included in the CMInject Python package.

\begin{widetext}
\begin{lstlisting}[language=Python, label=code:example-setup, caption=Python script
   (\texttt{example\_setup.py}) that defines a simple example setup. See also
   lib/cminject/setups/example.py in the code repository.]
#!/usr/bin/env python3
# We need argparse for parsing arguments from the command line
import argparse

# Import the most basic definitions for defining a Setup which returns an Experiment
# and has arguments parsed by a SetupArgumentParser
from cminject.experiment import Experiment
from cminject.base import Setup
from cminject.utils.args import SetupArgumentParser, distribution_description

# Import some concrete class implementations to be able to define a simple setup:
from cminject.utils.distributions import constant, GaussianDistribution  # Required types of distributions
from cminject.sources import VariableDistributionSource  # Generates particles from property distributions
from cminject.detectors import SimpleZDetector  # A detector which is positioned at some Z position
# A device for fluid flow based on an HDF5 file, together with FlowType, an enumeration of the possible models to use
from cminject.devices.fluid_flow import FluidFlowDevice, FlowType


class ExampleSetup(Setup):
    """
    A simple setup which will simulate a single flow field with particles starting from a fixed position and velocity
    distribution and radius. The only parameters available are the flow field filename and the density (density) of the
    particle material. The density is there just to show how additional arguments can be added.
    """
    @staticmethod
    def construct_experiment(main_args: argparse.Namespace, args: argparse.Namespace) -> Experiment:
        dt = 1e-5
        experiment = Experiment(number_of_dimensions=2, time_step=dt, time_interval=(0, 1))
        experiment.add_source(VariableDistributionSource(
            number_of_particles=main_args.nof_particles,
            position=args.pos,
            velocity=[GaussianDistribution(mu=0.0, sigma=1e-3), GaussianDistribution(mu=2, sigma=0.5)],
            radius=constant(5e-9), density=constant(args.rho)
        ))
        experiment.add_device(FluidFlowDevice(filename=args.filename, flow_type=FlowType.STOKES, brownian_motion=True))
        for z in [0.0, 0.01, 0.02]:
            experiment.add_detector(SimpleZDetector(z))

        return experiment

    @staticmethod
    def get_parser() -> SetupArgumentParser:
        parser = SetupArgumentParser()
        parser.add_argument('-f', '--filename', help='The filename of the flow field (HDF5).', type=str)
        parser.add_argument('--rho', help='The density of the particle material [kg/m^3].', type=float, default=1050.0)
        parser.add_argument('--pos', help='The position distributions in x/z space [m]', type=distribution_description,
                            nargs=2, default=[GaussianDistribution(0, 1e-3), constant(0.0)])
        return parser
\end{lstlisting}
\end{widetext}

\begin{lstlisting}[language=Bash, label=code:example-setup-run, caption=Example CMInject simulation
shell script (\texttt{example\textunderscore{}simulation.sh})]
#!/bin/bash
export PYTHONPATH=".:$PYTHONPATH"
cminject \
  `# Use ExampleSetup from example_setup.py` \
  -s example_setup.ExampleSetup \
  `# Use the 2D example field` \
  -f example_field.h5 \
  `# Simulate 100 particles` \
  -n 100 \
  `# Write the results to example-output.h5` \
  -o example_output.h5 \
  `# Track and store trajectories` \
  -T
\end{lstlisting}

\section{Code for the ALS simulation, analysis, and comparison to experiment}
\autoref{code:als-simulation} and \autoref{code:focus-curve} provide code that serves to reproduce
Figure 3 from the paper, i.e. the comparison figure between simulation and
experiment. \autoref{code:als-simulation} shows the command-line command to run the corresponding
simulation for $10^5$ particles and store the detected particle positions at virtual detectors
positioned at the same spots where experimental measurements were made. \autoref{code:focus-curve}
shows the Python code to do the same result data analysis and plotting procedure that generated
Figure 3 in the paper.

\begin{widetext}
\begin{lstlisting}[language=Bash, label=code:als-simulation, caption=Shell script
  (\texttt{als\textunderscore{}simulation.sh}) for the simulation that produced the focus curve
  shown in the paper.]
#!/bin/bash
export PYTHONPATH=".:$PYTHONPATH"
`# Use (the default OneFlowField setup with) als_field.h5`
cminject -f "als_field.h5"\
    `# Gold particles: Use appropriate density`\
    -rho 19320\
    `# Use 2 spatial dimensions`\
    -D 2\
    `# Normal distribution for x, constant for z position`\
    -p G[0,0.002] -0.128\
    `# Normal distributions for velocities in x and z`\
    -v G[0,1] G[1,0.1]\
    `# Normal distribution for the particle radius`\
    -r G[1.35e-8,1.91e-9]\
    `# Simulate the timespan from 0 to 1 seconds`\
    -t 0 1\
    `# Use time step of 10us`\
    -ts 1e-5\
    `# Insert the detectors at positions as in the experiment`\
    -d 0.001 0.00141 0.00181 0.00221 0.00261\
       0.00301 0.00341 0.00381 0.00421 0.00461\
    `# Simulate 10^5 particles`\
    -n 100\
    `# Write results to als_output.h5`\
    -o als_output.h5\
    `# Turn Brownian motion on for this simulation`\
    -B
\end{lstlisting}
\end{widetext}

\begin{widetext}
\begin{lstlisting}[language=Python, label=code:focus-curve, caption=Python script
  (\texttt{als\textunderscore{}analysis.py}) to produce the focus curve comparison plot (Figure 3 in
  the paper) from the results of the simulation given in
  \autoref{code:als-simulation}.]
#!/usr/bin/env python3
import sys
import matplotlib.pyplot as plt
import numpy as np
from cminject.result_storages import HDF5ResultStorage


# Global settings for LaTex output
plt.rcParams['text.usetex'] = True
plt.rcParams['text.latex.preamble'] = '\DeclareUnicodeCharacter{2212}{-}'
plt.rcParams['font.family'] = 'serif'
plt.rcParams['font.size'] = 16


# Create figure and axes
fig = plt.figure(figsize=(6, 4))
ax = fig.gca()


# Simulation
filename = sys.argv[1] if len(sys.argv) > 1 else './als_output.h5'
with HDF5ResultStorage(filename) as rs:
    detectors = rs.get_detectors()

    # Get x positions and retrieve 70th percentiles of x distribution for each detector
    hits = np.array(list(detectors.values()))
    xs = hits['position'][:, :, 0]
    x_percentiles = 2 * np.percentile(np.abs(xs), 70, axis=1)

    # Get z positions as the last dimension of the first entry of each detector
    # This is reasonable to do since we know our detectors detect at exactly one Z position
    zs = hits['position'][:, 0, -1]

    # Plot accordingly scaled values (z in mm, x percentiles in um)
    ax.plot(zs * 1e3, x_percentiles * 1e6, '-o', label='Simulation', zorder=-1)
    xs_ = np.abs(xs)


# Experiment
x_spheres = [1.41, 1.81, 2.21, 2.61, 3.01, 3.41, 3.81, 4.21, 4.61]
y_spheres = [49.9, 42.6, 38.3, 35.8, 35.8, 44.44, 50.62, 64.2, 69.1]
yerr_spheres = [1.84, 1.85, 1.74, 3.49, 3.49, 3.03, 3.49, 6.98, 9.24]
ax.errorbar(x_spheres, y_spheres, yerr_spheres, label='Experiment')


# Labeling, x/y limits and grid
ax.legend(loc='lower right')
ax.set_xlabel('$z$ distance from ALS exit [mm]')
ax.set_ylabel('70th percentile of $x$ positions [$\mu$m]')
ax.set_xlim(1.1, 4.9)
ax.set_ylim(0, 100)
ax.grid(axis='y')


# Show and save figure
plt.show()
fig.savefig('focus-curve-gold-27nm-uniform-zoom.pdf', bbox_inches="tight")
\end{lstlisting}
\end{widetext}

\section{Photophoretic force model}
This section provides a brief summary of the photophoretic force
model~\cite{Desyatnikov:OptExp17:8201} provided before~\cite{Welker:thesis:2019}. The photophoretic
force is implemented for a monochromatic Laguerre-Gaussian beam of order 1, a so-called LG01 beam.
Since such a beam's profile is radially symmetrical, its intensity distribution can be described in
$(r, z)$ coordinates:
\begin{equation}
  I(r, z) = \frac{P r^2}{\pi w(z)^4}\exp\left(-\frac{r^2}{w(z)^2}\right)
\end{equation}
where $P$ is the total beam power, $w(z)=w_0 \sqrt{1 + z^2/z_0^2}$ is the ring radius at $z$ for a
given beam waist radius $w_0$, and $z_0 = 2\pi w_0^2 / \lambda$ is the diffraction length for the
wavelength of the beam $\lambda$.

\begin{figure*}
  \includegraphics[width=0.75\linewidth]{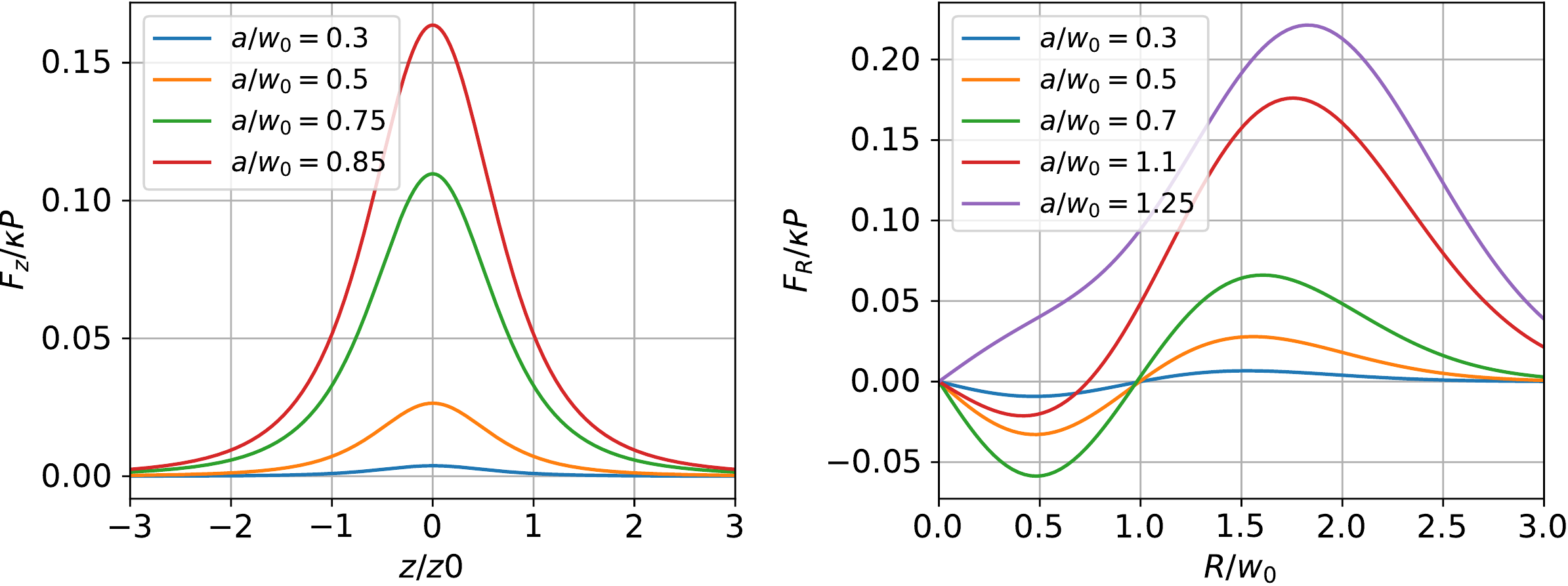}
  \caption{The axial and transverse components of the photophoretic force model by Desyatnikov et
     al~\cite{Desyatnikov:OptExp17:8201} as multiples of $\kappa P$. $a$ is the particle radius,
     $w_0$ is the beam waist radius, $R$ is the magnitude of the radial offset relative to the beam
     axis, and $z_0$ is the diffraction length of the Laguerre-Gaussian beam.}
  \label{fig:pp-force-curves}
\end{figure*}

Through various approximations~\cite{Desyatnikov:OptExp17:8201}, two expressions for the force's
axial and transverse components can be given. First, a phenomenological coefficient
$\kappa [m^{-1}]$~\cite{Desyatnikov:OptExp17:8201}, which absorbs thermal and optical parameters of
the particle and the surrounding gas is defined. $J_1$ is a scalar parameter measuring a
particle-specific asymmetry factor of the force, and is assumed to be $-\sfrac{1}{2}$ from here on.
$\mu_g$ is the dynamic viscosity of the gas, $a$ is the particle radius, $k_{p,g}$ refer to the
thermal conductivity of the particle and the gas, respectively, and $T_g$ and $\rho_g$ are the
temperature and density of the gas.

\begin{align}
  \label{eq:pp-kappa}
  \kappa := -J_1 \frac{9\mu_g^2}{2a\rho_g T_g(k_p + 2k_g)}
\end{align}
The transverse force component $F_R$ is then approximated as:
\begin{align}
  \label{eq:pp-fz}
  F_Z(z) = \kappa P f_Z\left(\frac{a^2}{w(z)^2}\right)\\
  f_Z(t) = 1 - (1+t)\exp(-t)
\end{align}
The approximation for the axial force component $F_Z$ is as follows:
\begin{align}
  \label{eq:pp-fr}
  F_R(r, z) &= -\frac{2 \kappa P}{\pi} \int_{0}^{a}\int_{-\sqrt{a^2 - x^2}}^{\sqrt{a^2 - x^2}}
              f_R(x, y, z, r, a)
              \,dy \,dx\\
  f_R(x, y, z, r, a) &= y \frac{x^2 + (y+r)^2}{w(z)^4} \frac{\exp\left(- \frac{x^2 + (y+r)^2}{w(z)^2}\right)}{\sqrt{a^2 - x^2 - y^2}}
\end{align}
Implementing this force, curves for varying particle radii $a$ and radial offsets $R$ are drawn in
\autoref{fig:pp-force-curves}, which match the ones published by Desyatnikov et
al~\cite{Desyatnikov:OptExp17:8201}.

\bibliography{string,cmi}